\documentclass[12pt,preprint]{aastex}
\usepackage[backref=page]{hyperref}

\newcommand{\noprint}[1]{}
\newcommand{\figsetstart}{{\bf Fig. Set} }
\newcommand{\figsetend}{}
\newcommand{\figsetgrpstart}{}
\newcommand{\figsetgrpend}{}
\newcommand{\figsetnum}[1]{{\bf #1.}}
\newcommand{\figsettitle}[1]{ {\bf #1} }
\newcommand{\figsetgrpnum}[1]{\noprint{#1}}
\newcommand{\figsetgrptitle}[1]{\noprint{#1}}
\newcommand{\figsetplot}[1]{\noprint{#1}}
\newcommand{\figsetgrpnote}[1]{\noprint{#1}}

\usepackage{apjfonts}
\usepackage[all]{hypcap}

\usepackage{float}
\usepackage{graphicx}
\usepackage{epsfig}
\usepackage[perpage,symbol*]{footmisc}
\usepackage{natbib}
\usepackage[figureright]{rotating}
\usepackage{longtable}
\usepackage{multirow}
\usepackage{paralist}
\usepackage{url}
\usepackage{epstopdf}
\usepackage{longtable}
\usepackage{lscape}

\shorttitle{A 6.7 GHz Methanol Maser Survey}
\shortauthors{Yang et al.}

\begin{document}

\title{A 6.7 GHz Methanol Maser Survey \uppercase\expandafter{\romannumeral2}. Low Galactic Latitudes}

\author{Kai Yang\altaffilmark{1,2}, Xi Chen\altaffilmark{1,3,4}, Zhi-Qiang Shen\altaffilmark{1,4}, Xiao-Qiong Li\altaffilmark{1,2}, Jun-Zhi Wang\altaffilmark{1,4}, Dong-Rong Jiang\altaffilmark{1,4}, Juan Li\altaffilmark{1,4}, Jian Dong\altaffilmark{1,4}, Ya-Jun Wu\altaffilmark{1,4}, Hai-Hua Qiao\altaffilmark{5,1}}

\altaffiltext{1}{Shanghai Astronomical Observatory, Chinese Academy of Sciences, 80 Nandan Road, Shanghai 200030, China; yangkai@shao.ac.cn, chenxi@shao.ac.cn, zshen@shao.ac.cn.}
\altaffiltext{2}{University of Chinese Academy of Sciences, 19A Yuquanlu, Beijing 100049, China.}
\altaffiltext{3}{Center for Astrophysics, Guangzhou University, Guangzhou 510006, China; chenxi@gzhu.edu.cn.}
\altaffiltext{4}{Key Laboratory of Radio Astronomy, Chinese Academy of Sciences, China.}
\altaffiltext{5}{National Time Service Center, Chinese Academy of Sciences, Xi'An, Shaanxi, 710600, China}

\begin{abstract}

 We report the results of our systematic survey for Galactic 6.7 GHz Class \uppercase\expandafter{\romannumeral2} CH$_3$OH maser emission toward a sample of young stellar objects. The survey was conducted with the Shanghai Tianma Radio Telescope (TMRT). The sample consists  of 3348 sources selected from the all-sky \emph{Wide-Field Infrared Survey Explorer (WISE)} point source catalog. We have discussed the selection criteria in detail and the detection results of those at high Galactic latitudes (i.e. $|b|>$ 2$^\circ$) in a previous paper (paper \uppercase\expandafter{\romannumeral1}). Here, we present the results from the survey of those at low Galactic latitudes, i.e. $|b|<$ 2$^\circ$. Of 1875 selected \emph{WISE} point sources, 291 positions that were actually associated with 224 sources were detected with CH$_3$OH maser emission. Among them, 32 are newly detected. Majority of the newly detected sources are associated with bright WISE sources. The majority of the detected sources (209/224 = 93.3\%) are quite close to the Galactic Plane ($|b|<$ 1$^\circ$) and lie on the inner spiral arms with positive LSR velocities. Detection rate and the color-color distribution of our detection are all matched with our anticipation. Combining with detections from previous surveys, we compile a catalogue of 1085 sources with 6.7 GHz CH$_3$OH maser emission in our Galaxy.

\end{abstract}


\section{Introduction}

Methanol (CH$_3$OH) masers are tightly associated with massive star-forming regions in our Galaxy. There are two classes of methanol masers, Class \uppercase\expandafter{\romannumeral1} and Class \uppercase\expandafter{\romannumeral2} \citep[]{Bat1987}. Unlike Class \uppercase\expandafter{\romannumeral1} methanol masers (e.g., transitions at 44 GHz and 95 GHz) which are pumped by collision excitation at different positions separated by $\sim$1 pc from high-mass young stellar objects \citep[YSOs,][]{Vor2010,Vor2014}, Class \uppercase\expandafter{\romannumeral2} methanol masers, pumped by infrared radiation, are commonly found close to YSOs \citep[]{Cas2010}. The 5$_1$ $\rightarrow$ 6$_0$ A$^+$ 6.7 GHz methanol maser is a well-known Class \uppercase\expandafter{\romannumeral2} methanol maser. It is widespread in the Milky Way and is the second strongest known maser, only exceeded by the 22 GHz water maser \citep[]{Men1991}. Different from other species of masers (OH, H$_2$O and SiO masers), the 6.7 GHz CH$_3$OH maser is exclusively associated with massive star-forming regions \citep[]{Min2003,Ell2006,Xu2008}, making it an excellent tool for studying massive star-forming regions \citep[e.g.][]{Ell2007}. In addition, the VLBI astrometry of the 6.7 GHz methanol maser can obtain the model-independent distance measurements for the study on the structures of our Galaxy \citep[]{Reid2009}.

Methanol masers at the 6.7 GHz transition have been detected towards more than 1,000 sources to date. Many surveys, including unbiased surveys and targeted surveys, have been conducted to search for the 6.7 GHz methanol maser \citep[e.g.][]{Mac1992,Cas1995,Cas1996,Ell1996,van1996,Wal1997,Ell2007}. As an unbiased survey, the Parkes Methanol Multibeam (MMB) Survey of a region of 186$^\circ$ $<|l|<$ 60$^\circ$ and $|b|<$ 2$^\circ$\citep[]{Cas2010,Gre2010,Cas2011,Gre2012,Bre2015} detected 954 sources, including 344 new detections.

Since 2015, we have performed a systematic 6.7 GHz Class \uppercase\expandafter{\romannumeral2} methanol maser survey toward targets selected from the all-sky \emph{Wide-Field Infrared Survey Explorer (WISE)} point source catalog. This WISE point catalogue is a relatively new database and covers the entire sky \citep[]{Wright2010}, which makes it ahead of several other infrared surveys and best candidate for us to construct a targeted sample. Our survey was conducted with the Shanghai Tianma Radio Telescope (TMRT). In a previous paper \citep[hereafter Paper \uppercase\expandafter{\romannumeral1}]{Yang2017}, we reported the survey results for the selected sources at high Galactic latitude ($|b|>$ 2$^\circ$); there are 3 new sources among the 12 detected sources. Here we present the remaining 6.7 GHz methanol maser survey results for those at low Galactic latitudes ($|b|<$ 2$^\circ$). We briefly describe the sample and observations in Section 2. The results are presented in Section 3, followed by discussions in Section 4 and a summary in Section 5.

\section{Sample and Observations}

The sample selection is fully detailed in Paper \uppercase\expandafter{\romannumeral1}. Here, we just recap some key points in constructing the sample from the WISE point catalog. We first pick out 473 WISE point sources which are associated with the MMB 6.7 GHz methanol maser catalog sources, with all the magnitudes available in the 4 WISE-bands. Because 265 sources from \citet{Bre2015} in the region of 20$^\circ$ $<|l|<$ 60$^\circ$ and $|b|<$ 2$^\circ$ were not published at that time, the MMB 6.7 GHz methanol maser catalog only contains 684 sources locate in the region of 186$^\circ$ $<|l|<$ 20$^\circ$ and $|b|<$ 2$^\circ$. Statistical analysis of the magnitude and color-color of these 473 sources shows that most (455/473=96\%) of the known CH$_3$OH masers meet the criteria of magnitude: $[3.4] < 14$ mag, $[4.6] < 12$ mag, $[12] < 11$ mag and $[22] < 5.5$ mag and the majority of them (330/473=73\%) meet the criteria of color: $[3.4]-[4.6]>2$, and $[12]-[22]>2$. 

Applying these selection criteria to the all-sky WISE catalog, we found about 13,000 candidate \emph{WISE} sources that may be associated with CH$_3$OH masers. After excluding those already detected in the Parkes MMB survey region (186$^\circ$ $<|l|<$ 20$^\circ$, $|b|<$ 2$^\circ$), we finally built up a sample of 3348 WISE sources with declination above $-$30$^\circ$. The sample can be divided into 2 sub-samples. One sub-sample includes 1473 sources which locate at high Galactic latitude region with $|b|>$ 2$^\circ$, and the results of the 6.7 GHz methanol survey have been published in Paper \uppercase\expandafter{\romannumeral1}. The other sub-sample consists of 1873 sources, which are located at the low Galactic latitude region with $|b|<$ 2$^\circ$ (Table 1).
 
We performed this 6.7 GHz methanol maser survey at low Galactic latitudes between 2016 June and 2018 January with the TMRT. The TMRT is a newly built and fully steerable radio telescope with a diameter of 65 m in Shanghai, China. A cryogenically cooled C-band receiver with a frequency range of 4$-$8 GHz and the Digital Backend System (DIBAS) were used to record signals. A spectral window of a bandwidth of 23.4 MHz was used to cover the rest frequency of the CH$_3$OH maser line, 6.6685192 GHz. This window has 16,384 channels with a spectral resolution of 1.431 kHz (or a velocity resolution is about 0.09 km s$^{-1}$). The system temperature is 20$-$30 K and the aperture efficiency of the TMRT is $\sim$ 55\%, resulting in a sensitivity of 1.5 Jy K$^{-1}$. The beam size is $\sim$ $3^{\prime}$ (HPBW) at the frequency of 6.7 GHz.
 
In our observations, position-switching mode was used and each source was observed in two ON$-$OFF cycles, with each cycle of 4 minutes. The OFF position was set to (0.0$^\circ$, $-$0.4$^\circ$) from the ON position in (R.A., decl.). We observed W3(OH) and NGC7538 as flux density calibrators with an uncertainty of less than 20\%.
 
The GILDAS/CLASS package was used to conduct the data reduction. We fitted and subtracted the linear baseline of the spectrum. The root-mean-square (rms) noise is about 50$-$80 mJy.
 
For those detected methanol maser sources with a wide velocity range of $>$30 km s$^{-1}$ (eight in total, see Section 3), we further carried out the on-the-fly (OTF) \citep{Don2016,Don2018a,Don2018b} mapping observation to determine positions of separated CH$_3$OH masers (see Section 3.2).

\section{Results} 

Single point observations with the TMRT detected 291 sources with the 6.7 GHz CH$_3$OH maser emission. In our survey, eight sources (G28.287$-$0.348, G30.788+0.203, G30.789+0.232, G30.823+0.134, G31.221+0.020, G31.253+0.003, G33.092$-$0.073, and G33.143$-$0.088) have a wide velocity range of $>$30 km s$^{-1}$ with methanol spectral features clearly separated. We further conducted OTF observations towards these sources to determine the maser positions of the different emission shown in their spectra. Results from the single point observations are described in Section 3.1, and the OTF observations in Section 3.2.

 \subsection{Detections from the single-point observations}

In total, TMRT survey of those at low Galactic latitudes detected 6.7 GHz CH$_3$OH masers from 291 positions which are actually from 224 sources because some close positions show similar methanol spectral features and thus are identified with a same origin. Comparison with previous surveys showed that 32 sources are newly detected. Figure 1 displays the spectra of these newly-identified CH$_3$OH masers and their properties are detailed in Table 2. The spectra and properties of previously detected sources (192 in total) are shown in Figure 2 and Table 3, respectively.

For the eight sources from our survey with a very wide velocity range ($>$30 km s$^{-1}$, see above), we can clearly see that the left and right parts of their spectra locate at separated positions from different regions after taking our OTF observation into account (see Section 3.2), so we only discuss their separated parts instead of their whole spectra and we name the different part of their spectra by adding the position in the spectra after their Galactic coordinate as one source, like G31.221+0.020(Left).

Each velocity component of the 6.7 GHz CH$_3$OH maser usually has a velocity range of 0.1$-$3 km s$^{-1}$ \citep[]{Bar2016}, so sources showing wide velocity ranges may have more components. The smallest velocity range of the detected sources is 0.47 km s$^{-1}$ (G43.089$-$0.011 and G75.010+0.274) with a single feature while the largest is 24.4 km s$^{-1}$ (G43.148+0.013). The median velocity range is 5.5 km s$^{-1}$ and 11 sources have a velocity range larger than 16 km s$^{-1}$.

The median peak flux density of our 224 detected sources is 3.0 Jy. G23.010$-$0.410 has the strongest peak flux density of 406.2 Jy. The weakest peak flux density is 0.23 Jy from source G30.980+0.216. There are 36 sources having a peak flux density higher than 20 Jy and 59 sources with a peak flux density lower than 1.0 Jy. Among the 32 newly detected sources, the strongest source is G82.308+0.729 and the weakest source is G25.177+0.211 with a peak flux density of 58.4 and 0.31 Jy, respectively. Most of them (28/32) have a peak flux density lower than 2.0 Jy.

 Among the 192 previously detected sources, 167 sources were detected by the Parkes MMB survey \citep{Bre2015}. The remaining 25 sources were detected by other surveys. Since our observations were conducted with the single-dish telescope, the positions are not exactly same as the previous detections. Thus, we only present the information on the variations of their spectral profiles. The details of their spectral profile changes are listed in Table 4.

  \subsection{Maser positions determined from the OTF observations}

 Eight sources (described above in Section 3.1) show a wide velocity range in their methanol spectra, we observed them with the OTF mode in order to figure out the maser distribution due to their obviously separated spectral features in left and right parts. Among them, the left and right parts of G30.788+0.203 and G30.789+0.232, G31.221+0.020 and G31.253+0.003, G33.092$-$0.073 and G33.143$-$0.088 share similar features respectively, so we observed toward five positions with the OTF observations. The center coordinates and the side length of the square regions chosen for the OTF observations towards five positions are listed in Table 5. We also list the maser positions determined from the OTF mapping observations in each region along with the positional offset of associated MMB sources decided by ATCA in Table 5. Taking the pointing error of TMRT ($\sim$10$^{\prime\prime}$) and the fitting error from the OTF observation ($\sim$10$^{\prime\prime}$) into consideration, our TMRT OTF observations have a positional accuracy better than 20$^{\prime\prime}$. Comparing with the positions determined by the ATCA observations, we found that the positional accuracy achieved from our TMRT OTF observations is better than 10$^{\prime\prime}$ towards six sources (see Table 5). The other three maser sources show larger position offsets (10$^{\prime\prime}$ $\sim$ 20$^{\prime\prime}$) with regard to the ATCA measurements, but this is consistent with the estimated positional accuracy of less than 20$^{\prime\prime}$ from the TMRT OTF mapping observations. Figure 3 shows the velocity-integrated intensity map from the OTF observation. We can see that the left and right parts (middle parts) of each source are all from two (three) different positions and we will discuss them as different sources.

\section{Discussions}

 \subsection{Newly detected sources}
 Although the Parkes MMB survey is an unbiased survey,  we still detected 18 new sources towards the MMB region (20$^\circ$ $<|l|<$ 60$^\circ$, $|b|<$ 2$^\circ$) also covered by the MMB survey. The variability of the 6.7 GHz methanol masers may contribute to the new detections. Several proposed mechanisms may explain its variability: (i) pulsation of a young high-mass star \citep[]{Ina2013, San2015}, (ii) rotating spiral shocks from hot and dense material in the central gap of the circumbinary accretion disc \citep[]{Par2014}, (iii) periodic accretion of a protostar or accretion disc \citep[]{Ara2010}, (iv) a colliding-wind binary (CWB) system \citep[]{Wal2009, Wal2016}, (v) an eclipsing binary \citep[]{Mas2015}. Recently, there are several luminous bursts in high-mass star formation \citep[e.g.][]{Car2017, Hun2017, Hun2018, Szy2018}, episodic accretion due to disk fragmentation \citep[]{Mey2017} may be a dominant mechanism for such bursts. Notably, the spectral profile of the 6.7 GHz CH$_3$OH maser can even change in a timescale of a few days \citep[]{Cas1995b}. In addition, these new sources may also be due to a better sensitivity of the TMRT survey with a rms noise of 0.05 Jy compared to a typical rms noise level of 0.07 Jy in the MMB survey and 0.06 Jy to 4 Jy of other previous surveys.
 
 For the 32 newly detected sources, we show their infrared WISE three-color images in Figure 4. Five sources (G24.362$-$0.146, G31.253+0.003(Left), G35.225$-$0.360, G49.537$-$0.904, and G75.010+0.274) locate in the infrared dark cloud. Most of the newly detected sources, $27/32=84.4\%$, are associated with bright WISE sources. There are also six sources (G76.093+0.158, G78.969+0.541, G84.193+1.439, G84.951$-$0.691, G84.984$-$0.529, and G124.015$-$0.027) that are bright in the infrared environment but not in the infrared dark cloud. The remaining 21 sources have bright point counterparts in the infrared dark cloud.
 
The ATLASGAL (APEX Telescope Large Area Survey of the Galaxy) survey imaged the Galactic plane in a region of $-$60$^\circ$ $<|l|<$ 60$^\circ$ and $|b|<$ 1.5$^\circ$ at 870 $\mu$m with the APEX Telescope in its first step \citep[]{Sch2009}. \citet[]{Urq2014} combined the ATLASGAL survey and the Red Midcourse Space Experiment Source (RMS) surveys, and identified a sample which contains $\sim$ 1300 clumps associated with $\sim$ 1700 embedded massive young protostars. We have 18 maser sources located in the overlapped region of 20$^\circ$ $<|l|<$ 60$^\circ$, $|b|<$ 1.5$^\circ$ from the 32 newly detected sources. Among these 18 sources, 12 sources have been found to be associated clumps (see Table 2). After excluding extended emission with a multiscale decomposition tool and using a Gaussian source-fitting algorithm (MRE-GCL), \citet[]{Cse2014} identified 10861 compact submillimeter sources also from the ATLASGAL survey in the region of $|l|<$ 60$^\circ$, $|b|<$ 1.5$^\circ$ and $-$80$^\circ$ $< l <$ $-$60$^\circ$, $-$2$^\circ$ $< b <$ 1$^\circ$. 14 sources are found associated sources among the 18 sources. In addition to this, \citet[]{Urq2013} presented a sample of molecular clumps containing compact and ultracompact H \uppercase\expandafter{\romannumeral2} regions in the area of 10$^\circ$ $< l <$ 60$^\circ$, $b <$ 1$^\circ$ by combining the ATLASGAL survey and the CORNISH (Co-Ordinated Radio 'N' Infrared Survey for High-mass star formation) project. But we have not found any associated regions. There have been many surveys conducted by the Very Large Array (VLA) at 6cm \citep[]{Law2008, Urq2009} searching for compact sources, and several observations conducted by the Atacama Large Millimetre/submllimetre Array (ALMA) towards the sources from the ATLASGAL survey which are considered to be young massive clumps \citep[]{Chi2017, Cse2017a, Cse2017b, Cse2018}, but we have not found sources associated with the sources we newly detected. In conclusion, our newly detected sources may be in the early evolutionary stage of star-formation, and the successive observation of their variation will also give us a better understanding of this period of star forming.
 
 \subsection{Detection rate versus Galactic longitude and WISE color}
 
 \subsubsection{Detection rate}
 In our survey, a detection rate of $224/1875=11.9\%$ was achieved towards the WISE$-$selected sources with magnitudes and colors described in Section 2. We can derive an expected detection rate using the methanol maser and WISE-selected source data in the MMB survey region. For a singe dish survey, such as the Parkes MMB survey, any targets that lie within the FWHM of the telescope could be detected by a single point observation. Moreover, since the complicated background emission around the Galactic center might affect WISE source selection, we exclude the WISE sources towards the Galactic center region with $|l|<10^\circ$. As such, the number of selected WISE sources is about 3,700 in the above-mentioned MMB survey region (186$^\circ$ $<|l|<$ 20$^\circ$, $|b|<$ 2$^\circ$, and without the Galactic center) after taking the beam coverage factor into account. Within such a region, there are 280 CH$_3$OH maser sources detected by the MMB survey which meet the WISE selection criteria, while the total number of the 6.7 GHz CH$_3$OH maser sources which have WISE sources with 4-band flux density measurements is 388. The total number of the 6.7 GHz CH$_3$OH masers in this region is 528. Thus we can derive a detection rate of $280/3700\times528/388=10.3\%$ towards the WISE sources with the criteria. Therefore, the actual detection rate ($11.9\%$) in our survey is quite consistent with the expected rate ($10.3\%$) using the blind MMB survey data.
  
Our survey and the Parkes MMB survey have an overlap region, 20$^\circ$ $<|l|<$ 60$^\circ$, $|b|<$ 2$^\circ$. The Parkes MMB survey detected 265 sources in this region. Among the 189 sources we detected in this region, there are 167 sources associated with MMB detections. From our sample selection in Paper \uppercase\expandafter{\romannumeral1}, almost all (96\%) of the known methanol masers meet the magnitude criteria and 73\% of the them meet the color criteria, the proportion of our detected sources that are associated with MMB sources in the MMB sources is $96\%\times73\%=70.1\%$. The true rate is $167/265=63.0\%$ which is less than the estimated rate. 22 new sources we detect are mainly due to our better sensitivity. If considering these 22 sources, the actual detection rate will be $189/265=71.3\%$, which is well consistent with the estimated rate.

 \subsubsection{Detection rate versus Galactic longitude}
To investigate the relationship between the detection rate and Galactic longitude distribution of our survey, we divided our 1875 sample sources and 224 detected CH$_3$OH masers along their Galactic longitudes and calculated the detection rate every 5$^\circ$ bin of longitude. A histogram of sample source number and detection number are shown in the upper panel of Figure 5. The corresponding detection rate is given in the lower panel of Figure 5. Overall, it is clearly seen that the detection rate of methanol maser decreases with the Galactic longitude from 20$^\circ$ to 180$^\circ$. The sources in the Galactic center region of 20$^\circ$ $-$ 40$^\circ$ are located in the 1st Quad molecular cloud where the inner spiral arms lie on \citep[]{Dame2001}, thus having strong star formation activity and near distance will result in high appearance chance of methanol maser. The number becomes much lower in other regions. The Cygnus X and Cas A molecular clouds around the Galactic longitude of 80$^\circ$ and 110$^\circ$ contribute to the appearance of several methanol sources and the increase of the detection rate. These all show that the detection rate and the distribution of CH$_3$OH maser sources are both in good agreement with the presence of molecular clouds where high mass star-forming regions rise. Due to the small number of sample sources at Galactic longitude of $>$$120^\circ$, we do not discuss the detection rate.
 
 \subsubsection{Detection rate versus WISE color}
 Since our sample sources are selected with a certain criteria of the WISE color, $[3.4]-[4.6]>2$ and $[12]-[22]>2$, we can compare the color-color diagram of the detected sources and the sample sources in the upper panel of Figure 6. From the distribution of the black dots (sample sources) and the red dots (detected sources), we found they almost share the similar infrared characteristic. In order to study their infrared environment of our detected sources, the numbers of sample sources and detected sources and detection rate along every 0.2 color are shown in the lower panels of Figure 6. The numbers of sample sources decrease with [3.4]$-$[4.6] color, but have a peak at $\sim$3.3 with [12]$-$[22] color. The number of detected sources show the similar distribution of the number of sample sources. The detection rates along [3.4]-[4.6] color have a fluctuation around 15\% and those along [12]-[22] color show a rising trend.

 \subsection{Galactic distribution}
 
 \subsubsection{Our detection}
 
 The Galactic latitude distribution of our detected 224 sources is shown in Figure 7. Because the region of $186^\circ$ $\leq$ $l$ $\leq$ $20^\circ$  had been included by the Parkes MMB survey before our sample selection, we observed the rest region with 20$^\circ$ $<|l|<$ 186$^\circ$ and $|b|<$ 2$^\circ$. Most of the detected sources, $208/224=92.9\%$, locate at regions of $|b|<1^\circ$. Only 16 sources have a Galactic latitude larger than 1$^\circ$. Among them, 3 sources (G84.193+1.439, G89.930+1.669 and G99.070+1.200) are newly detected. In the MMB survey, there are $893/954=93.6\%$ sources with a Galactic latitude of $|b|<1^\circ$, which is consistent with our survey.
 
 The LSR velocities of our detected sources versus Galactic longitude are shown in Figure 8 (middle, blue and red dots). As 6.7 GHz methanol maser is only associated with high-mass star-forming regions and thus is a good tracer of spiral structures in the Milky Way \citep[]{Reid2014, Reid2016}, the detected sources can be derived from different spiral arms. After combining the estimated kinematic distances derived from the Galactic rotation model in \citet[]{Reid2014}, we found that sources in the $20^\circ$ $-$ $60^\circ$ longitude region have positive LSR velocities and they are mostly stay on the spiral arms of Perseus, Carina-Sagittarius, Crux-Scutum and Norma. According to the estimated distances, the positions of these detected sources are plotted in Figure 9. Sources in the $60^\circ$ $-$ $186^\circ$ longitude region mostly have negative LSR velocities and they are mainly locate on the spiral arms of Perseus and Norma.

 \subsubsection{All known 6.7 GHz methanol maser sources}
 
 Many Galactic 6.7 GHz CH$_3$OH maser surveys have been conducted in the last 20 years. By combining survey results from the literature \citep[]{Pes2005,Ell2007,Pan2007,Xu2008,Cyg2009,Cas2010,Cas2011,Gre2010,Gre2012,Szy2012,Olm2014,Bre2015,Yang2017}, we compiled a catalogue of 1085 sources at 6.7 GHz Class \uppercase\expandafter{\romannumeral2} CH$_3$OH masers (Table 6). Their galactic distribution, LSR velocity as a function of Galactic longitude, and the source number count as a function of Galactic latitude and longitude are presented in Figure 8.
 
 Among these 1085 6.7 GHz methanol masers, there are 404 sources located in the $0^\circ$ $-$ $40^\circ$ longitude region, and 407 in the opposite longitude $320^\circ$ $-$ $360^\circ$ region ($-40^\circ$ $-$ $0^\circ$). As a result, most of the 6.7 GHz methanol masers, $811/1085=74.7\%$, are located in the above two regions and the maser distribution is the most abundance in the regions. From the distribution of the LSR velocity versus Galactic longitude in Fig. 8 (middle), we find that most of the sources within the $0^\circ$ $-$ $40^\circ$ longitude region show positive velocities and most sources in the $-40^\circ$ $-$ $0^\circ$ longitude region have negative velocities. Sources in these two regions show peak velocities at Galactic longitude of $\sim$$30^\circ$ and $\sim$$-30^\circ$, respectively. These two opposite Galactic longitude regions locate near the Galactic center including the inner and the most compact parts of some main spiral arms in our Galaxy. Sources with the peak velocity at Galactic longitude of $\sim$$30^\circ$ and $\sim$$-30^\circ$ locate on the Crux-Scutum arm and Norma arm, respectively. The peak may trace the interacting region of the Galactic bar and the Crux-Scutum arm and Norma arms. Due to the 344 newly detected sources in the Parkes MMB survey \citep[]{Cas2010,Gre2010,Cas2011,Gre2012,Bre2015}, there are now more sources in the Galactic region of $-20^\circ$ $-$ $20^\circ$, compared to the Galactic longitude distribution in \citet[]{Pes2005}.
 
 Most of the known 6.7 GHz methanol masers (1068/1085=98.4\%) are located in the regions of $|b|<2^\circ$, and they are mostly found around $|b|<1^\circ$, revealing that 6.7 GHz methanol masers prefer to being associated with the star formation in the Galactic Plane. Besides the previous surveys were mostly conducted towards the Galactic Plane, the star forming cluster environment is much more complicated at low latitudes, hence much more gases and dusts which are essential for star-forming activity. Therefore, it should be the case that the most 6.7 GHz CH$_3$OH masers are located at low latitudes around the Galactic Plane. We simulated a Gaussian fit with an FWHM of 0.56$^\circ$ to the Galactic latitude distribution of methanol masers, which is consistent with the Gaussian fit with an FWHM of 0.52$^\circ$ in \citet[]{Pes2005}.

 \section{Summary}
 With the newly built TMRT, we performed a systematic survey of 6.7 GHz CH$_3$OH masers towards 1875 sources at low Galactic latitudes ($|b|<2^\circ$). These sources are selected from the all-sky WISE point catalog with a certain criteria of magnitude and color. There are 32 new detections among 224 detected sources. Their spectra have peak flux densities of 0.23 Jy to 406.2 Jy and their velocity ranges have a range of 0.47 km s$^{-1}$ to 24.4 km s$^{-1}$. The source number along the Galactic latitude shows that most of them are located at even lower Galactic latitude, $|b|<1^\circ$. Their distribution and LSR velocities are associated with the presence of spiral arms and molecular clouds. The detection rates along the Galactic longitude and color meet our anticipation. We also present a compiled catalog containing all 1085 6.7 GHz methanol masers known to date. 
 
\section*{Acknowledgements}

We are thankful for the assistance from the operators of the TMRT during the observations. This work was supported by the National Natural Science Foundation of China (11590780, 11590781, 11590784 and 11873002), the Knowledge Innovation Program of the Chinese Academy of Sciences (Grant No. KJCX1-YW-18), the Scientific Program of Shanghai Municipality (08DZ1160100), and Key Laboratory for Radio Astronomy, CAS. K. Yang would like to thank the China Scholarship Council (CSC) for the support. H.-H. Q. is partially supported by the Special Funding for Advanced Users, budgeted and administrated by Center for Astronomical Mega-Science, Chinese Academy of Sciences (CAMS-CAS) and CAS ``Light of West China" Program.

\newpage
\begin{sidewaystable}[h]
\centering
{\scriptsize
\caption{WISE$-$Selected sources for the TMRT survey of methanol maser}
\centerline{\begin{tabular}{ccccccccccc}
\hline
 Number & l & b & R.A. & Decl. & w1 (3.4 $\mu m$) & w2 (4.6 $\mu m$) & w3 (12 $\mu m$) & w4 (22 $\mu m$) & w1$-$w2 & w3$-$w4 \\
 & ($^{\circ}$) & ($^{\circ}$) & (J2000) & (J2000) & (mag) & (mag) & (mag) & (mag) & (mag) & (mag) \\
 &  &  & (h m s) & ($^{\circ}$ $^{\prime}$ $^{\prime\prime}$) &  &  &  &  &  &  \\
 (1) & (2) & (3) & (4) & (5) & (6) & (7) & (8) & (9) & (10) & (11) \\
\hline
 1 & 20.234 & 0.085 & 18:27:39.95 & $-$11:14:32.3 & 12.940 & 8.873 & 6.925 & 3.186 & 4.067 & 3.739 \\
 2 & 20.363 & $-$0.014 & 18:28:16.02 & $-$11:10:25.4 & 13.167 & 10.206 & 5.805 & 1.167 & 2.961 & 4.638 \\
 3 & 20.762 & $-$0.064 & 18:29:12.19 & $-$10:50:36.0 & 6.458 & 4.428 & 0.890 & $-$1.821 & 2.030 & 2.711 \\
 4 & 20.926 & $-$0.050 & 18:29:27.80 & $-$10:41:28.8 & 11.144 & 7.433 & 4.516 & 1.576 & 3.711 & 2.940 \\
 5 & 21.023 & $-$0.063 & 18:29:41.54 & $-$10:36:42.5 & 13.182 & 9.106 & 4.360 & 1.017 & 4.076 & 3.343 \\
 6 & 21.370 & $-$0.226 & 18:30:56.20 & $-$10:22:48.7 & 9.588 & 7.408 & 5.919 & 2.130 & 2.180 & 3.789 \\
 7 & 22.050 & 0.211 & 18:30:38.52 & $-$09:34:28.2 & 14.712 & 10.219 & 7.869 & 3.357 & 4.493 & 4.512 \\
 8 & 22.355 & 0.066 & 18:31:44.05 & $-$09:22:17.3 & 9.174 & 7.028 & 3.763 & 0.036 & 2.146 & 3.727 \\
 9 & 23.010 & $-$0.410 & 18:34:40.25 & $-$09:00:38.2 & 11.475 & 6.966 & 6.050 & 0.072 & 4.509 & 5.978 \\
 10 & 23.185 & $-$0.380 & 18:34:53.23 & $-$08:50:27.4 & 13.677 & 10.738 & 8.320 & 5.712 & 2.939 & 2.608 \\
  &  &  &  &  & ............ &  &  &  &  &  \\
\hline 
\end{tabular}}}
\tablecomments {
\footnotesize
  Column 1: number; Columns 2 - 5: positions (gl, gb, R.A., decl.); Columns 6 - 9: magnitudes in the 4 \emph{WISE} bands; Columns 10 - 11: the \emph{WISE} color. (This table is available in its entirety in machine-readable form.)}
\end{sidewaystable}
\clearpage

{
\tiny
\begin{landscape}
  \begin{longtable}{ccccccccccc}
  \caption{32 newly detected 6.7 GHz methanol masers.}
  \tabletypesize{\tiny}
  \tablewidth{0pt} \\
  \hline 
   Name & R.A. & DEC & $\Delta V$ & $v_p$ & $S_p$ & $S_i$ & Epoch & Distance & Only one & Other name \\
   (gl, gb)  & (J2000) & (J2000) & (km s$^{-1}$) & (km s$^{-1}$) & (Jy) & (Jy km s$^{-1}$) & yy/mm/dd & kpc & features? & \\
   ($^{\circ}$, $^{\circ}$)  & (h m s) & ($^{\circ}$ $^{\prime}$ $^{\prime\prime}$) &  &  &  &  &  &  &  & \\
   (1) & (2) & (3) & (4) & (5) & (6) & (7) & (8) & (9) & (10) & (11) \\
  \hline
 G24.273$-$0.137 & 18:36:02.31 & $-$07:45:48.5 & 88.7, 101.4 & 100.9 & 1.04 & 0.44 & 16/08/07 &  &  & \\
 G24.313$-$0.154 & 18:36:10.51 & $-$07:44:06.1 & 99.6, 101.5 & 100.9 & 2.03 & 0.90 & 16/08/07 & 5.91 & N & AGAL024.311$-$00.154/G024.3123$-$0.1543 \\
  & & & & & & & & & & \\ 
 G24.362$-$0.146 & 18:36:14.34 & $-$07:41:19.3 & 54.8, 55.9 & 55.3 & 1.86 & 0.70 & 16/08/07 & 3.67 & Y & \\
  & & & & & & & & & & \\
 G24.633+0.153 & 18:35:40.11 & $-$07:18:34.8 & 112.7, 118.7 & 114.3 & 1.22 & 0.45 & 16/09/05 & 7.36 & N & AGAL024.633+00.152/G024.6316+0.1526 \\
  & & & & & & & & & & \\ 
 G25.177+0.211 & 18:36:28.20 & $-$06:48:00.7 & 99.9, 102.1 & 100.5 & 0.31 & 0.24 & 16/09/05 & 8.96 & N & AGAL025.178+00.211/G025.1774+0.2102 \\
  & & & & & & & & & & \\
 G28.804$-$0.023 & 18:43:58.88 & $-$03:40:59.6 & 100.3, 102.6 & 101.3 & 0.54 & 0.62 & 16/09/16 & 7.96 & N & AGAL028.802$-$00.022/G028.8030$-$0.0214 \\
  & & & & & & & & & & \\ 
 G29.281$-$0.330 & 18:45:56.85 & $-$03:23:56.4 & 91.6, 93.1 & 92.4 & 1.86 & 1.09 & 16/09/16 & 4.84 & Y & AGAL029.281$-$00.331/G029.2819$-$0.3313 \\
  & & & & & & & & & & \\ 
 G31.221+0.020(Left) & 18:48:14.70 & $-$01:30:49.1 & 79.6, 80.3 & 80.1 & 0.41 & 0.15 & 16/08/03 & 4.66 & Y & AGAL031.221+00.021/G031.2210+0.0203 \\ 
  & & & & & & & & & & \\ 
 G31.253+0.003(Left) & 18:48:21.93 & $-$01:29:36.0 & 36.7, 42.0 & 37.4 & 0.50 & 0.25 & 16/08/03 & 2.06 & N & AGAL031.251+00.002/G031.2511+0.0040 \\ 
  & & & & & & & & & & \\
 G33.229$-$0.018 & 18:52:02.71 & +00:15:20.6 & 92.1, 98.2 & 92.4 & 0.48 & 0.15 & 16/08/19 & 6.22 & N & G033.2278$-$0.0179 \\
  & & & & & & & & & & \\ 
 G33.425$-$0.315 & 18:53:27.46 & +00:17:41.7 & 45.3, 46.4 & 46.0 & 0.55 & 0.28 & 16/09/22 & 2.20 & Y & \\
  & & & & & & & & & & \\   
 G35.149+0.809 & 18:52:36.09 & +02:20:31.6 & 71.7, 72.5 & 72.0 & 0.68 & 0.23 & 16/09/22 & 4.66 & Y & AGAL035.149+00.809/G035.1493+0.8099 \\
  & & & & & & & & & & \\ 
 G35.225$-$0.360 & 18:56:54.24 & +01:52:35.7 & 59.1, 60.0 & 59.7 & 0.62 & 0.23 & 16/09/22 & 10.21 & Y & AGAL035.226$-$00.359/G035.2266$-$0.3587 \\
  & & & & & & & & & & \\   
 G43.089$-$0.011 & 19:10:09.53 & +09:01:26.8 & 8.80, 9.27 & 9.06 & 0.33 & 0.10 & 16/11/14 & 11.46 & Y & \\
  & & & & & & & & & & \\ 
 G49.537$-$0.904 & 19:25:42.16 & +14:18:22.1 & 39.2, 41.9 & 39.6 & 0.63 & 0.20 & 16/09/11 & 3.25 & N & G049.5444$-$0.8830 \\
  & & & & & & & & & & \\ 
 G53.485+0.521 & 19:28:19.07 & +18:27:22.2 & 54.0, 54.5 & 54.2 & 0.32 & 0.10 & 16/09/29 & 4.04 & Y & AGAL053.496+00.522/G053.4850+0.5221 \\
  & & & & & & & & & & \\ 
 G54.371$-$0.613 & 19:34:18.73 & +18:41:13.8 & 34.9, 38.2 & 37.9 & 0.81 & 0.32 & 17/03/06 & 3.92 & N & AGAL054.373$-$00.614/G054.3721$-$0.6140 \\
  & & & & & & & & & & \\ 
 G59.436+0.820 & 19:39:34.20 & +23:48:25.5 & $-$50.4, $-$47.5 & $-$50.2 & 0.35 & 0.10 & 16/09/29 & 11.95 & N & \\
  & & & & & & & & & & \\ 
 G59.498$-$0.236 & 19:43:42.45 & +23:20:13.8 & 27.1, 28.5 & 27.5 & 0.62 & 0.50 & 16/08/19 & 3.49 & N & AGAL059.497$-$00.236/G059.4972$-$0.2358 \\
  & & & & & & & & & & \\ 
 G62.310+0.114 & 19:48:35.35 & +25:56:41.8 & 23.4, 25.9 & 23.6 & 0.46 & 0.13 & 16/08/04 & 3.19 & N & \\
  & & & & & & & & & & \\ 
 G74.098+0.110 & 20:17:56.32 & +35:55:24.3 & $-$6.63, 1.03 & $-$0.22 & 5.19 & 2.85 & 16/09/24 & 4.25 & N & \\
  & & & & & & & & & & \\  
 G75.010+0.274 & 20:19:49.29 & +36:46:09.4 & 3.06, 3.53 & 3.36 & 0.53 & 0.11 & 16/09/25 & 3.84 & Y & \\
  & & & & & & & & & & \\  
 G76.093+0.158 & 20:23:23.65 & +37:35:34.3 & 4.27, 6.90 & 4.92 & 0.49 & 0.22 & 16/10/02 & 3.72 & N & \\
  & & & & & & & & & & \\  
 G78.969+0.541 & 20:30:22.77 & +40:09:23.2 & 4.01, 5.48 & 4.74 & 1.24 & 0.88 & 16/09/16 & 2.87 & N & \\
  & & & & & & & & & & \\ 
 G81.794+0.911 & 20:37:47.39 & +42:38:39.0 & $-$4.73, 7.54 & 7.19 & 0.59 & 0.14 & 16/08/18 & 2.87 & N & \\
  & & & & & & & & & & \\
 G82.308+0.729 & 20:40:16.72 & +42:56:28.6 & 10.0, 11.3 & 10.3 & 58.4 & 21.1 & 16/07/30 & 2.55 & Y & \\
 G82.317+0.689 & 20:40:28.99 & +42:55:24.5 & 10.0, 11.3 & 10.4 & 1.74 & 0.60 & 16/07/30 &  &  & \\
  & & & & & & & & & & \\ 
 G84.193+1.439 & 20:43:36.36 & +44:51:52.1 & $-$2.80, $-$2.14 & $-$2.38 & 0.49 & 0.09 & 16/08/18 & 2.54 & Y & \\
  & & & & & & & & & & \\ 
 \hline
   Name & R.A. & DEC & $\Delta V$ & $v_p$ & $S_p$ & $S_i$ & Epoch & Distance & Only one & Other name \\
   (gl, gb)  & (J2000) & (J2000) & (km s$^{-1}$) & (km s$^{-1}$) & (Jy) & (Jy km s$^{-1}$) & yy/mm/dd & kpc & features? & \\
   ($^{\circ}$, $^{\circ}$)  & (h m s) & ($^{\circ}$ $^{\prime}$ $^{\prime\prime}$) &  &  &  &  &  &  &  & \\
   (1) & (2) & (3) & (4) & (5) & (6) & (7) & (8) & (9) & (10) & (11) \\
  \hline
 G84.951$-$0.691 & 20:55:32.50 & +44:06:10.2 & $-$37.0, $-$25.0 & $-$36.5 & 5.35 & 1.52 & 16/09/05 & 5.35 & N & \\
  & & & & & & & & & & \\ 
 G84.984$-$0.529 & 20:54:58.33 & +44:13:58.3 & $-$39.1, $-$37.9 & $-$38.5 & 0.53 & 0.25 & 16/09/05 & 5.39 & N & \\
  & & & & & & & & & & \\   
 G89.930+1.669 & 21:04:15.41 & +49:24:27.4 & $-$70.6, $-$69.6 & $-$69.9 & 0.53 & 0.18 & 16/08/03 & 7.93 & Y & \\
  & & & & & & & & & & \\
 G99.070+1.200 & 21:49:40.65 & +55:24:51.8 & $-$67.2, $-$66.1 & $-$66.6 & 0.48 & 0.27 & 16/09/07 & 6.31 & N & \\
  & & & & & & & & & & \\ 
 G124.015$-$0.027 & 01:00:55.42 & +62:49:29.5 & $-$41.0, $-$39.7 & $-$40.3 & 1.54 & 0.71 & 16/08/05 & 2.62 & N & \\
  & & & & & & & & & & \\ 
 G149.076+0.397 & 04:01:36.49 & +53:19:41.4 & $-$39.7, $-$37.9 & $-$38.6 & 1.75 & 0.68 & 16/08/14 & 4.73 & N & \\
 \hline
  \end{longtable}
   \tablecomments {Column 1: source name; Columns 2 - 3: the targeted positions for the TMRT observations; Column 4: the velocity interval of the maser emission; Column 5: the velocity of peak emission; Column 6: the peak flux density; Column 7: the integrated flux density; Column 8: the epoch of observation; Column 9: estimated kinematic distances by using the Galactic rotation model (http://bessel.vlbi-astrometry.org/bayesian); Column 10: whether the spectra of the sources only have one feature? Y: yes; N: no; Column 11: name of associated source in \citet[]{Urq2014,Cse2014}.}
\end{landscape}
}

\newpage
\begin{deluxetable}{cccccccccc}
\tabletypesize{\tiny}
\tablecolumns{10} 
\tablewidth{0pt}
\tablecaption{192 already detected 6.7 GHz methanol masers. \label{table: already detected}}
\tablehead{
 \colhead{Name} & \colhead{R.A.} & \colhead{DEC} & \colhead{$\Delta V$} & \colhead{$v_p$} & \colhead{$S_p$} & \colhead{$S_i$} & \colhead{Epoch} & \colhead{Distance} & \colhead{Ref.} \\
 \colhead{(gl, gb)}  & \colhead{(J2000)} & \colhead{(J2000)} & \colhead{(km s$^{-1}$)} & \colhead{(km s$^{-1}$)} & \colhead{(Jy)} & \colhead{(Jy km s$^{-1}$)} & \colhead{yy/mm/dd} & \colhead{kpc} & \colhead{} \\
 \colhead{($^{\circ}$, $^{\circ}$)}  & \colhead{(h m s)} & \colhead{($^{\circ}$ $^{\prime}$ $^{\prime\prime}$)} & \colhead{} & \colhead{} & \colhead{} & \colhead{} & \colhead{} & \colhead{} & \colhead{}\\
 \colhead{(1)} & \colhead{(2)} & \colhead{(3)} & \colhead{(4)} & \colhead{(5)} & \colhead{(6)} & \colhead{(7)} & \colhead{(8)} & \colhead{(9)} & \colhead{(10)} 
 }
\startdata
 G20.234+0.085 & 18:27:39.95 & $-$11:14:32.3 & 59.7, 77.7 & 73.2 & 22.9 & 14.1 & 16/07/04 & 5.21 & Parkes  \\
 G20.315+0.072 & 18:27:52.06 & $-$11:10:35.8 & 71.4, 74.2 & 73.3 & 1.05 & 0.55 & 16/07/04 &  &  \\
  & & & & & & & & & \\
 G20.363$-$0.014 & 18:28:16.02 & $-$11:10:25.4 & 49.6, 59.6 & 56.2 & 5.05 & 9.65 & 16/07/04 & 3.47 & Parkes  \\
  & & & & & & & & & \\
 G20.762$-$0.064 & 18:29:12.19 & $-$10:50:36.0 & 53.7, 64.0 & 61.2 & 2.09 & 8.58 & 16/07/04 & 3.75 & Parkes  \\
  & & & & & & & & & \\ 
 G20.926$-$0.050 & 18:29:27.80 & $-$10:41:28.8 & 24.7, 30.4 & 25.9 & 8.91 & 15.3 & 16/07/04 & 13.94 & Parkes  \\
 \enddata
\tablecomments{Column 1: source name; Columns 2 - 3: targeted positions for the TMRT observations; Column 4: velocity interval of the maser emission; Column 5: velocity of the peak emission; Column 6: peak flux density; Column 7: integrated flux density; Column 8: epoch of observation; Column 9: estimated kinematic distance by using the Galactic rotation model \citep[][http://bessel.vlbi-astrometry.org/bayesian]{Reid2009,Reid2014}; Column 10: the discovery references Parkes \citep[]{Bre2015}, Szy12 \citep[]{Szy2012}.}
\tablecomments{Table 3 is published in its entirety in the electronic 
edition of the {\it Astrophysical Journal}.  A portion is shown here 
for guidance regarding its form and content.}
\end{deluxetable}
\clearpage

\newpage
\begin{deluxetable}{cccl}
\tabletypesize{\tiny}
\tablecolumns{4} 
\tablewidth{0pt}
\tablecaption{Comparison between the spectral profiles of our survey and the previous surveys. \label{table: Comparison}}
\tablehead{
 \colhead{Name} & \colhead{Name in the previous surveys} & \colhead{Ref.} & \colhead{Notes}  \\
 \colhead{(1)} & \colhead{(2)} & \colhead{(3)} & \colhead{(4)} 
 }
\startdata
 G20.234+0.085 & G20.237+0.065/G20.239+0.065 & Parkes & Both show strong features at 71.9 km s$^{-1}$ and 73.2 km s$^{-1}$.  \\
  & & & \\
 G20.363$-$0.014 & G20.364$-$0.013 & Parkes & Has not changed a lot. \\
  & & & \\
 G20.762$-$0.064 & G20.733$-$0.059 & Parkes & With an angular separation of 105.5$^{\prime\prime}$ away from G20.762$-$0.064, \\
  &  &  & G20.733$-$0.059 only had one strong feature at 60.7 km s$^{-1}$. \\
  & & & \\ 
 G20.926$-$0.050 & G20.926$-$0.050 & Parkes & Show similar features at 25.6 km s$^{-1}$, 25.9 km s$^{-1}$ and 27.4 km s$^{-1}$. \\
  & & & \\ 
 G21.023$-$0.063 & G21.023$-$0.063 & Parkes & Same features.  \\
  & & & \\ 
 G21.370$-$0.226 & 21.407$-$0.254 & Szy12 & The feature at 95.2 km s$^{-1}$ disappears and a feature rises at 93.0 km s$^{-1}$  \\
  & & & in our TMRT survey. \\   
  & & & \\                          
 G22.050+0.211 & G22.039+0.222 & Parkes & With an angular separation of 60.3$^{\prime\prime}$, a new feature at 54.6 km s$^{-1}$ is confirmed. \\
  & & & \\ 
 G22.355+0.066 & G22.356+0.066 & Parkes & A new feature at 78.8 km s$^{-1}$. \\
  & & & \\ 
 G23.010$-$0.410 & G23.010$-$0.410 & Parkes & Strong features at similar velocities. \\
  & & & \\  
 G23.185$-$0.380 & G23.207$-$0.377 & Parkes & A new feature at 82.9 km s$^{-1}$. \\
  & & & \\  
 G23.271$-$0.256 & G23.257$-$0.241 & Parkes & Similar features at the same velocities. \\
  & & & \\ 
 G23.389+0.185 & G23.389+0.185 & Parkes & Similar. \\
  & & & \\ 
 G23.436$-$0.184 & G23.437$-$0.184/G23.440$-$0.182 & Parkes & Both show features at 96.6, 97.6, 98.0, 102.9, 104.0, 107.0 km s$^{-1}$. \\
  & & & \\ 
 G23.653$-$0.143 & G23.657$-$0.127 & Parkes & Similar features at the same velocity range. \\
  & & & \\ 
 G23.680$-$0.189 & G23.706$-$0.198 & Parkes & The feature at 58.2 km s$^{-1}$ seems obscure in our observation. \\
  & & & \\
 G23.899+0.065 & G23.885+0.060/G23.901+0.077 & Parkes & Share features at 31.6, 35.7, 44.0 and 44.9 km s$^{-1}$, but a new feature  \\  
  & & &at 42.8 km s$^{-1}$. \\ 
  & & & \\
 G23.965$-$0.110 & G23.966$-$0.109/G23.986$-$0.089/G23.996$-$0.100 & Parkes & Share features at 65.1, 67.7, 68.2 and 71.0 km s$^{-1}$. \\
  & & & \\ 
 G24.148$-$0.009 & G24.148$-$0.009 & Parkes & Similar features at the same velocities. \\
  & & & \\ 
 G24.328+0.144 & G24.329+0.144 & Parkes & Both show features at 110.3, 111.9, 112.9, 115.4 and 119.8 km s$^{-1}$. \\
  & & & \\ 
 G24.485+0.180 & G24.461+0.198 & Parkes & The features at the velocity range of 119.9 to 126.0 km s$^{-1}$ are much weaker now. \\
  & & & \\ 
 G24.528+0.337 & G24.541+0.312 & Parkes & The feature at 106.8 km s$^{-1}$ in our TMRT detection is much weaker. \\
  & & & \\
 G24.634$-$0.323 & G24.634$-$0.324 & Parkes & A new feature at 38.0 km s$^{-1}$. \\
  & & & \\ 
 G24.790+0.084 & G24.790+0.083a/G24.790+0.083b & Parkes & Similar features at the same velocities. \\  
  & & & \\ 
 G24.943+0.074 & G24.920+0.086/G24.943+0.074 & Parkes & Similar features but one at 47.7 km s$^{-1}$ is not seen by the TMRT survey. \\
  & & & \\ 
 G25.256$-$0.446 & G25.270$-$0.4349 & Parkes & Both show features at the same velocities. \\
  & & & \\ 
 G25.346$-$0.189 & G25.382$-$0.182/G25.407$-$0.170 & Parkes & Similar features. \\
  & & & \\ 
 G25.395+0.033 & 25.39+0.03 & Szy12 & Only one feature at 95.7 km s$^{-1}$. \\  
  & & & \\ 
 G25.410+0.105 & G25.411+0.105 & Parkes & A feature at 93.4 km s$^{-1}$ is not seen in our survey. \\
  & & & \\
 G25.498+0.069 & G25.494+0.062 & Parkes & A new weak feature at 92.0 km s$^{-1}$ appeared in our survey. \\
  & & & \\ 
 G25.709+0.044 & G25.710+0.044 & Parkes & This source has not changed. \\  
  & & & \\
 G25.613+0.226 & G25.613+0.226 & Parkes & A feature at 109.8 km s$^{-1}$ seems disappeared in our TMRT survey. \\
  & & & \\ 
 G25.649+1.050 & G25.650+1.049 & Parkes & This source remains stable. \\
  & & & \\ 
 G25.837$-$0.378 & G25.838$-$0.378 & Parkes & Both have weak features at the same velocities. \\
  & & & \\
 G26.421+1.686 & G26.422+1.685 & Parkes & A new feature at 30.4 km s$^{-1}$ from our TMRT survey. \\
  & & & \\ 
 G26.545+0.423 & G26.545+0.42 & Parkes & Only one feature at 82.5 km s$^{-1}$. \\
  & & & \\ 
 G26.598$-$0.024 & G26.598$-$0.024 & Parkes & This source remains unchanged. \\  
  & & & \\ 
 G26.623$-$0.259 & G26.601$-$0.221 & Parkes & Due to the separation of $2.5^{\prime}$, G26.601$-$0.221 has many much stronger features. \\
  & & & \\ 
 G26.645+0.021 & G26.648+0.018 & Parkes & Features at 104.0, 107.0, 111.7 and 114.3 km s$^{-1}$ are not shown in our TMRT detection. \\  
  & & & \\
 G27.220+0.261 & G27.220+0.261 & Parkes & A weak feature at 6.20 km s$^{-1}$ is not seen in our detection. \\
  & & & \\ 
 G27.222+0.136 & G27.221+0.136 & Parkes & Share features at a velocity range of 112.0 to 120.8 km s$^{-1}$, but a feature at 110.2 km s$^{-1}$  \\
  & & & seems disappeared in our TMRT detection. \\ 
  & & & \\  
 G27.287+0.154 & G27.286+0.151 & Parkes & G27.286+0.151 has a much larger velocity range of 23.4 to 36.7 km s$^{-1}$ than the velocity \\  
  & & &range of 34.0 to 36.9 km s$^{-1}$ from our TMRT observation.  \\ 
  & & & \\  
 G27.725+0.037 & G27.757+0.050 & Parkes & Share weak features at the same velocity range. \\
  & & & \\ 
 G27.784+0.057 & G27.784+0.057 & Parkes & Similar. \\
  & & & \\ 
 G27.795$-$0.277 & G27.783$-$0.259 & Parkes & A feature at 93.6 km s$^{-1}$ has disappeared in our TMRT observation. \\
  & & & \\ 
 G28.147$-$0.004 & G28.146$-$0.005 & Parkes & Similar features. \\
  & & & \\ 
 G28.180$-$0.093 & G28.201$-$0.049 & Parkes & Features at 94.5, 100.2 and 111.9 km s$^{-1}$ have disappeared in our TMRT detection. \\
  & & & \\ 
 G28.287$-$0.348(Left) & G28.282$-$0.359 & Parkes & Share similar features. \\ 
  & & & \\
 G28.287$-$0.348(Right) & G28.305$-$0.387 & Parkes & Similar. \\ 
  & & & \\ 
 G28.320$-$0.012 & G28.321$-$0.011 & Parkes & Both have features at 96.7, 97.5, 102.7 and 104.7 km s$^{-1}$. \\
  & & & \\ 
 G28.393+0.085 & G28.397+0.081 & Parkes & Similar. \\
  & & & \\ 
 G28.532+0.129 & G28.523+0.127/G28.532+0.129 & Parkes & Show features at 24.8, 25.0 and 39.8 km s$^{-1}$. \\
  & & & \\ 
 G28.609+0.017 & G28.608+0.018 & Parkes & Our TMRT detection has a new feature at 104.6 km s$^{-1}$. \\  
  & & & \\
 G28.832$-$0.250 & G28.832$-$0.253 & Parkes & Our detection has some new features at a velocity range of 99.7 to 103.6 km s$^{-1}$. \\
  & & & \\ 
 G28.843+0.494 & G28.842+0.493 & Parkes & Both have features at the same velocities. \\
  & & & \\ 
 G28.862+0.066 & G28.861+0.065 & Parkes & G28.861+0.065 has a much weaker feature at 105.3 km s$^{-1}$. \\
  & & & \\ 
 G29.320$-$0.162 & G29.320$-$0.162 & Parkes & A feature at 40.3 km s$^{-1}$ is not detected by the TMRT survey. \\
  & & & \\ 
 G29.835$-$0.012 & G29.863$-$0.044 & Parkes & Our TMRT detection shows new features at 93.8, 96.0 and 99.6 km s$^{-1}$. \\
  & & & \\  
 G29.927+0.054 & G29.955$-$0.016 & Parkes & The feature at 100.2 km s$^{-1}$ in the spectrum of G29.955$-$0.016 disappeared in  \\ 
  & & & the TMRT survey.\\ 
  & & & \\
 G29.941$-$0.070 & G29.978$-0.047$ & Parkes & Similar features. \\  
  & & & \\
 G30.004$-$0.265 & G29.993$-$0.282/G30.010$-$0.273 & Parkes & Both have two features at the same velocities. \\
  & & & \\
 G30.250$-$0.232 & 30.30$-$0.20 & Szy12 & The profile of our TMRT detection is much more complex with a wider velocity range. \\
  & & & \\ 
 G30.370+0.483 & G30.370+0.482 & Parkes & Both have features at 12.3, 16.7 and 19.4 km s$^{-1}$. \\
  & & & \\
 G30.403$-$0.297 & G30.400$-$0.296 & Parkes & Similar features but one at 105.0 km s$^{-1}$ has become stronger. \\
  & & & \\ 
 G30.419$-$0.232 & G30.419$-$0.232 & Parkes & A feature at 111.2 km s$^{-1}$ disappears now. \\
  & & & \\   
 G30.536$-$0.004 & G30.542+0.011 & Parkes & Similar features at the same velocity range. \\
  & & & \\ 
 G30.589$-$0.043 & G30.589$-$0.043 & Parkes & Share similar features. \\  
  & & & \\ 
 G30.662$-$0.139 & G30.703$-$0.068 & Parkes & The feature at 86.0 km s$^{-1}$ is not seen by the TMRT survey. \\
  & & & \\
 G30.807+0.080 & G30.774+0.078 & Parkes & Share features at the same velocity range. \\
  & & & \\
 G30.770$-$0.804 & G30.771$-$0.804 & Parkes & Similar. \\
  & & & \\ 
 G30.789+0.232(Left) & G30.780+0.230 & Parkes & They share the same spectral profiles. \\ 
  & & & \\ 
 G30.788+0.203(Right) & G30.788+0.204 & Parkes & Share similar features at the same velocity range. \\ 
  & & & \\  
 G30.810$-$0.050 & G30.822$-$0.053/G30.818$-$0.057 & Parkes & The features at 91.3 and 103.4 km s$^{-1}$ are difficult to find in the TMRT survey. \\ 
  & & & \\  
 G30.819+0.273 & G30.818+0.273 & Parkes & Both have features at 100.5, 101.1, 102.4, 104.8 and 110.2 km s$^{-1}$. \\
  & & & \\   
 G30.866+0.114 & G30.851+0.123 & Parkes & Show similar profile but the feature at 37.2 km s$^{-1}$ is difficult to distinguish now. \\  
  & & & \\   
 G30.823+0.134(Right) & G30.788+0.204 & Parkes & Similar features. \\  
  & & & \\ 
 G30.897+0.163 & G30.898+0.161 & Parkes & This source has not changed. \\
  & & & \\ 
 G30.959+0.086 & G30.960+0.086 & Parkes & A feature at  36.0 km s$^{-1}$ seems gone now. \\  
  & & & \\
 G30.972$-$0.141 & G30.972$-$0.142 & Parkes & Similar features within a velocity range of 73.6 to 80.6 km s$^{-1}$. \\
  & & & \\ 
 G30.973+0.562 & G30.973+0.562 & Parkes & Only one feature at 20.0 km s$^{-1}$. \\
  & & & \\ 
 G30.980+0.216 & G30.980+0.216 & Parkes & The feature at 110.0 km s$^{-1}$ has become much weaker. \\
  & & & \\ 
 G31.076+0.458 & G31.076+0.457 & Parkes & One feature at 26.6 km s$^{-1}$ seems disappeared now. \\
  & & & \\
 G31.159+0.058 & G31.158+0.046 & Parkes & Similar features at the same velocity range. \\  
  & & & \\ 
 G31.237+0.067 & G31.281+0.061 & Parkes & G31.281+0.061 has similiar features with much stronger flux density. \\
  & & & \\
 G31.413+0.308 & G31.412+0.307 & Parkes & The feature at 105.7 km s$^{-1}$ seems obscure now. \\  
  & & & \\ 
 G31.579+0.076 & G31.581+0.077 & Parkes & Similar features at the same velocity range. \\  
  & & & \\ 
 G32.045+0.059 & G32.045+0.059 & Parkes & Almost the same spectral profile. \\  
  & & & \\
 G32.118+0.090 & G32.082+0.078 & Parkes & A feature rises at 104.2 km s$^{-1}$ from our TMRT survey. \\
  & & & \\
 G32.773$-$0.059 & G32.744$-$0.057 & Parkes & G32.744$-$0.057 does not have feature at 35.0 km s$^{-1}$. \\  
  & & & \\
 G32.798+0.190 & G32.802+0.1931 & Parkes & Similar. \\  
  & & & \\ 
 G32.828$-$0.315 & G32.825$-$0.328 & Parkes & The feature at 84.0 km s$^{-1}$ in G32.825$-$0.328 cannot be seen in the TMRT survey. \\
  & & & \\
 G32.992+0.034 & G32.992+0.03 & Parkes & Almost the same features at the same velocity range. \\
  & & & \\ 
 G33.143$-$0.088(Left) & G33.133$-$0.092 & Parkes & Similar features. \\ 
  & & & \\
 G33.092$-$0.073(Right) & G33.093$-$0.073 & Parkes & The features at 95.7, 98.3 and 106.0 km s$^{-1}$ in the spectrum of G33.093$-$0.073  \\ 
  & & &have become much weaker now. \\ 
  & & & \\ 
 G33.322$-$0.364 & G33.317$-$0.360 & Parkes & The feature at 30.9 km s$^{-1}$ seems disappear now. \\
  & & & \\  
 G33.393+0.010 & G33.393+0.010 & Parkes & A new feature at 107.5 km s$^{-1}$. \\
  & & & \\  
 G33.638$-$0.035 & G33.634$-$0.021 & Parkes & One weak peak at 102.9 km s$^{-1}$. \\
  & & & \\
 G33.641$-$0.228 & G33.641$-$0.228 & Parkes & Has features at 58.8, 59.6, 60.3, 61.8, 62.7 and 63.2 km s$^{-1}$. \\
  & & & \\ 
 G33.726$-$0.119 & G33.725$-$0.120 & Parkes & A weak feature at 57.4 km s$^{-1}$ is not seen in our TMRT survey. \\
  & & & \\ 
 G34.096+0.018 & G34.096+0.018 & Parkes & Features at 54.9 and 57.3 km s$^{-1}$ are weaker now. \\
  & & & \\
 G34.229+0.133 & G34.244+0.133 & Parkes & There are new features at 52.1 and 62.7 km s$^{-1}$. \\
  & & & \\  
 G34.411+0.235 & G34.396+0.222 & Parkes & Similar features. \\ 
  & & & \\ 
 G34.789$-$1.392 & G34.791$-$1.387 & Parkes & Share features at 44.1, 45.0, 46.3 and 47.2 km s$^{-1}$. \\
  & & & \\
 G34.757+0.025 & G34.82+0.35 & Parkes & The only one feature at 76.6 km s$^{-1}$ is stronger now. \\
  & & & \\ 
 G34.974+0.365 & G35.025+0.350 & Parkes & The spectrum of G35.025+0.350 has much stronger and more complex features at a wider  \\
  & & & velocity range of 41.1 to 47.0 km s$^{-1}$.\\  
  & & & \\ 
 G35.141$-$0.750 & G35.132$-$0.744 & Parkes & The peak feature at 35.4 km s$^{-1}$ has nearly disappeared now. \\
  & & & \\ 
 G35.194$-$1.725 & G35.200$-$1.736 & Parkes & Both spectra show similar features at a velocity range of 39.9 to 45.7 km s$^{-1}$. \\
  & & & \\  
 G35.197$-$0.729 & G35.197$-$0.743/G35.197$-$0.743n & Parkes & Similar features. \\ 
  & & & \\
 G35.247$-$0.237 & G35.247$-$0.237 & Parkes & A weak feature at 72.9 km s$^{-1}$ seems disappear now. \\
  & & & \\ 
 G35.398+0.025 & G35.397+0.025 & Parkes & The feature at 90.3 km s$^{-1}$ in our TMRT detection is new. \\
  & & & \\
 G35.578+0.048 & G35.588+0.060 & Parkes & The feature at 52.0 km s$^{-1}$ has disappeared. \\ 
  & & & \\ 
 G35.792$-$0.174 & G35.793$-$0.175 & Parkes & Same features at a velocity range of 58.5 to 63.1 km s$^{-1}$. \\
  & & & \\ 
 G36.137+0.564 & G36.115+0.552 & Parkes & Both surveys show features at 70.4, 71.8, 72.8, 74.7, 75.8, 81.6, 82.2 and 84.0 km s$^{-1}$. \\
  & & & \\ 
 G36.634$-$0.203 & G36.634$-$0.203 & Parkes & Share the only one feature at 77.4 km s$^{-1}$. \\
  & & & \\ 
 G36.705+0.096 & G36.705+0.096 & Parkes & The spectrum at a velocity range of 51.9 to 55.9 km s$^{-1}$ has less features now but is much  \\
  & & & more complex at a velocity range of 61.2 to 63.9 km s$^{-1}$.\\
  & & & \\ 
 G36.833$-$0.031 & G36.839$-$0.022 & Parkes & Both spectra have features at 54.7, 55.4, 57.0, 61.2, 62.0 and 63.6 km s$^{-1}$. \\
  & & & \\
 G36.919+0.483 & G36.918+0.483 & Parkes & One feature at the same velocity with a similar flux density. \\ 
  & & & \\ 
 G37.043$-$0.035 & G37.030$-$0.038/G37.043$-$0.035 & Parkes & Both spectra show features at 78.4, 80.0, 80.8, 83.7 and 84.8 km s$^{-1}$. \\
  & & & \\ 
 G37.430+1.517 & G37.430+1.518 & Parkes & Both surveys show features at the same velocity range. \\
  & & & \\ 
 G37.479$-$0.105 & G37.546$-$0.112 & Parkes & Both surveys find features at 50.1 and 52.6 km s$^{-1}$. \\
  & & & \\ 
 G37.554+0.201 & G37.554+0.201 & Parkes & The feature at 84.9 km s$^{-1}$ has grown stronger than the feature at 83.7 km s$^{-1}$ now. \\
  & & & \\ 
 G37.602+0.428 & G37.598+0.425 & Parkes & The feature at 85.8 km s$^{-1}$ is slightly weak now. \\
  & & & \\
 G37.763$-$0.215 & G37.763$-$0.215 & Parkes & Similar features at the same velocity range of 54.1 to 70.0 km s$^{-1}$. \\
  & & & \\ 
 G38.076$-$0.265 & 38.038$-$0.300 & Szy12 & Both spectra show features at 55.8 and 57.2 km s$^{-1}$. \\
  & & & \\ 
 G38.119$-$0.229 & G38.119$-$0.229 & Parkes & The features at 77.3 and 79.2 km s$^{-1}$ have become weak after comparing with \\ 
  & & & the strongest feature. \\
  & & & \\
 G38.202$-$0.068 & G38.203$-$0.067 & Parkes & Almost the same profile at the same velocity range of 77.8 to 85.1 km s$^{-1}$. \\
  & & & \\ 
 G38.255$-$0.200 & G38.255$-$0.200 & Parkes & Both have several weak features at the same velocity range. \\
  & & & \\ 
 G38.258$-$0.074 & G38.258$-$0.074 & Parkes & Share features at 7.09, 12.3 and 15.5 km s$^{-1}$. \\
  & & & \\ 
 G38.598$-$0.213 & G38.598$-$0.212 & Parkes & The feature at 69.2 km s$^{-1}$ seems to have disappeared now. \\
  & & & \\ 
 G38.933$-$0.361 & G38.916$-$0.353 & Parkes & The peak feature at the same velocity range is weaker now. \\
  & & & \\
 G39.100+0.491 & G39.100+0.491 & Parkes & Both surveys detect features at 14.6, 15.8, 17,7, 24.8 and, 28.7 km s$^{-1}$,  \\
  & & & but the feature at 14.6 km s$^{-1}$ is much weaker now. \\ 
  & & & \\ 
 G39.387$-$0.141 & G39.388$-$0.141 & Parkes & Our TMRT detection shows a new feature at 72.0 km s$^{-1}$.\\ 
  & & & \\  
 G40.282$-$0.220 & G40.282$-$0.219 & Parkes & The spectrum of G40.282$-$0.219 has much more and stronger features with \\
  & & & a much larger velocity range. \\ 
  & & & \\ 
 G40.425+0.700 & G40.425+0.700 & Parkes & Same spectral profile. \\ 
  & & & \\
 G40.597$-$0.719 & G40.597$-$0.719 & Parkes & Similar. \\
  & & & \\ 
 G40.622$-$0.138 & G40.622$-$0.138 & Parkes & Both surveys show features at 30.2, 31.3, 31.9 and 36.3 km s$^{-1}$. \\ 
  & & & \\
 G40.964$-$0.025 & G40.934$-$0.041 & Parkes & Both have three features at 36.9, 37.5 and 41.1 km s$^{-1}$. \\
  & & & \\ 
 G41.121$-$0.107 & G41.121$-$0.107 & Parkes & The feature at 36.6 km s$^{-1}$ is weaker and the feature at 37.2 km s$^{-1}$ is \\
  & & & stronger now. \\
  & & & \\ 
 G41.307$-$0.169 & G41.348$-$0.136 & Parkes & Features at a velocity range of 6.8 to 9.4 km s$^{-1}$ are not seen now. \\
  & & & \\ 
 G42.035+0.191 & G42.034+0.190 & Parkes & Similar profile. \\
  & & & \\ 
 G42.692$-$0.129 & G42.698$-$0.147 & Parkes & Similar spectral profile but our TMRT detection is much weaker. \\
  & & & \\ 
 G43.037$-$0.453 & G43.038$-$0.453 & Parkes & Unchanged. \\
  & & & \\ 
 G43.076$-$0.078 & G43.074$-$0.077 & Parkes & Almost the same profile at the same velocity range of 9.70 to 11.3 km s$^{-1}$. \\
  & & & \\ 
 G43.148+0.013 & G43.149+0.013/G43.165+0.013/G43.167$-$0.004 & Parkes & Similar features in a quite large velocity range. \\
  & /G43.171+0.005/G43.175$-$0.015 & & \\ 
  & & & \\ 
 G43.178$-$0.519 & G43.180$-$0.518 & Parkes & Five features at 55.8, 58.1, 58.9, 59.5 and 65.1 km s$^{-1}$. \\ 
  & & & \\  
 G43.808$-$0.080 & G43.795$-$0.127 & Parkes & Almost same. \\
  & & & \\ 
 G43.890$-$0.790 & G43.890$-$0.784 & Parkes & One feature at 56.6 km s$^{-1}$ becomes stronger and the feature at 49.7 km s$^{-1}$  \\
  & & & is no longer there now. \\ 
  & & & \\ 
 G45.070+0.124 & G45.071+0.132 & Parkes & A new feature at 59.8 km s$^{-1}$. \\
  & & & \\ 
 G45.360$-$0.598 & G45.380$-$0.594 & Parkes & Only one feature at the same velocity. \\
  & & & \\
 G45.454+0.060 & G45.445+0.069/G45.467+0.053 & Parkes & The feature at 49.8 km s$^{-1}$ seems weaker. \\ 
  & & & \\ 
 G45.493+0.126 & G45.493+0.126/G45.473+0.134 & Parkes & The feature at 73.2 km s$^{-1}$ has disappeared now. \\
  & & & \\ 
 G45.804$-$0.356 & G45.804$-$0.356 & Parkes & Similar features at the same velocity range. \\
  & & & \\
 G48.905$-$0.261 & G48.902$-$0.273 & Parkes & G48.905$-$0.261 has three weak features at 72.2, 74.3 and, 75.2 km s$^{-1}$, \\
  & & & while G48.902$-$0.273 only has one feature at 71.8 km s$^{-1}$.\\
  & & & \\ 
 G48.991$-$0.299 & G48.990$-$0.299 & Parkes & Only one feature at 71.6 km s$^{-1}$. \\ 
  & & & \\ 
 G49.043$-$1.079 & G49.043$-$1.079 & Parkes & The spectral profile has changed. \\
  & & & \\  
 G49.265+0.311 & G49.265+0.311 & Parkes & The weak feature at $-$6.66 km s$^{-1}$ has become a little stronger. \\
  & & & \\
 G49.466$-$0.408 & G49.482$-$0.402/G49.489$-$0.369/G49.490$-$-.388 & Parkes & Our TMRT observation shows a new feature at 61.0 km s$^{-1}$. \\
  & & & \\ 
 G49.599$-$0.249 & G49.599$-$0.249 & Parkes & The feature at 66.7 km s$^{-1}$ has become stronger and the peak feature \\
  & & & at 63.1 km s$^{-1}$ in the MMB survey is much weaker. \\ 
  & & & \\ 
 G50.034+0.581 & G50.035+0.582 & Parkes & The feature at a velocity range of $-$11.2 to $-$8.25 km s$^{-1}$ has become much weaker. \\
  & & & \\ 
 G50.779+0.152 & G50.779+0.152 & Parkes & Similar features at the same velocity range of 48.1 to 50.8 km s$^{-1}$. \\
  & & & \\ 
 G51.678+0.719 & G51.679+0.719 & Parkes & The features at $-$3.89 and 2.00 km s$^{-1}$ seems have disappeared now. \\
  & & & \\ 
 G52.199+0.723 & G52.199+0.723 & Parkes & Similar features at a velocity range of 2.28 to 4.52 km s$^{-1}$. \\  
  & & & \\ 
 G52.663$-$1.092 & G52.663$-$1.092 & Parkes & Both surveys detected features at 56.3, 65.2, 65.9 and 66.4 km s$^{-1}$. \\
  & & & \\ 
 G52.922+0.414 & G52.922+0.414 & Parkes & Similar features. \\
  & & & \\ 
 G53.022+0.100 & G53.036+0.113 & Parkes & Only one feature at 10.0 km s$^{-1}$. \\ 
  & & & \\ 
 G53.141+0.071 & G53.142+0.071 & Parkes & Similar profile at a velocity range of 23.7 to 24.8 km s$^{-1}$. \\
  & & & \\ 
 G53.618+0.036 & G53.618+0.035 & Parkes & G53.618+0.035 has almost the same but much stronger features. \\
  & & & \\ 
 G56.963$-$0.234 & G56.963$-$0.235 & Parkes & Both have only one feature at 29.7 km s$^{-1}$. \\
  & & & \\ 
 G58.775+0.647 & G58.775+0.644 & Parkes & A new feature at 36.6 km s$^{-1}$. \\
  & & & \\ 
 G59.634$-$0.192 & G59.634$-$0.192 & Parkes & The features at 23.3 and 25.9 km s$^{-1}$ are new. \\
  & & & \\ 
 G59.785+0.068 & G59.783+0.065 & Parkes & Obviously the feature at 19.1 km s$^{-1}$ has become much stronger now. \\
  & & & \\  
 G59.833+0.672 & G59.833+0.672 & Parkes & This source remains the same profile. \\
  & & & \\  
 G69.543$-$0.973 & 69.540$-$0.976/ON1 & Szy12 & Their spectra show the same profile with features at $-$0.06. 1.13 and 14.6 km s$^{-1}$. \\
  & & & \\
 G71.522$-$0.385 & 71.52$-$0.38 & Szy12 & The feature at 10.2 km s$^{-1}$ seems much stronger. \\
  & & & \\ 
 G73.063+1.796 & 73.06+1.80 & Szy12 & The feature at $-$2.54 km s$^{-1}$ becomes brighter and there is a new weak feature at 0.34 km s$^{-1}$. \\
  & & & \\ 
 G75.770+0.344 & 75.78+0.34 & Szy12 & The features at $-$9.53 and $-$0.95 km s$^{-1}$ are not seen in our TMRT survey. \\
  & & & \\ 
 G78.882+0.723 & 78.89+0.71 & Szy12 & One single feature at the same velocity range of $-$7.41 to $-$6.51 km s$^{-1}$. \\  
  & & & \\ 
 G79.736+0.991 & 79.736+0.991 & Szy12 & Similar features at the same velocity range of $-$6.63 to $-$2.84 km s$^{-1}$. \\
  & & & \\
 G80.862+0.383 & 80.861+0.383/DR20 & Szy12 & Two new features at $-$11.2 and $-$2.02 km s$^{-1}$. \\
  & & & \\  
 G81.752+0.591 & 81.76+0.59/W75S & Szy12 & Similar features. \\
  & & & \\
 G81.871+0.779 & 81.87+0.78/W75N & Szy12 & Almost the same profile. \\ 
  & & & \\    
 G85.394$-$0.023 & 85.41+0.00 & Szy12 & Features at the same velocity range of $-$33.5 to $-$27.8 km s$^{-1}$. \\
  & & & \\ 
 G90.921+1.487 & 90.92+1.49 & Szy12 & Features at $-$71.2, $-$70.4 and $-$69.2 km s$^{-1}$ are detected. \\
  & & & \\
 G94.609$-$1.790 & 94.602$-$1.796 & Szy12 & Similar features at the same velocity range. \\
  & & & \\  
 G98.036+1.446 & 98.04+1.45 & Szy12 & Only one feature at the same velocity. \\  
  & & & \\
 G108.758$-$0.986 & 108.76$-$0.99 & Szy12 & Both show features at $-$56.0, $-$54.6, $-$47.3 and $-$45.6 km s$^{-1}$. \\
  & & & \\  
 G111.256$-$0.770 & 111.26$-$0.77 & Szy12 & Both have four features at $-$41.2, $-$38.7, $-$37.8 and, $-$36.9 km s$^{-1}$. \\
  & & & \\
 G111.532+0.759 & 111.542+0.777/NGC7538 & Szy12 & They all have similar profile at the same velocity range. \\  
  & & & \\  
 G121.329+0.639 & 121.298+0.659/L1287 & Szy12 & The feature at $-$27.0 km s$^{-1}$ seems obscure now. \\
  & & & \\
 G134.029+1.072 & 133.947+1.064/W3(OH) & Szy12 & Both detect this source at a velocity range of $-$48.5 to $-$41.6 km s$^{-1}$. \\
  & & & \\ 
 G136.859+1.165 & 136.84+1.15 & Szy12 & Only one feature at the same velocity range. \\
  & & & \\
 G174.205$-$0.069 & 174.201$-$0.071/AFGL5142 & Szy12 & The feature at 0.36 km s$^{-1}$ disappears in our TMRT survey. \\
  & & & \\ 
 G183.349$-$0.575 & 183.35$-$0.58 & Szy12 & Both spectra show features at $-$15.2, $-$14.5 and, $-$4.74 km s$^{-1}$. \\
 \enddata
\tablecomments{Column 1: source name in our survey; Columns 2: source name in the previous detected surveys; Column 3: the discovery references Parkes \citep[]{Bre2015}, Szy12 \citep[]{Szy2012}; Column 4: notes about the comparison between the spectral profiles of our survey and the previous surveys.}
\end{deluxetable}
\clearpage

\newpage
\begin{sidewaystable}[h!]
\caption{The OTF observations.}
\scriptsize
\centerline{\begin{tabular}{ccccccccccc}
\hline
Name & \multicolumn {4} {c} {OTF Parameters} & & \multicolumn {3} {c} {Properties of the Fitting Centers} & Associated & Difference \\
  \cline{2-5} \cline{7-9}
 & R.A. & Decl. & Side & Epoch & & Velocity ranges & R.A. & Decl. & MMB source & \\
 (gl, gb)  & (J2000) & (J2000) & Length & yy/mm/dd & & km s$^{-1}$ & (J2000) & (J2000) & & ($^{\prime\prime}$)  \\
 ($^{\circ}$, $^{\circ}$)  & (h m s) & ($^{\circ}$ $^{\prime}$ $^{\prime\prime}$) & ($^{\prime}$) & & & & (h m s) & ($^{\circ}$ $^{\prime}$ $^{\prime\prime}$) & & \\
 (1) & (2) & (3) & (4) & (5) & & (6) & (7) & (8) & (9) & (10) \\
\hline
 G28.287$-$0.348 & 18:44:11.48 & $-$04:17:47.1 & 10 & 17/11/21 & & 40.7, 43.1 & 18:44:13.15 & $-$04:18:00.3 & G28.282$-$0.359 & 4.9 \\ 
 & & & & & & 79.4, 94.2 & 18:44:22.08 & $-$04:17:42.9 & G28.305$-$0.387 & 4.7 \\ 
 G30.788+0.203 & 18:46:48.14 & $-$01:49:24.4 & 10 & 17/11/21 & & 47.3, 49.4 & 18:46:41.34 & $-$01:48:34.2 & G30.780+0.230 & 3.9 \\
  & & & & & & 75.0, 90.2 & 18:46:48.14 & $-$01:48:54.3 & G30.788+0.204 & 0.8 \\ 
 G30.823+0.134 & 18:47:06.88 & $-$01:47:57.9 & 15 & 18/01/08 & & 27.0, 29.4 & 18:47:12.35 & $-$01:47:48.8 & G30.851+0.123 & 2.6 \\
  & & & & & & 75.4, 90.3 & 18:46:47.68 & $-$01:48:57.7 & G30.788+0.204 & 1.0 \\ 
 G31.221+0.020 & 18:48:24.70 & $-$01:28:49.1 & 20 & 18/01/08 & & 40.6, 41.8 & 18:48:01.57 & $-$01:33:46.3 & &  \\
  & & & & & & 79.0, 80.3 & 18:48:16.23 & $-$01:31:16.2 & &  \\ 
  & & & & & & 102.3, 113.2 & 18:48:12.23 & $-$01:26:46.0 & G31.281+0.061 & 15.4 \\
 G33.143$-$0.088 & 18:52:09.17 & +00:08:33.0 & 10 & 17/11/29 & & 70.4, 81.8 & 18:52:07.88 & +00:08:24.0 & G33.133$-$0.092 & 11.2 \\ 
 & & & & & & 94.1, 107.2 & 18:51:58.54 & +00:06:33.9 & G33.093$-$0.073 & 15.0 \\
\hline
\end{tabular}}
\tablecomments {
\footnotesize
  Column 1: source name; Columns 2 - 5: the pointing center coordinates and the side length of each region for the TMRT OTF observations and the epoch of observation; Columns 6: the velocity interval of the maser emission of the fitting centers, Columns 7 - 8: the center positions determined from the fitting peak positions of masers, Column 9: name of associated MMB source in \citet[]{Bre2015}, Column 10: the separation between positions of the fitting centers and associated MMB sources decided by ATCA.}
\end{sidewaystable}
\clearpage

\newpage
\begin{sidewaystable}[h!]
\centering
{\scriptsize
\caption{A 6.7 GHz CH$_3$OH maser sources catalog.}
\centerline{\begin{tabular}{cccccccc}
\hline
 Number & Name & R.A. & Decl. & $\Delta V$ & $v_p$ & $S_p$ & Ref. \\
 & (gl, gb) & (J2000) & (J2000) & (km s$^{-1}$) & (km s$^{-1}$) & (Jy) &  \\
 & ($^{\circ}$, $^{\circ}$) & (h m s) & ($^{\circ}$ $^{\prime}$ $^{\prime\prime}$) &  &  &  & \\
 (1) & (2) & (3) & (4) & (5) & (6) & (7) & (8) \\
\hline
1 & 0.092$-$0.663 & 17 48 25.90 & $-$29 12 05.9 & 10, 25 & 23.5 & 24.8 & MMB \\
2 & 0.167$-$0.446 & 17 47 45.46 & $-$29 01 29.3 & 9.5, 17 & 13.8 & 4.44 & MMB \\
3 & 0.212$-$0.001 & 17 46 07.63 & $-$28 45 20.9 & 41, 50.5 & 49.3 & 3.47 & MMB \\
4 & 0.315$-$0.201 & 17 47 09.13 & $-$28 46 15.7 & 14, 27 & 19.4 & 72.16 & MMB \\
5 & 0.316$-$0.201 & 17 47 09.33 & $-$28 46 16.0 & 20, 22 & 21 & 0.6 & MMB \\
6 & 0.376+0.040 & 17 46 21.41 & $-$28 35 40.0 & 35, 40 & 37 & 2.32 & MMB \\
7 & 0.39$-$0.03 & 17 46 41.120 & $-$28 37 05.50 & 22, 31 & 28.7 & 5.8 & P05 \\
8 & 0.409-0.504 & 17 48 33.48 & -28 50 52.5 & 24.5, 27 & 25.3 & 2.77 & MMB \\
9 & 0.475-0.010 & 17 46 47.07 & -28 32 06.9 & 23, 31 & 28.8 & 3.43 & MMB \\
10 & 0.496+0.188 & 17 46 03.96 & -28 24 52.8 & -12, 2 & 0.8 & 32.14 & MMB \\
 &  &  & ...... &  &  &  & \\
\hline
\end{tabular}}}
\tablecomments {
\footnotesize
  Column 1: number; Column 2: source name; Columns 3 - 4: positions (R.A., decl.); Column 5: velocity interval of the maser emission; Column 6: velocity of the peak emission; Column 7: peak flux density; Column 8: the source references, MMB: the Parkes MMB survey, P05: \citet[]{Pes2005}, P07: \citet[]{Pan2007}, X08: \citet[]{Xu2008}, S12: \citet[]{Szy2012}, O14: \citet[]{Olm2014} (This table is available in its entirety in machine-readable form.)}
\end{sidewaystable}
\clearpage

\newpage
\begin{figure*}[h!]
\begin{center}
\begin{tabular}{l}
\includegraphics[width=160mm]{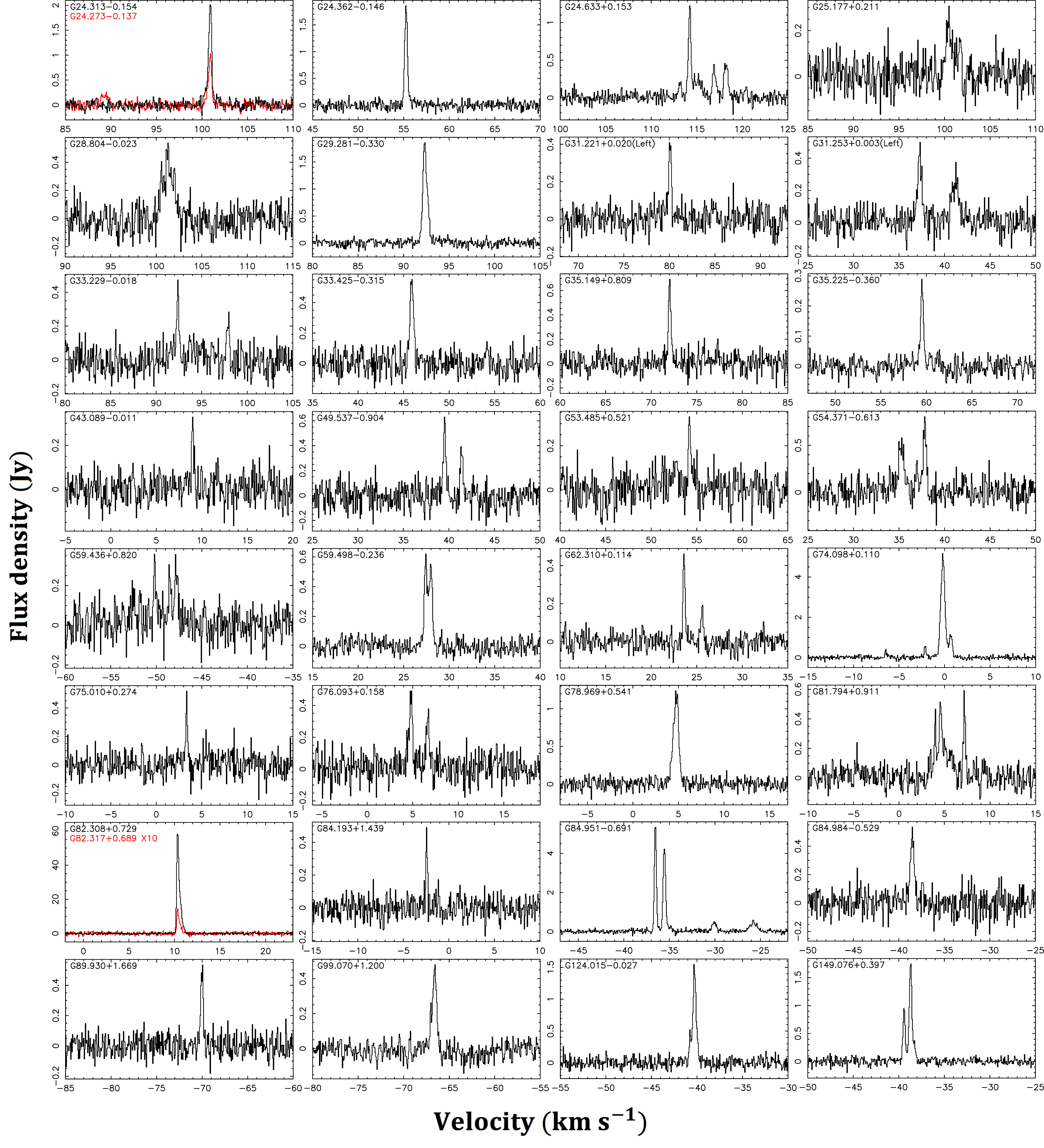}
\end{tabular}
\end{center}
\caption{The spectrum of the 32 newly detected CH$_3$OH maser sources by the TMRT. For the same source but detected at two different positions (G24.313$-$0.154 and G24.273$-$0.137), we show their spectra in different colors.}
\end{figure*}
\clearpage


\begin{figure*}[h!]
\begin{center}
\begin{tabular}{l}
\includegraphics[width=160mm]{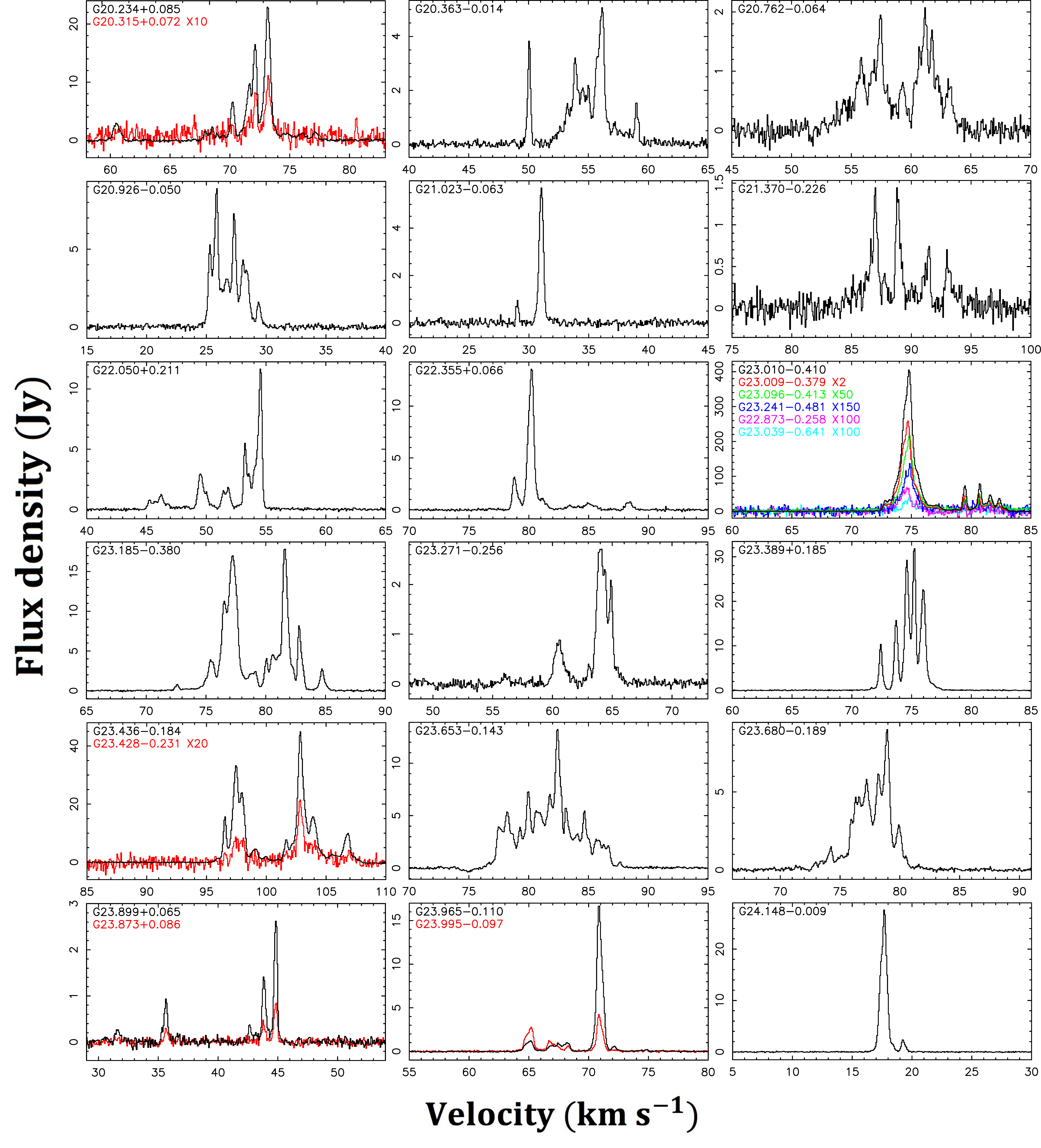}
\end{tabular}
\end{center}
\caption{The TMRT spectrum of 6.7 GHz CH$_3$OH maser towards 192 sources which have been previously CH$_3$OH maser detected. For the same sources but detected at two or more separated positions, we show their spectra in different colors.}
\end{figure*}
\clearpage

\begin{figure*}[h!]
\addtocounter{figure}{-1}
\begin{center}
\begin{tabular}{l}
\includegraphics[width=160mm]{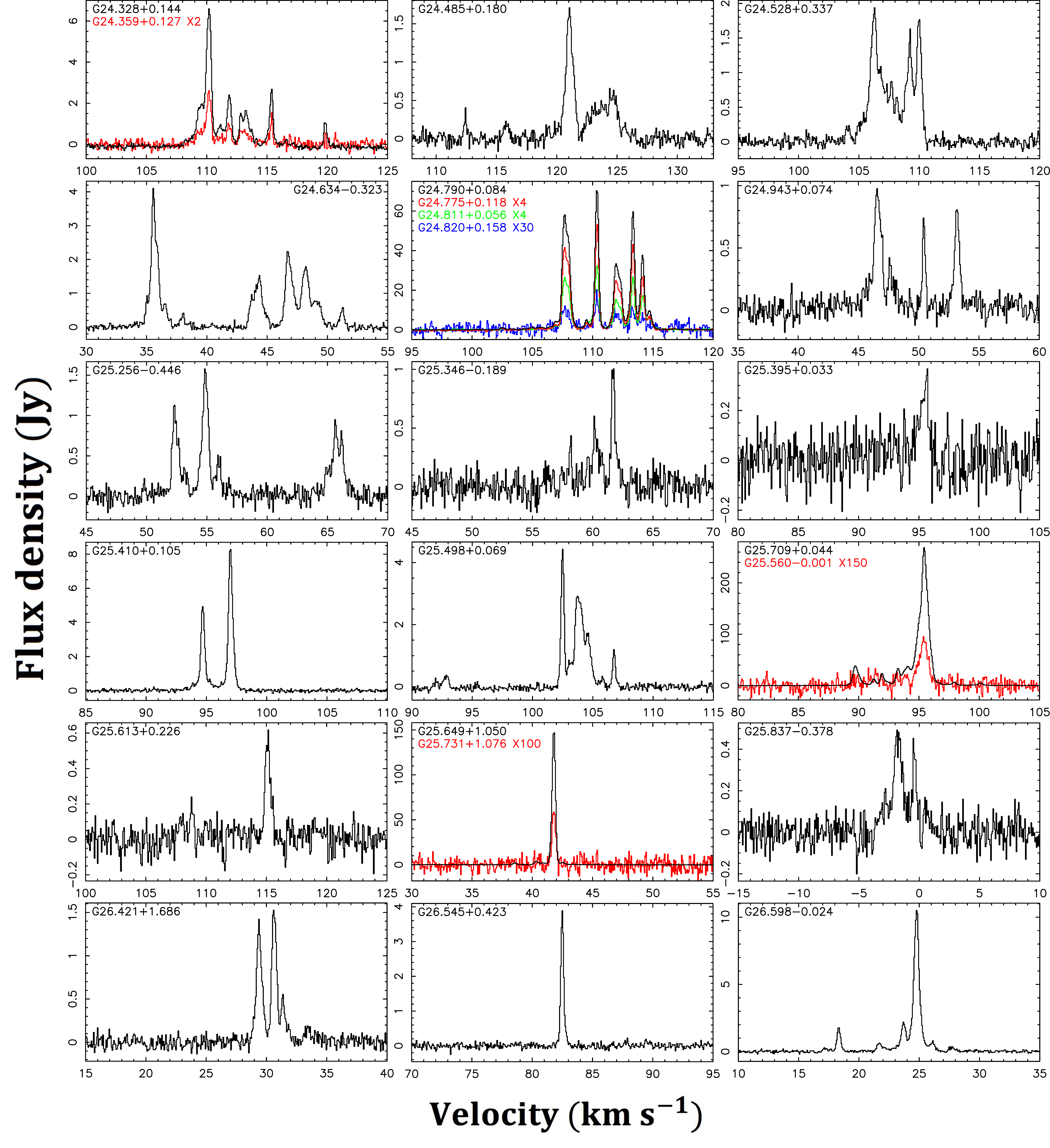}
\end{tabular}
\end{center}
\caption{Continued}
\end{figure*}
\clearpage

\begin{figure*}[h!]
\addtocounter{figure}{-1}
\begin{center}
\begin{tabular}{l}
\includegraphics[width=160mm]{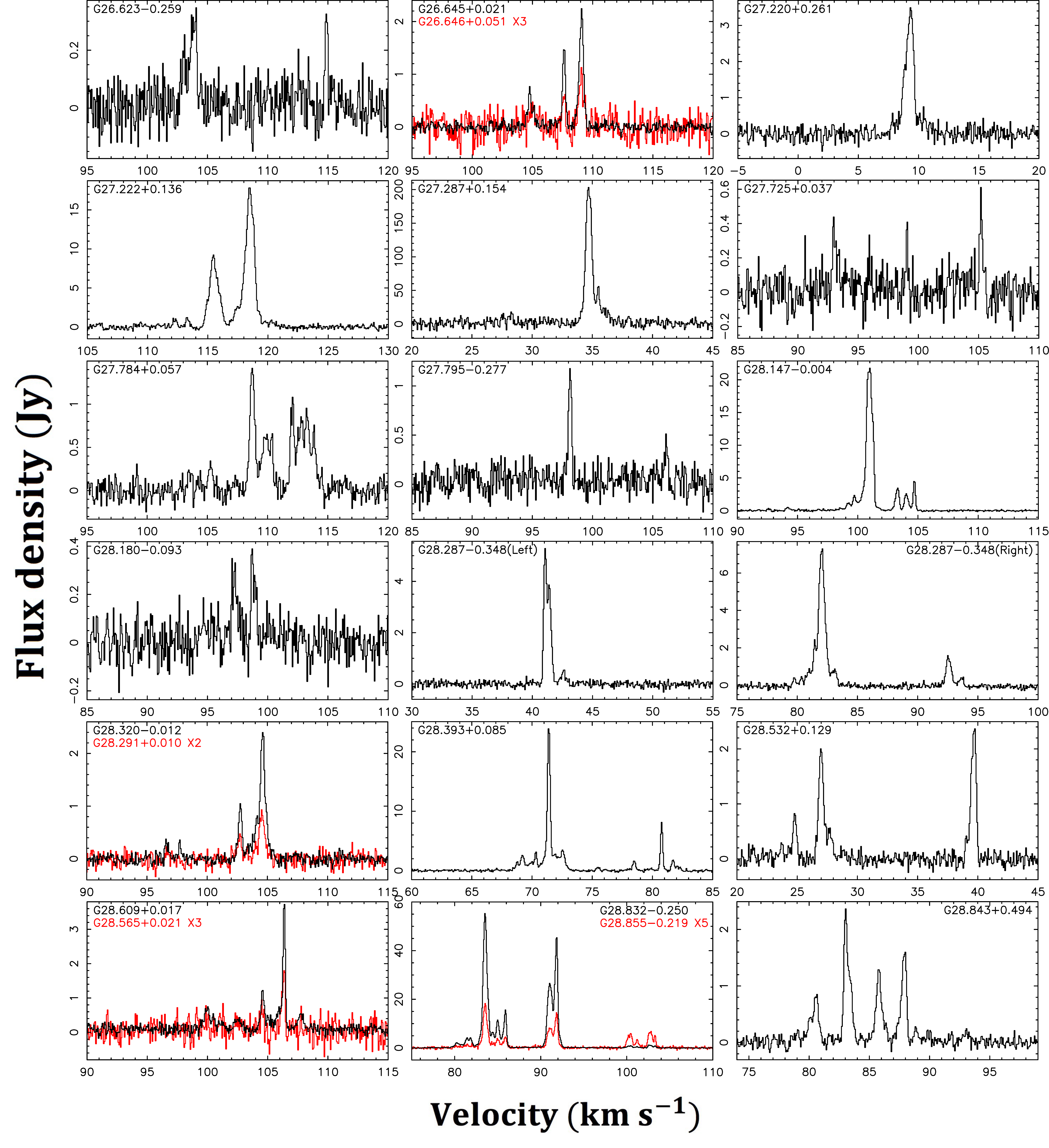}
\end{tabular}
\end{center}
\caption{Continued}
\end{figure*}
\clearpage

\begin{figure*}[h!]
\addtocounter{figure}{-1}
\begin{center}
\begin{tabular}{l}
\includegraphics[width=160mm]{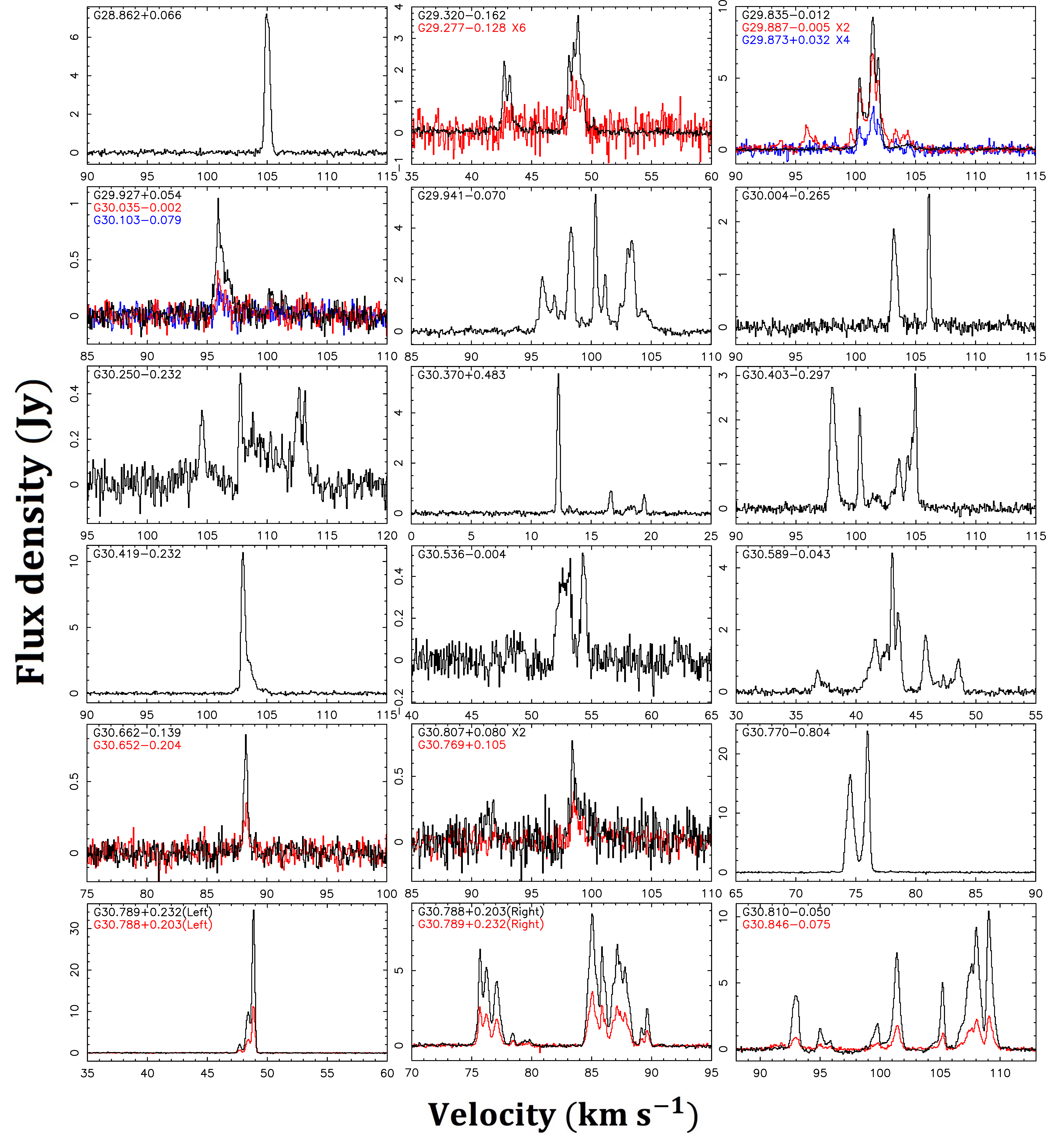}
\end{tabular}
\end{center}
\caption{Continued}
\end{figure*}
\clearpage

\begin{figure*}[h!]
\addtocounter{figure}{-1}
\begin{center}
\begin{tabular}{l}
\includegraphics[width=160mm]{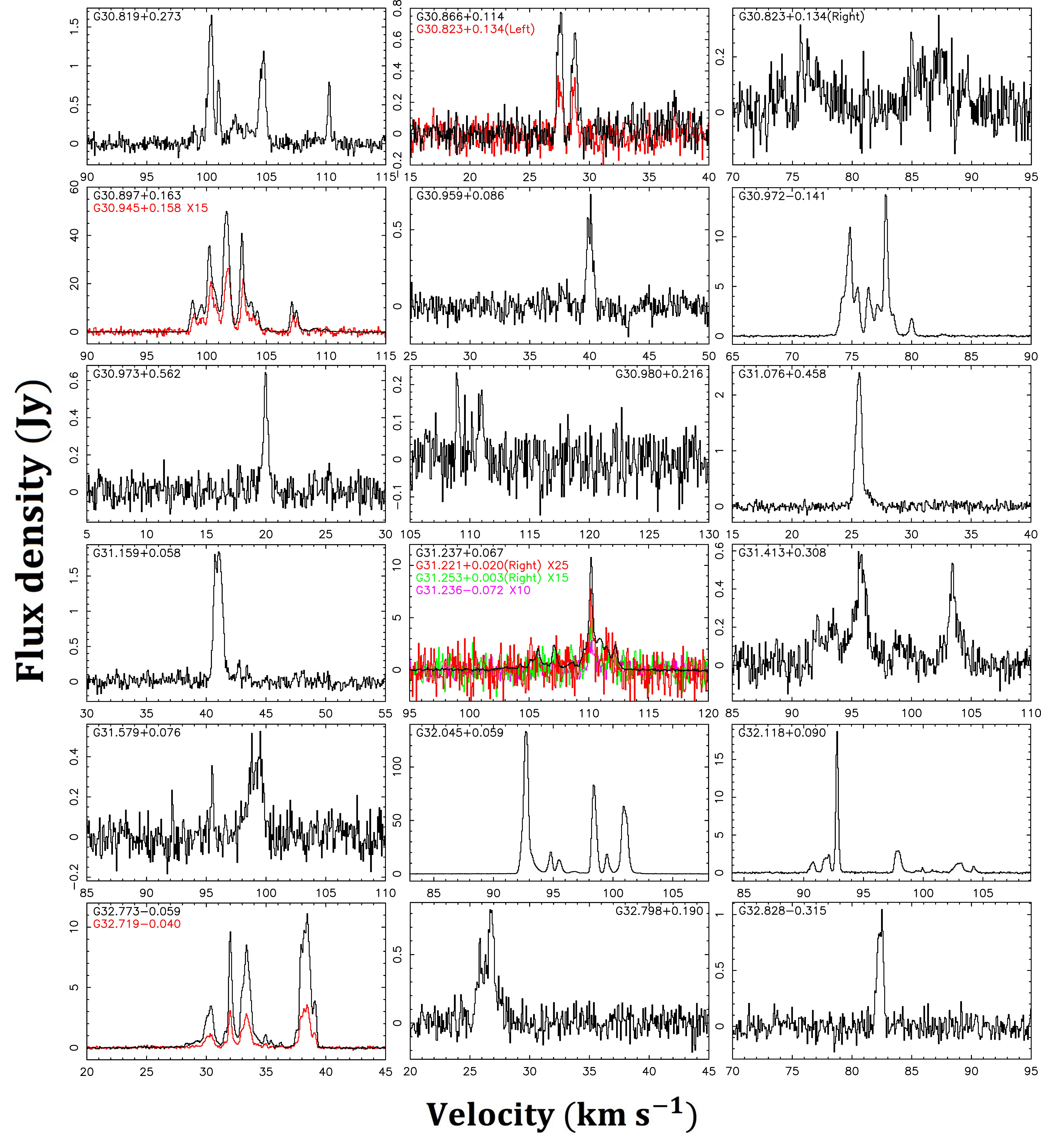}
\end{tabular}
\end{center}
\caption{Continued}
\end{figure*}
\clearpage

\begin{figure*}[h!]
\addtocounter{figure}{-1}
\begin{center}
\begin{tabular}{l}
\includegraphics[width=160mm]{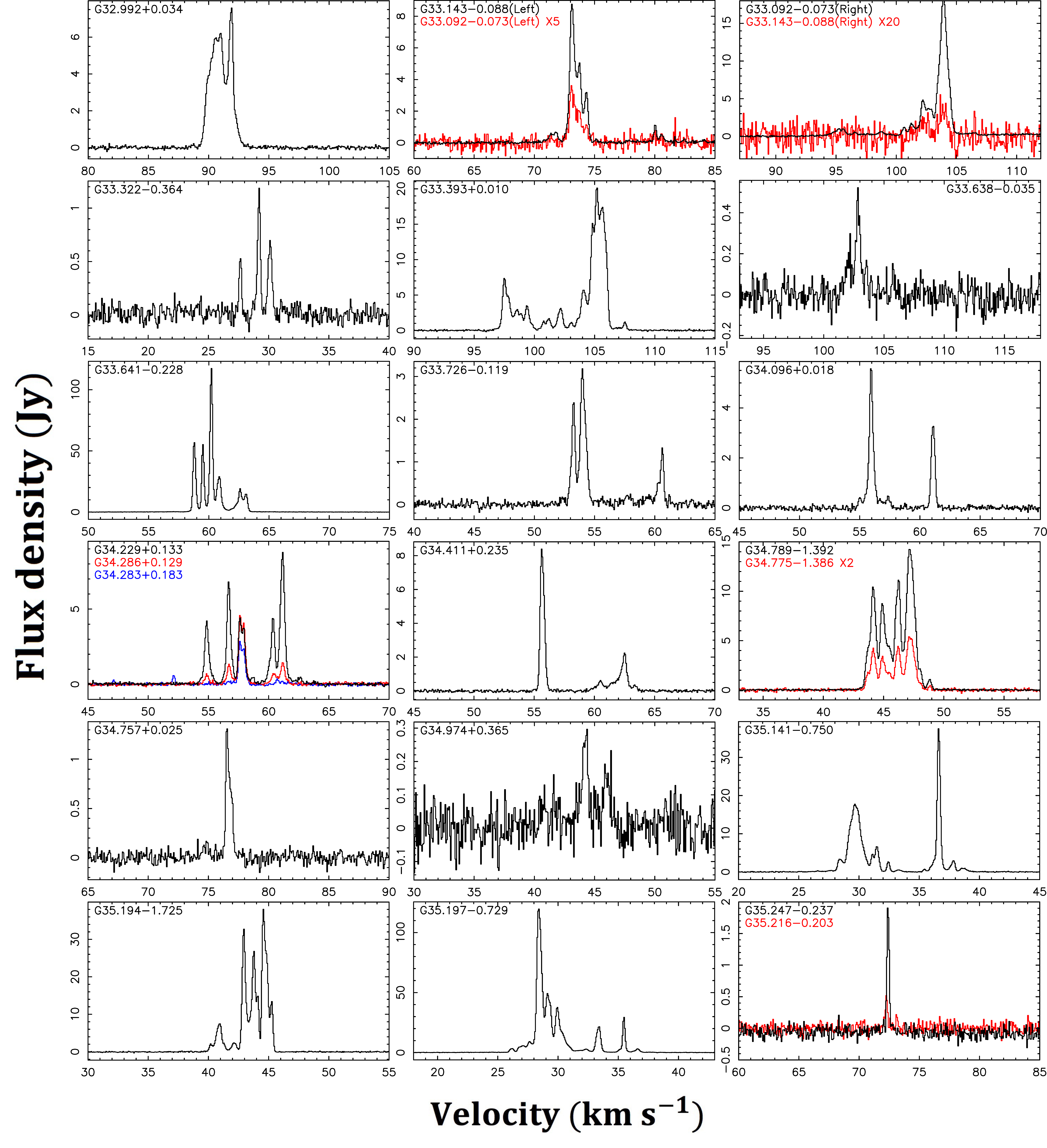}
\end{tabular}
\end{center}
\caption{Continued}
\end{figure*}
\clearpage

\begin{figure*}[h!]
\addtocounter{figure}{-1}
\begin{center}
\begin{tabular}{l}
\includegraphics[width=160mm]{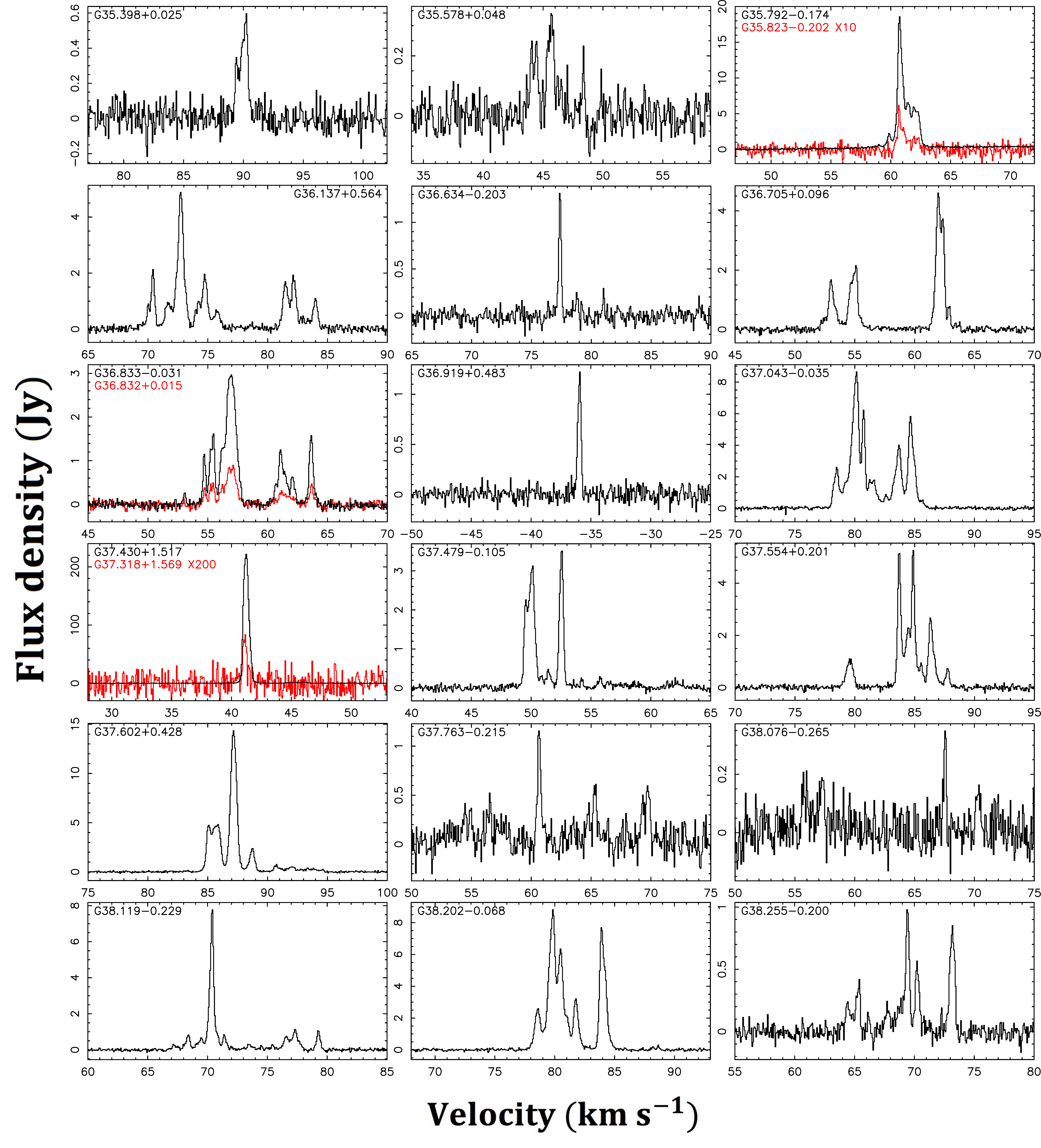}
\end{tabular}
\end{center}
\caption{Continued}
\end{figure*}
\clearpage

\begin{figure*}[h!]
\addtocounter{figure}{-1}
\begin{center}
\begin{tabular}{l}
\includegraphics[width=160mm]{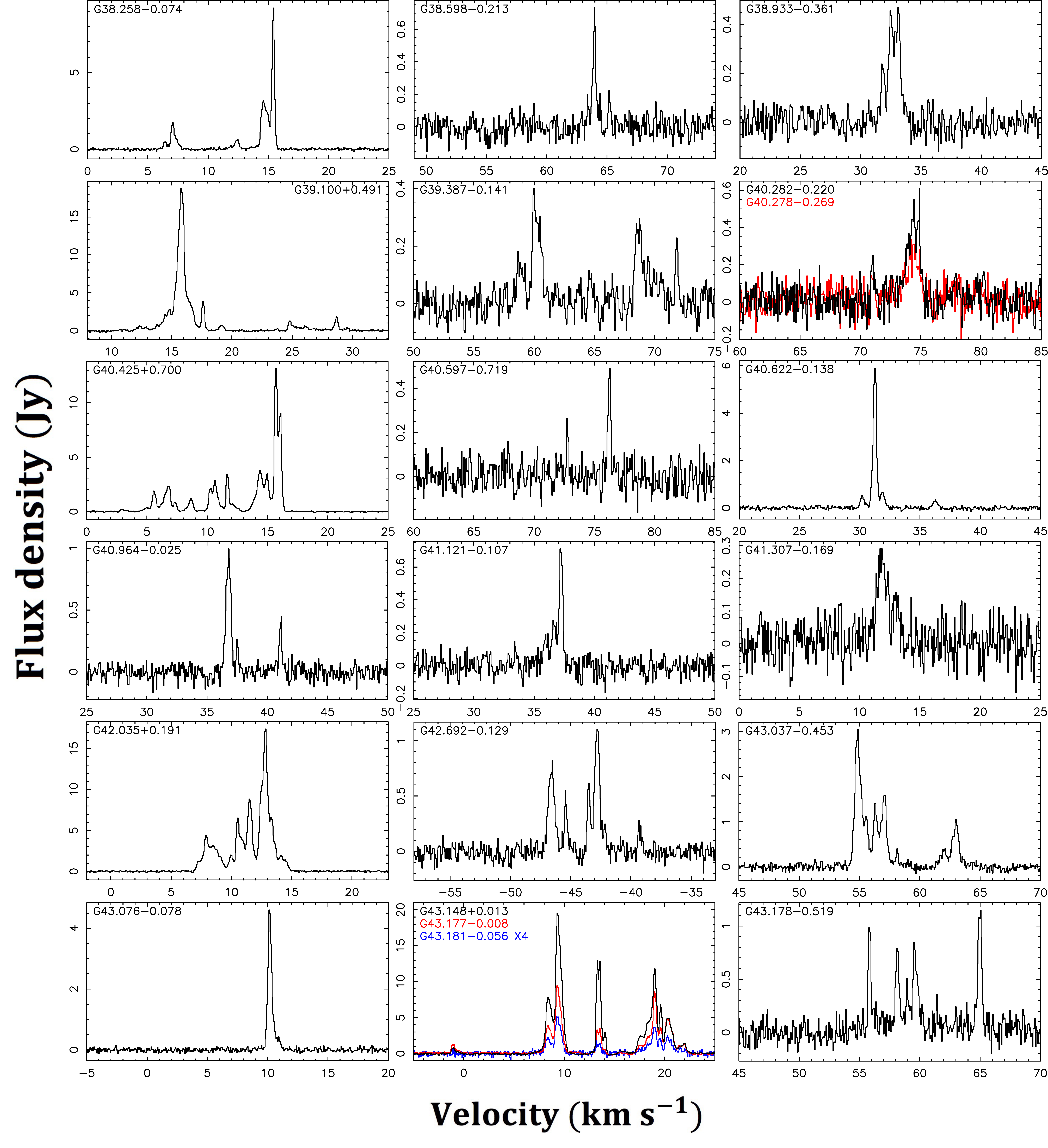}
\end{tabular}
\end{center}
\caption{Continued}
\end{figure*}
\clearpage

\begin{figure*}[h!]
\addtocounter{figure}{-1}
\begin{center}
\begin{tabular}{l}
\includegraphics[width=160mm]{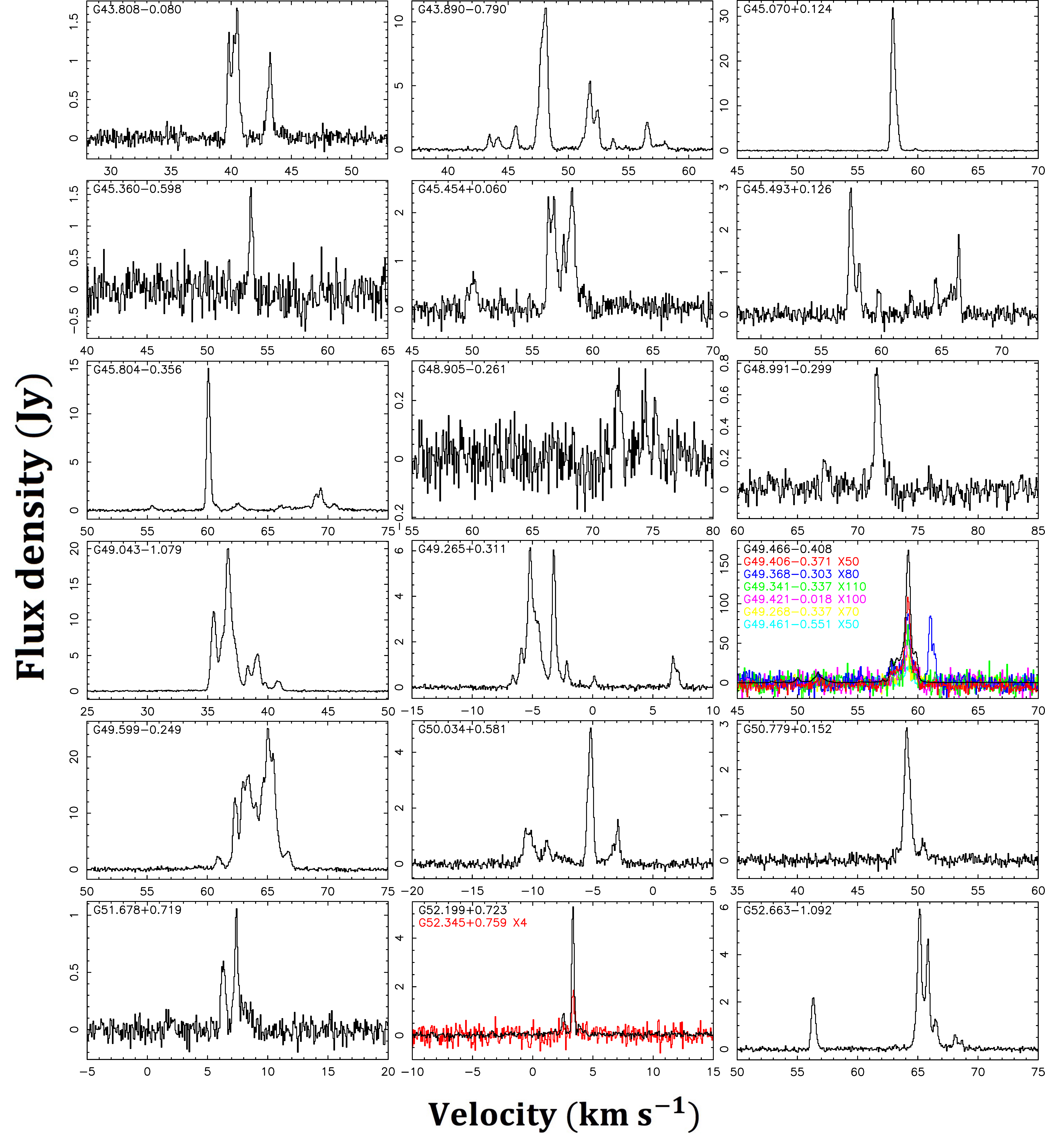}
\end{tabular}
\end{center}
\caption{Continued}
\end{figure*}
\clearpage

\begin{figure*}[h!]
\addtocounter{figure}{-1}
\begin{center}
\begin{tabular}{l}
\includegraphics[width=160mm]{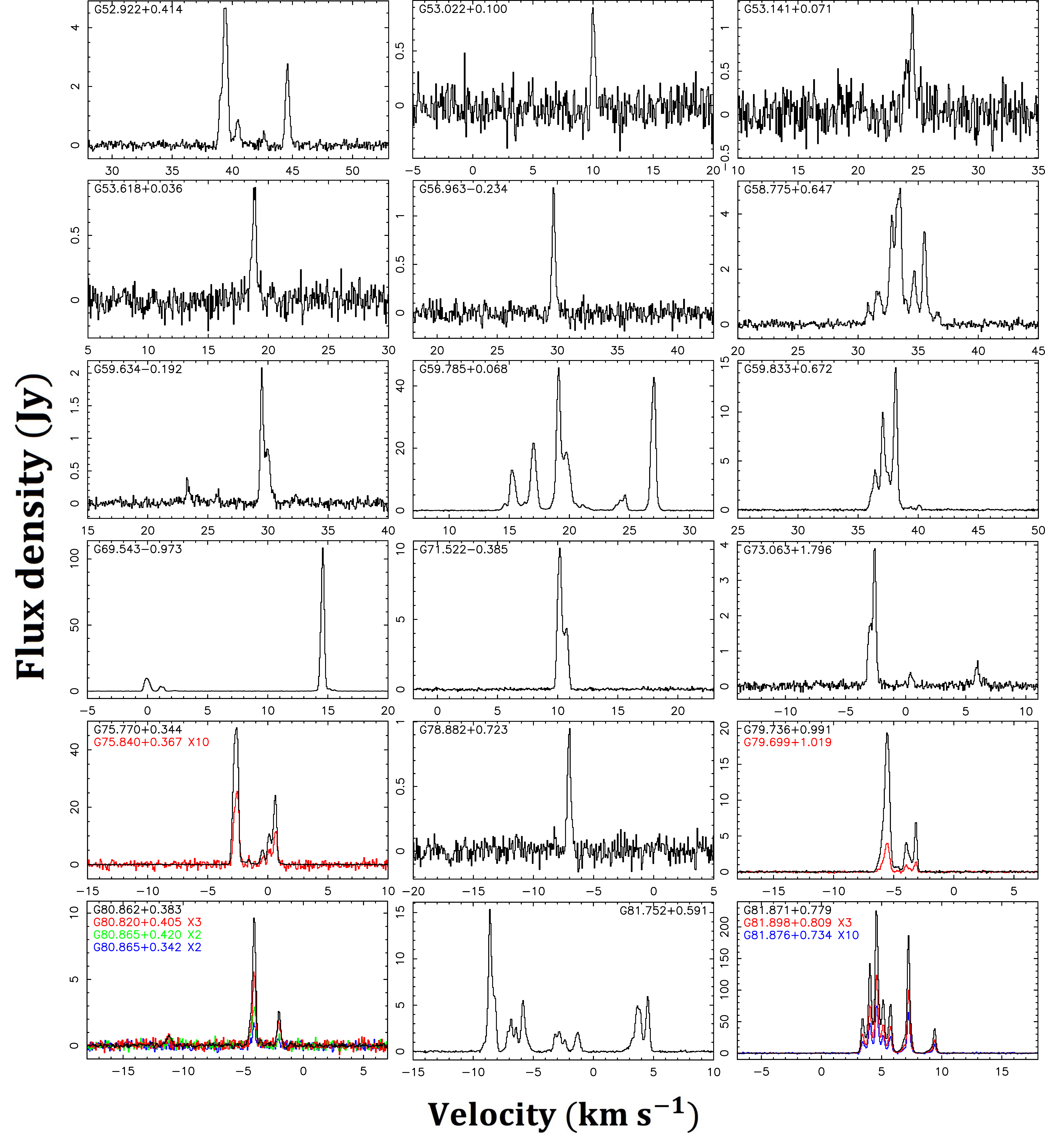}
\end{tabular}
\end{center}
\caption{Continued}
\end{figure*}
\clearpage

\begin{figure*}[h!]
\addtocounter{figure}{-1}
\begin{center}
\begin{tabular}{l}
\includegraphics[width=160mm]{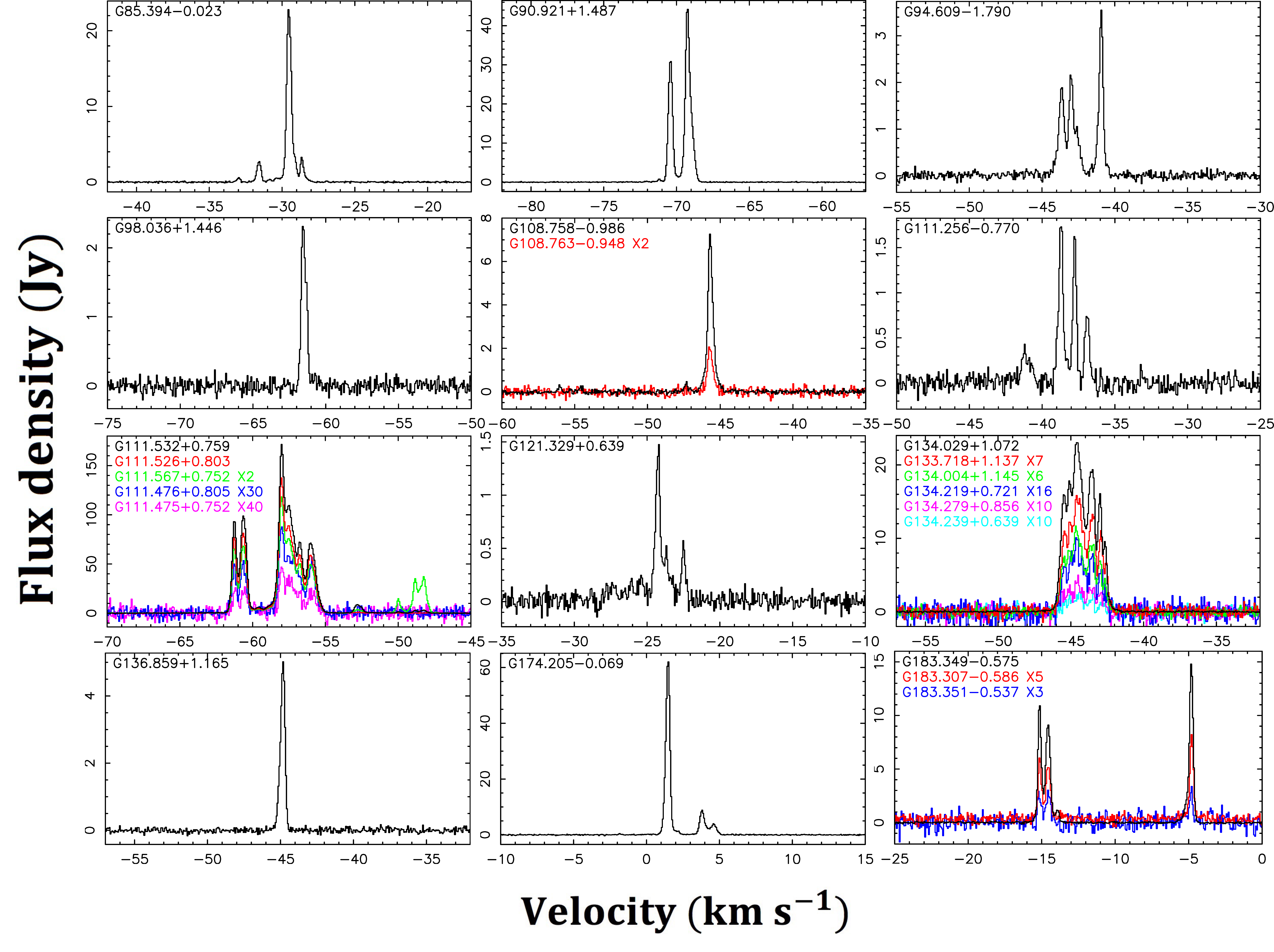}
\end{tabular}
\end{center}
\caption{Continued}
\end{figure*}
\clearpage

\figsetstart
\figsetnum{3}
\figsettitle{Velocity-integrated intensity maps obtained with TRMR OTF observations and the spectra of the 6.7 GHz CH$_3$OH methanol masers}

\figsetgrpstart
\figsetgrpnum{3.1}
\figsetgrptitle{G28.287-0.348}
\figsetplot{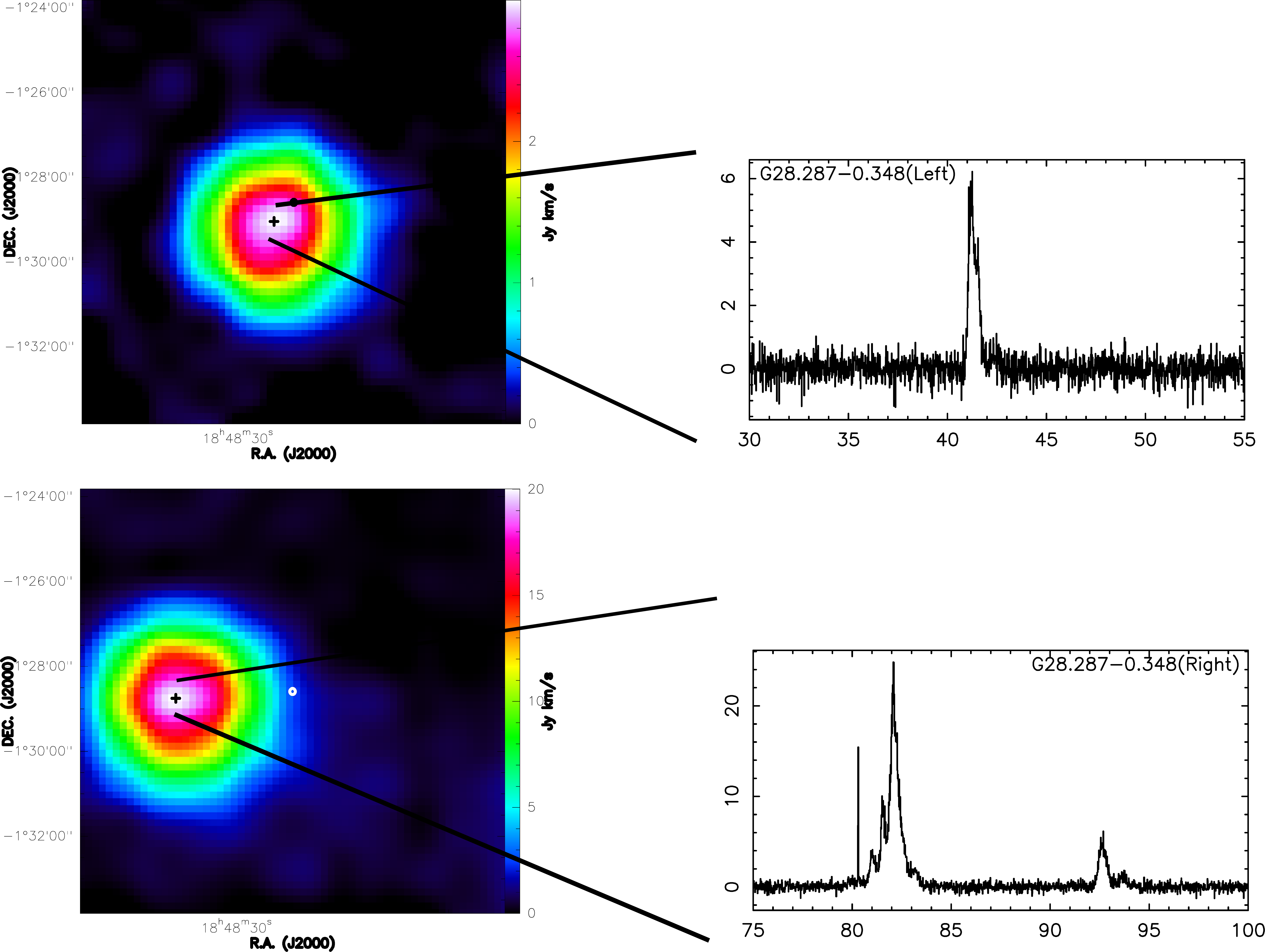}
\figsetgrpnote{The ``o" represents the position of the \emph{WISE} source in the single-point-survey, and the ``+" represents the fitted peak position of the CH$_3$OH emission from the TMRT OTF observation.}
\figsetgrpend

\figsetgrpstart
\figsetgrpnum{3.2}
\figsetgrptitle{G30.788+0.203}
\figsetplot{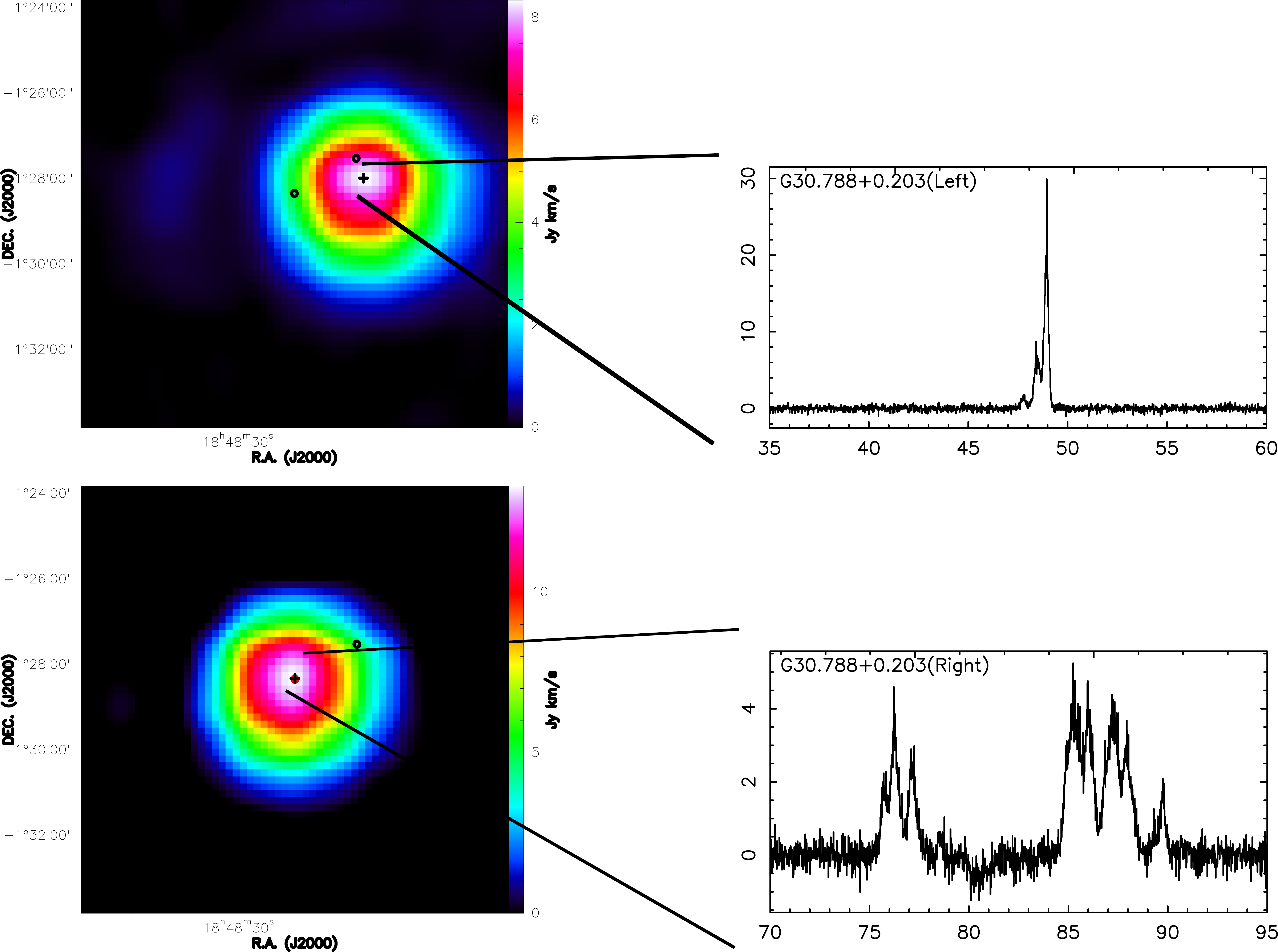}
\figsetgrpnote{The ``o" represents the position of the \emph{WISE} source in the single-point-survey, and the ``+" represents the fitted peak position of the CH$_3$OH emission from the TMRT OTF observation.}
\figsetgrpend

\figsetgrpstart
\figsetgrpnum{3.3}
\figsetgrptitle{G30.823+0.134}
\figsetplot{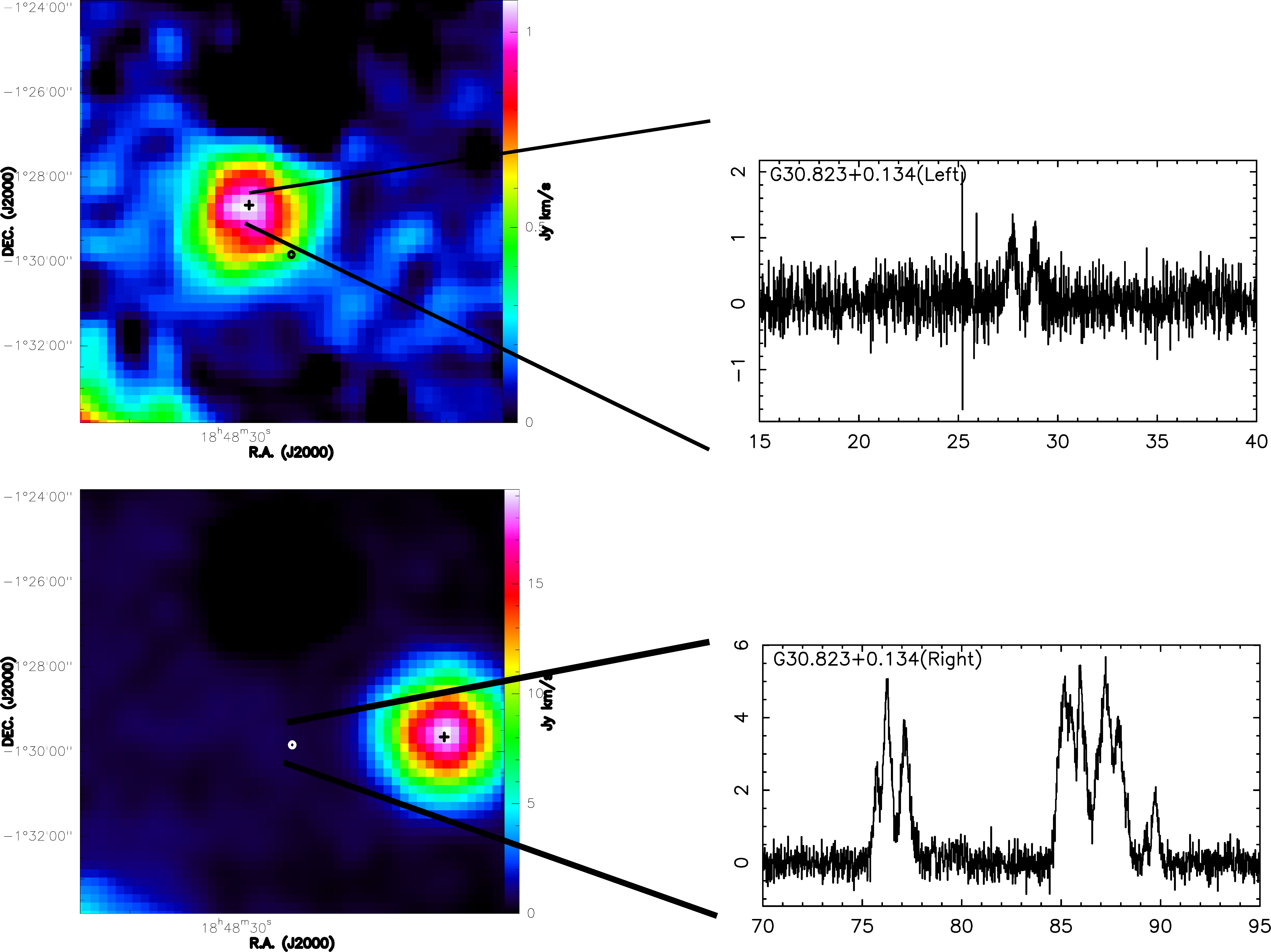}
\figsetgrpnote{The ``o" represents the position of the \emph{WISE} source in the single-point-survey, and the ``+" represents the fitted peak position of the CH$_3$OH emission from the TMRT OTF observation.}
\figsetgrpend

\figsetgrpstart
\figsetgrpnum{3.4}
\figsetgrptitle{G31.221+0.020}
\figsetplot{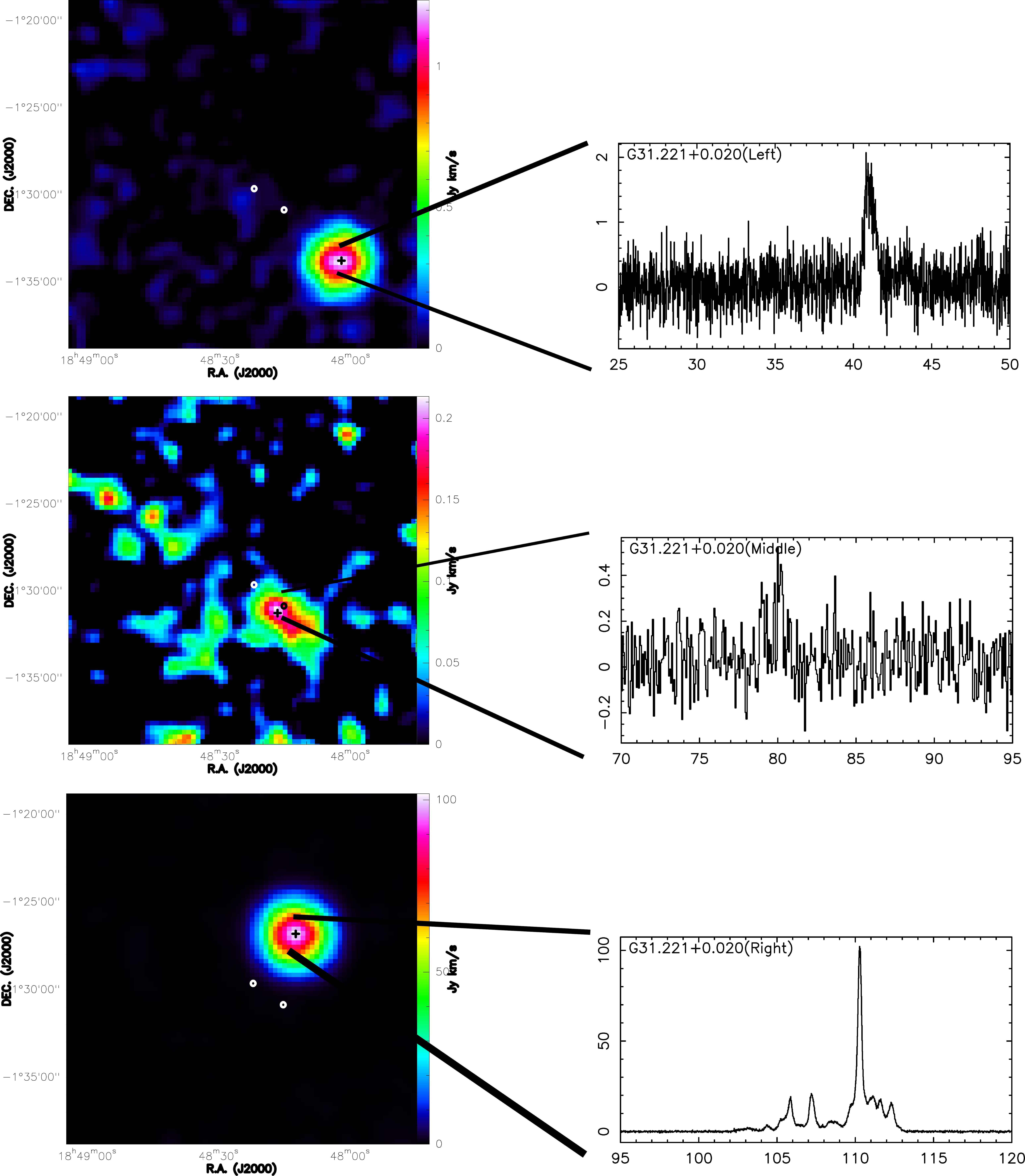}
\figsetgrpnote{The ``o" represents the position of the \emph{WISE} source in the single-point-survey, and the ``+" represents the fitted peak position of the CH$_3$OH emission from the TMRT OTF observation.}
\figsetgrpend

\figsetgrpstart
\figsetgrpnum{3.5}
\figsetgrptitle{G33.143-0.088}
\figsetplot{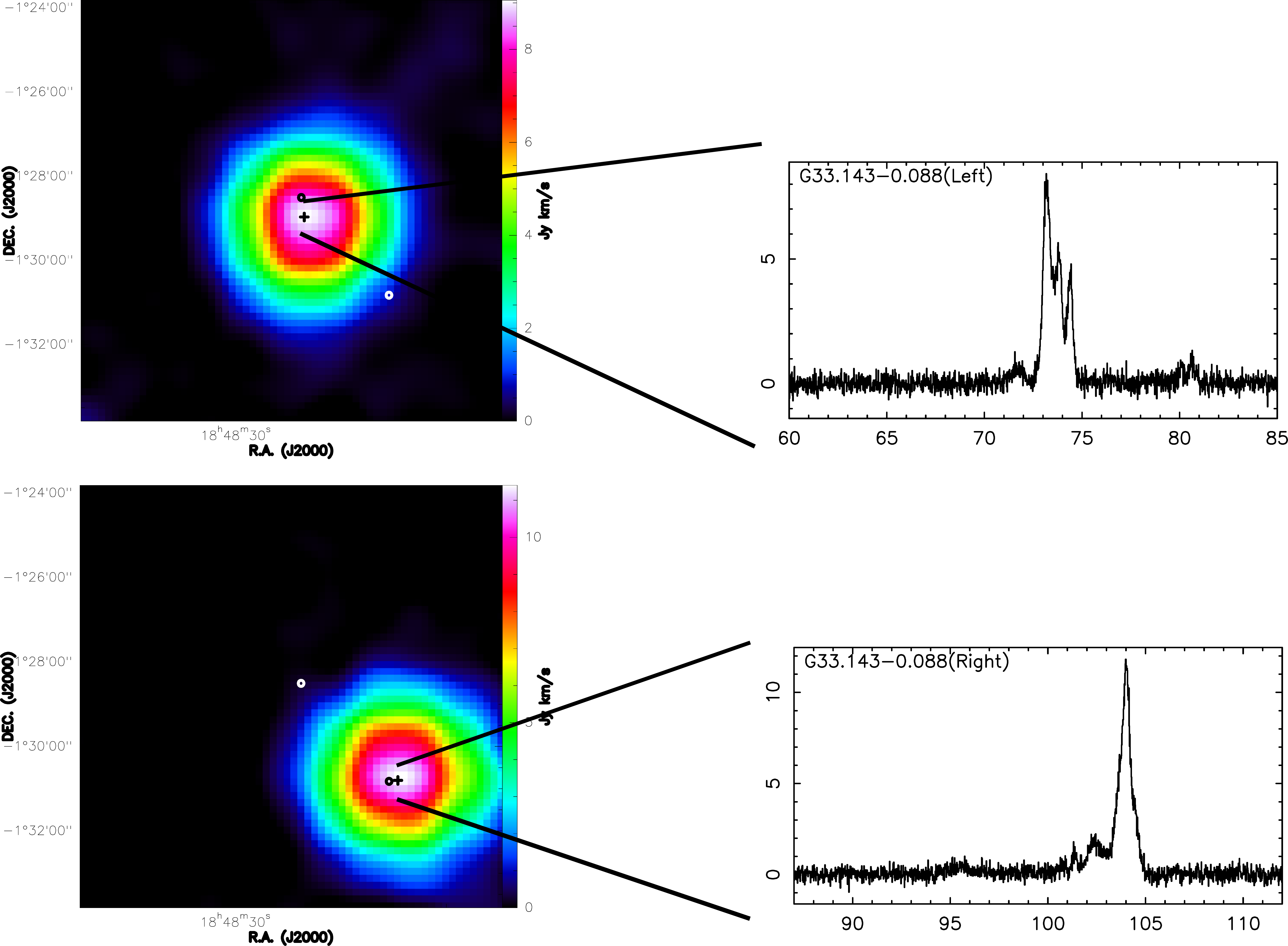}
\figsetgrpnote{The ``o" represents the position of the \emph{WISE} source in the single-point-survey, and the ``+" represents the fitted peak position of the CH$_3$OH emission from the TMRT OTF observation.}
\figsetgrpend

\figsetend

\newpage
\begin{sidewaysfigure}[h!]
\centering
\includegraphics[width=15cm]{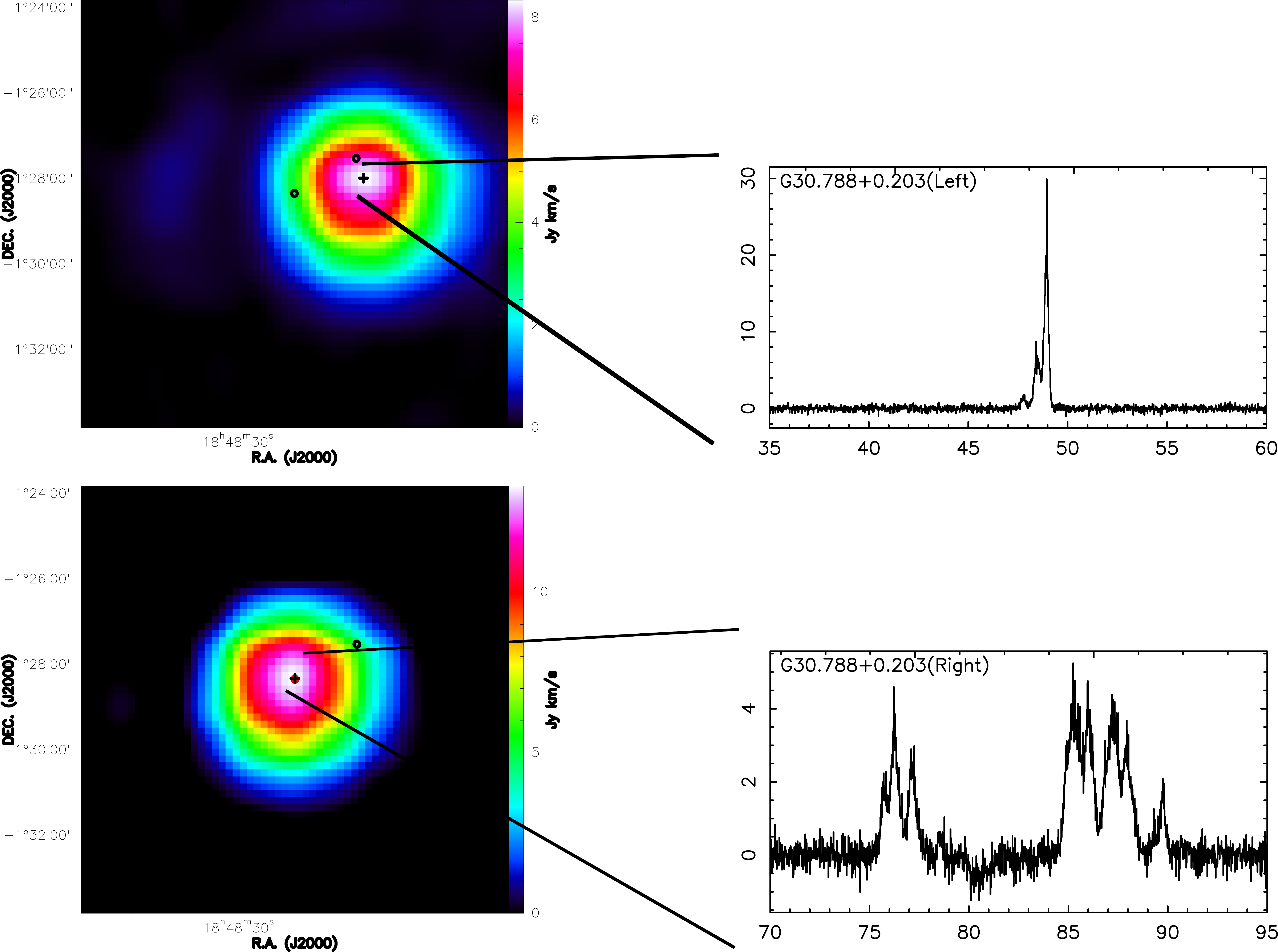}
\caption{Velocity-integrated intensity maps obtained with TRMR OTF observations and the spectra of the 6.7 GHz CH$_3$OH methanol masers from G30.788+0.203 (upper) and G30.788+0.203 (lower). The ``o" represents the position of the \emph{WISE} source in the single-point-survey, and the ``+" represents the fitted peak position of the CH$_3$OH emission from the TMRT OTF observation. The complete figure of 5 images is available online.}
\end{sidewaysfigure}
\clearpage

\newpage
\begin{figure*}[h!]
\begin{center}
\begin{tabular}{l}
\includegraphics[width=10cm]{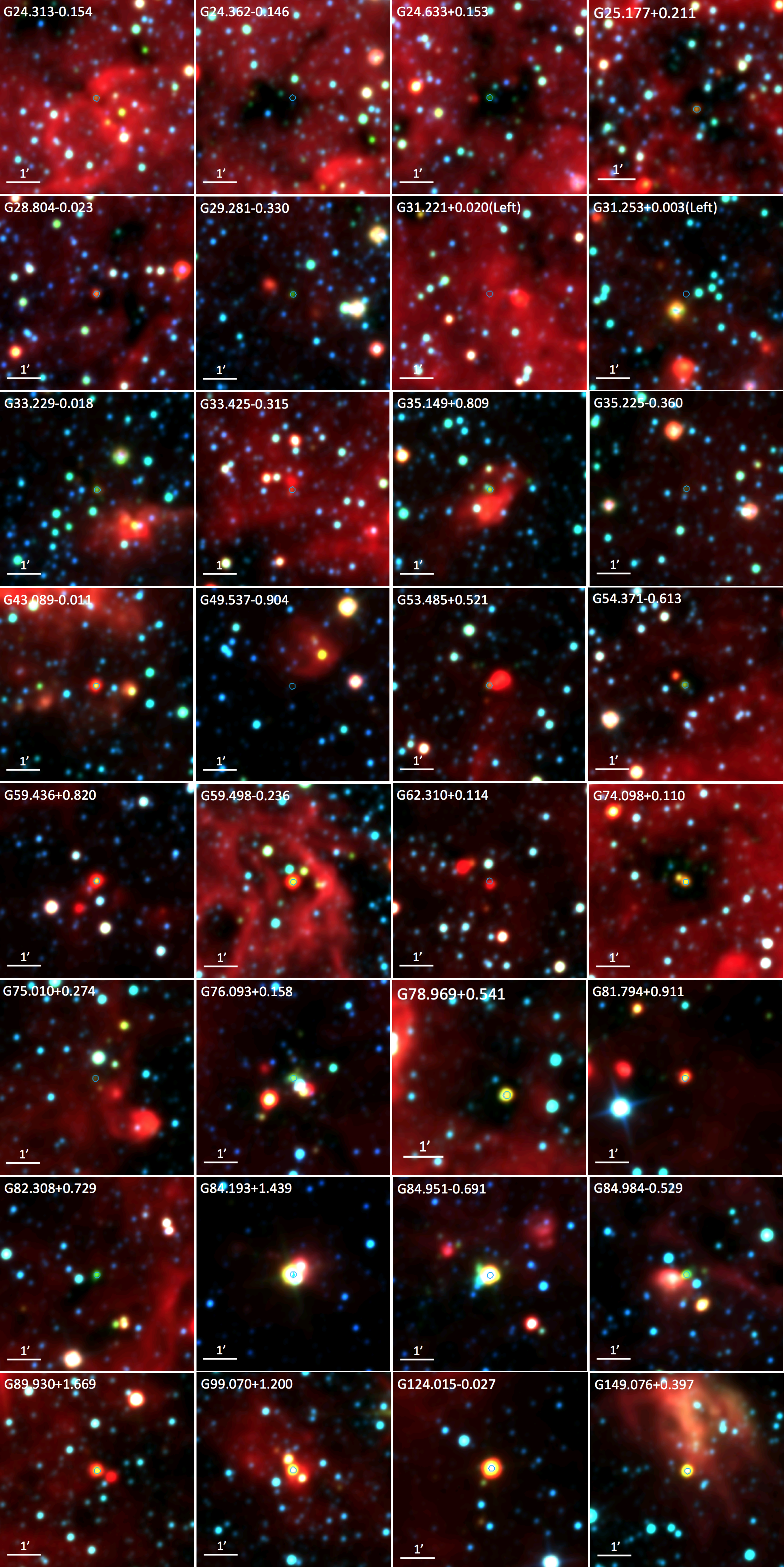}
\end{tabular}
\end{center}
\caption{Infrared three-color images of the 32 newly detected methanol sources.  The blue, green, and red colors represent 3.4, 4.6, and 12 $\mu$m bands in \emph{WISE}. The blue ``o" in the middle of each image represents the WISE source position used for the TMRT observations. The field of view of each region is a square with a side length of 6$^{\prime}$.}
\end{figure*}
\clearpage

\newpage
\begin{sidewaysfigure}[h]
\centering
\includegraphics[width=12cm]{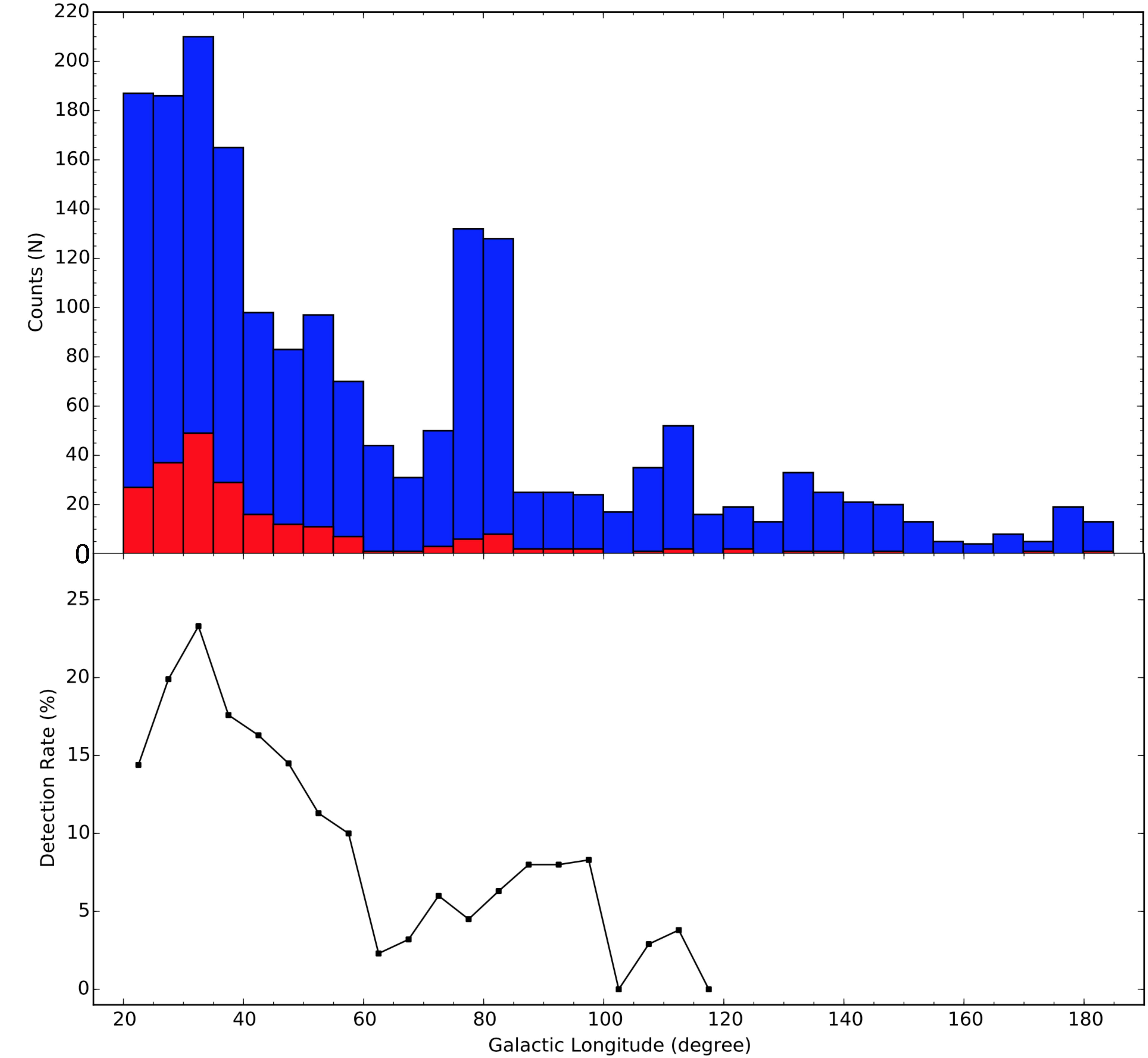}
\caption{Upper panel: Histogram of the observing sample source number (blue) and the detected source number (red). Lower panel: detection rate (line chart) along every 5$^\circ$ Galactic longitude.}
\end{sidewaysfigure}
\clearpage

\newpage
\begin{figure*}
\centering
\includegraphics[width=12cm]{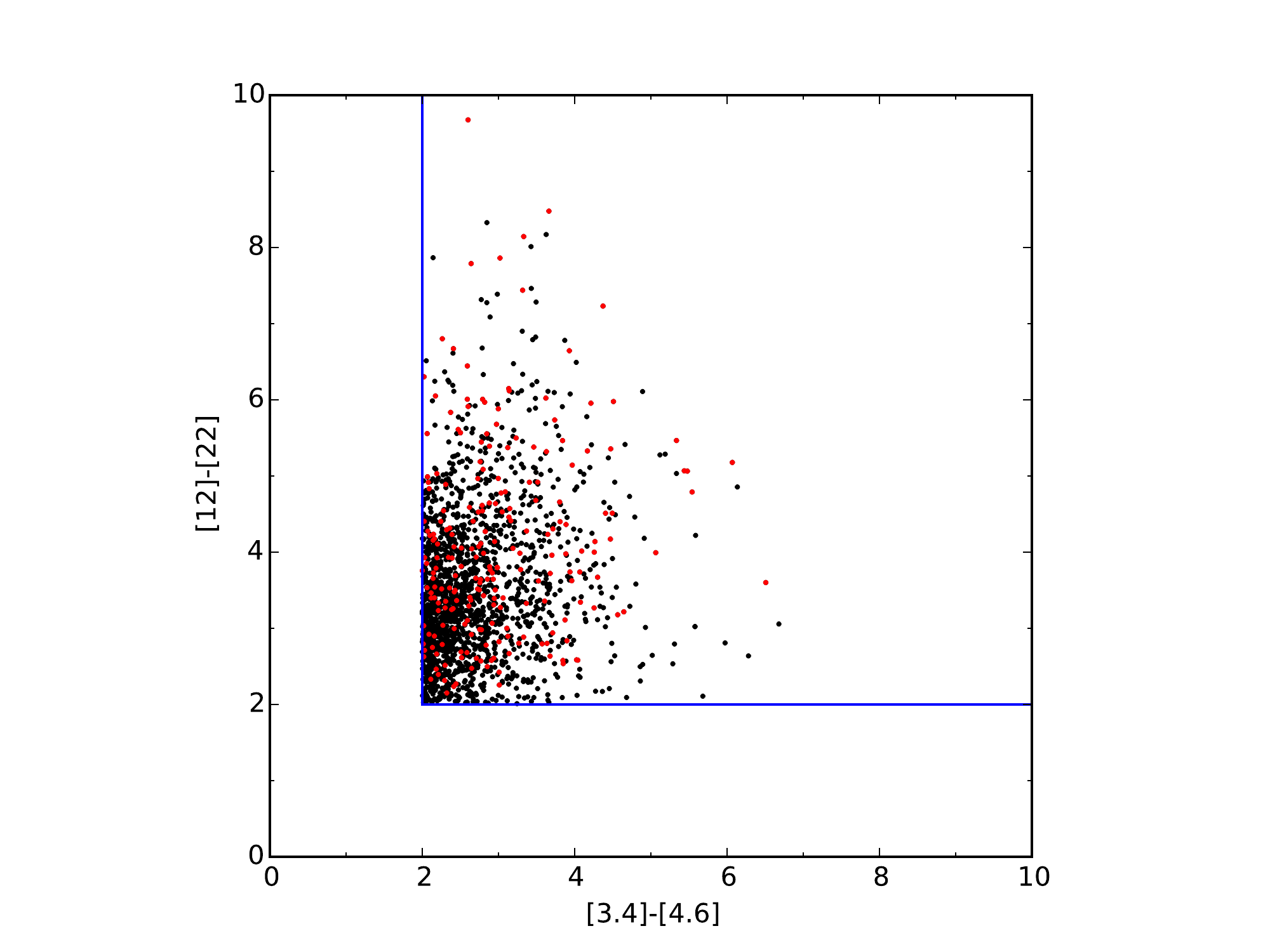}
\includegraphics[width=8.1cm]{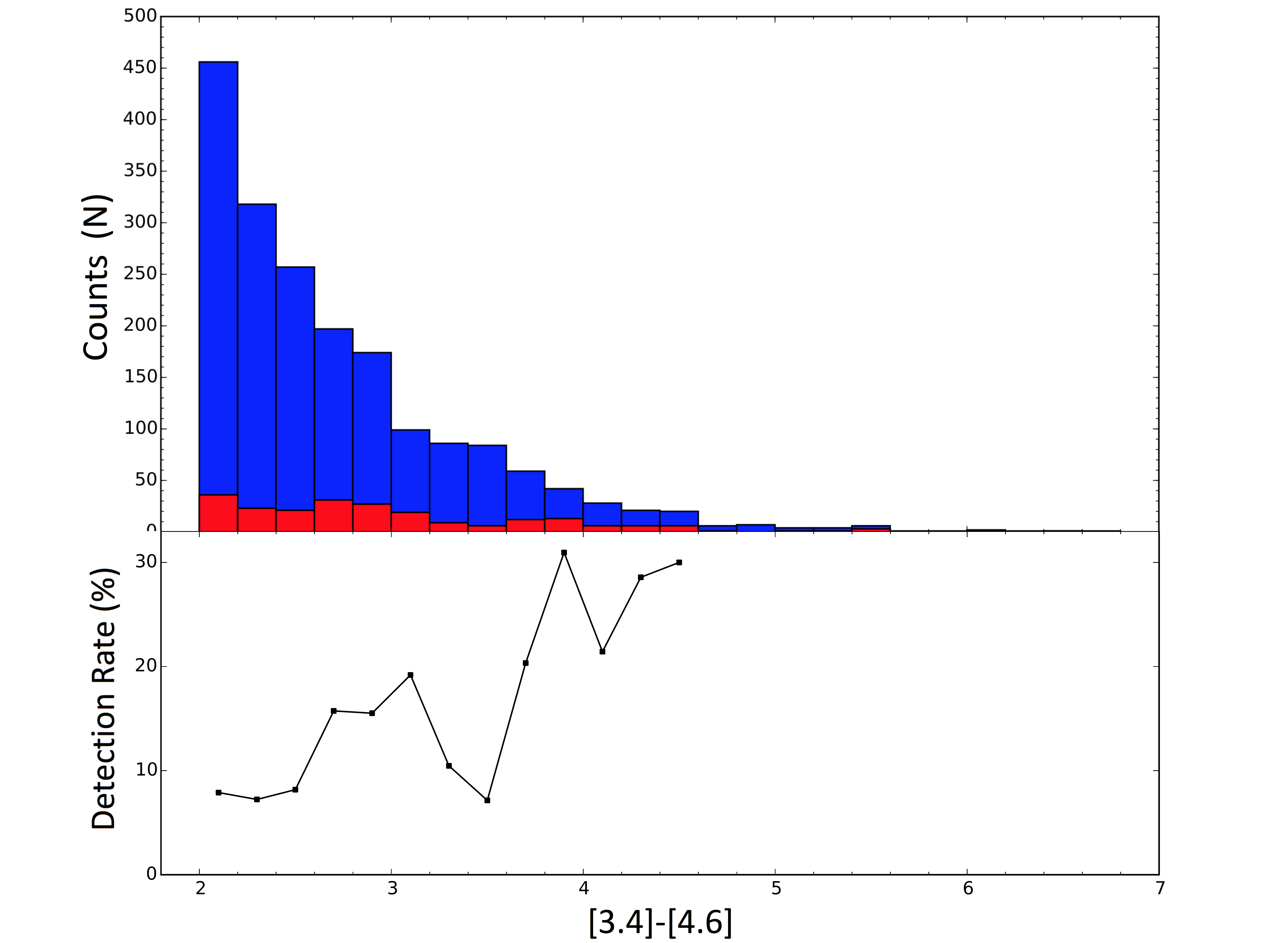}
\includegraphics[width=8.1cm]{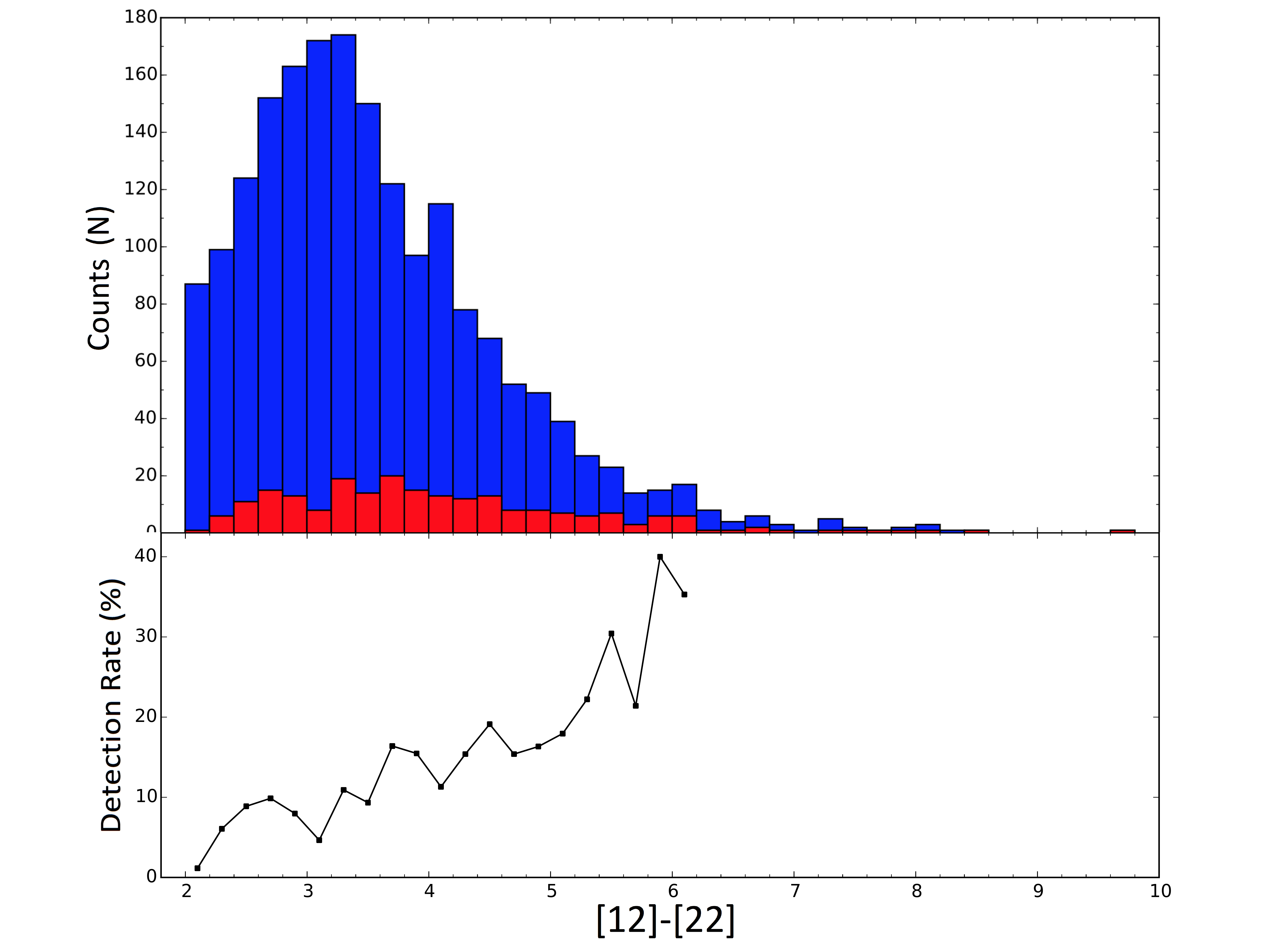}
\caption{Upper panel: [3.4]$-$[4.6] versus [12]$-$[22] color of detected sources (red dots) and the 1875 observing sample sources (black dots) in our survey. Lower panels: the sample source numbers (histogram, blue), the detected source numbers (histogram, red) and detection rate (line chart) along every 0.2 color of [3.4]$-$[4.6] (left) and [12]$-$[22] (right).}
\end{figure*}
\clearpage

\newpage
\begin{sidewaysfigure}[h]
\centering
\includegraphics[width=24cm]{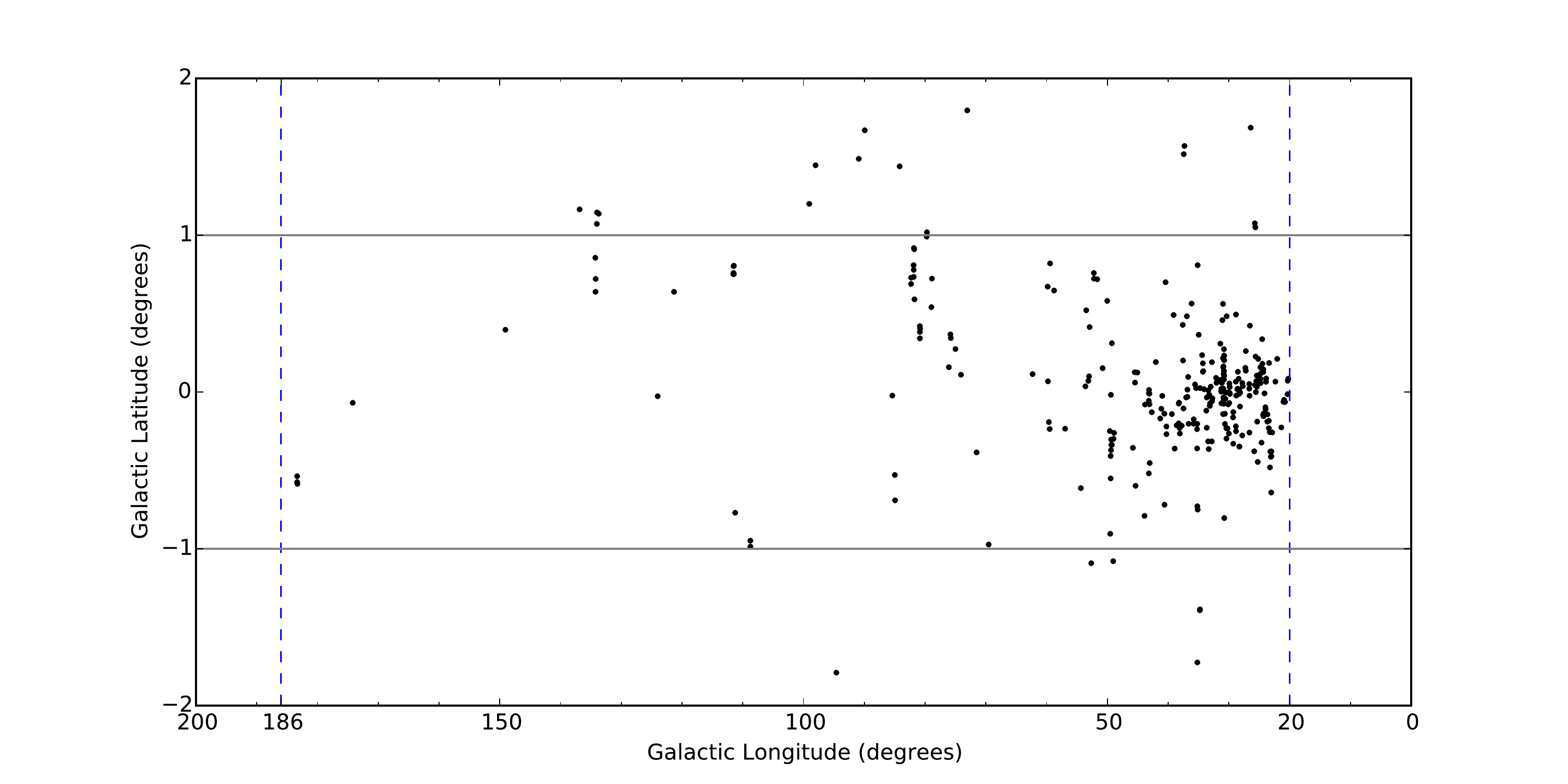}
\caption{Galactic latitude distribution of the 224 6.7 GHz methanol maser sources detected by our TMRT survey at low latitude region, 20$^\circ$ $<|l|<$ 186$^\circ$ and $|b|<$ 2$^\circ$.}
\end{sidewaysfigure}
\clearpage

\newpage
\begin{figure*}
\centering
\includegraphics[width=15cm]{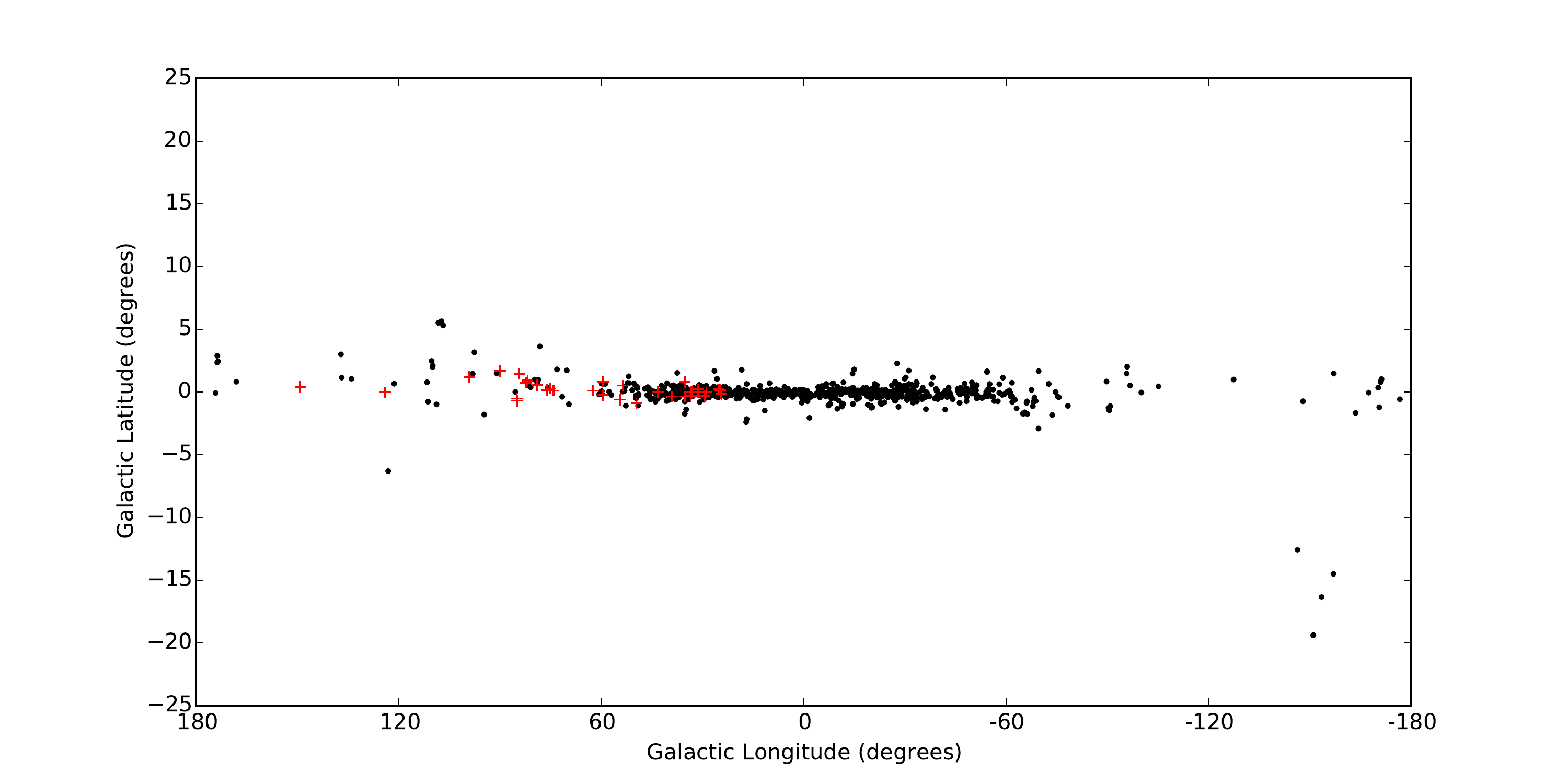}
\includegraphics[width=15cm]{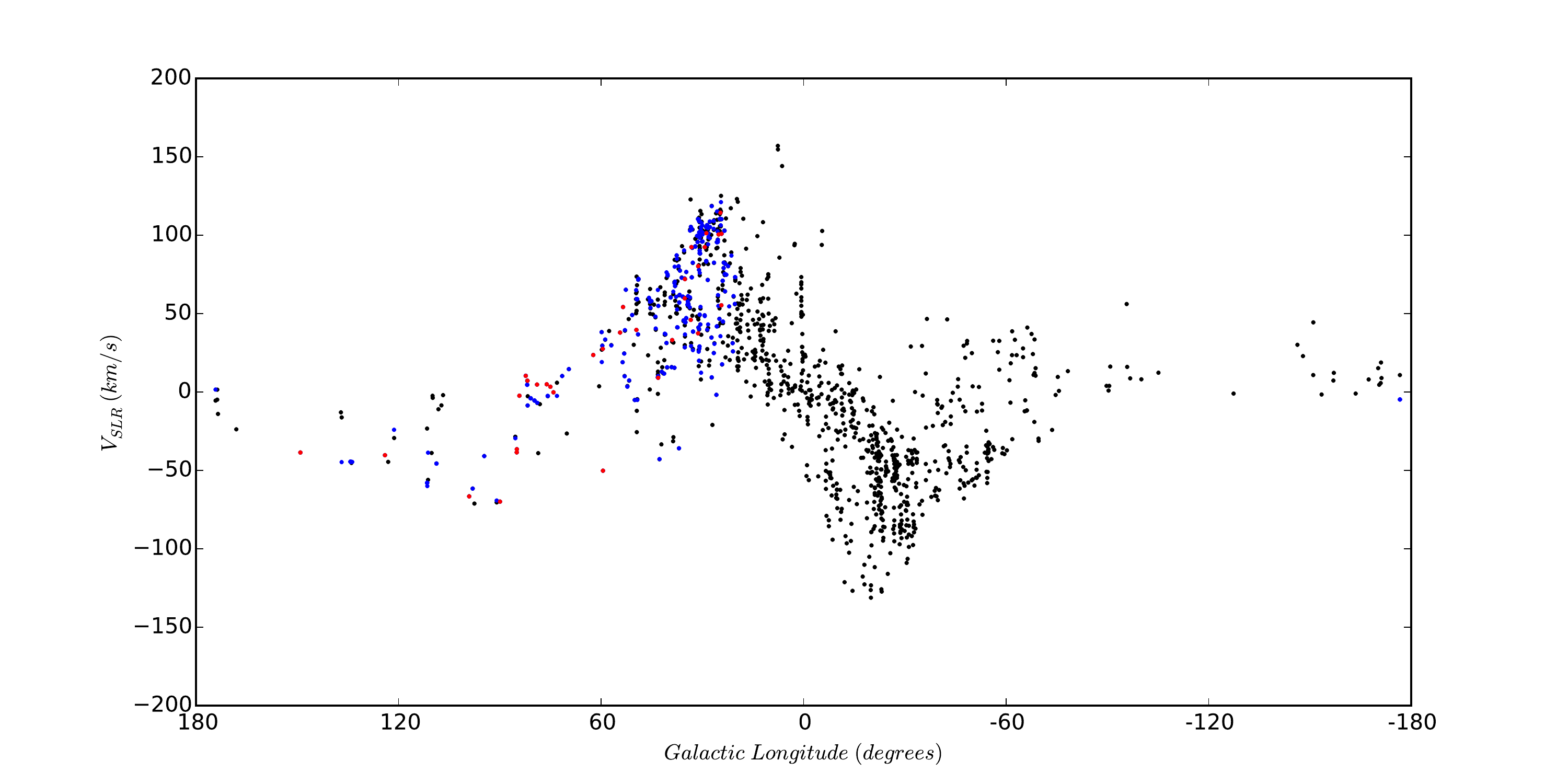}
\includegraphics[width=9cm]{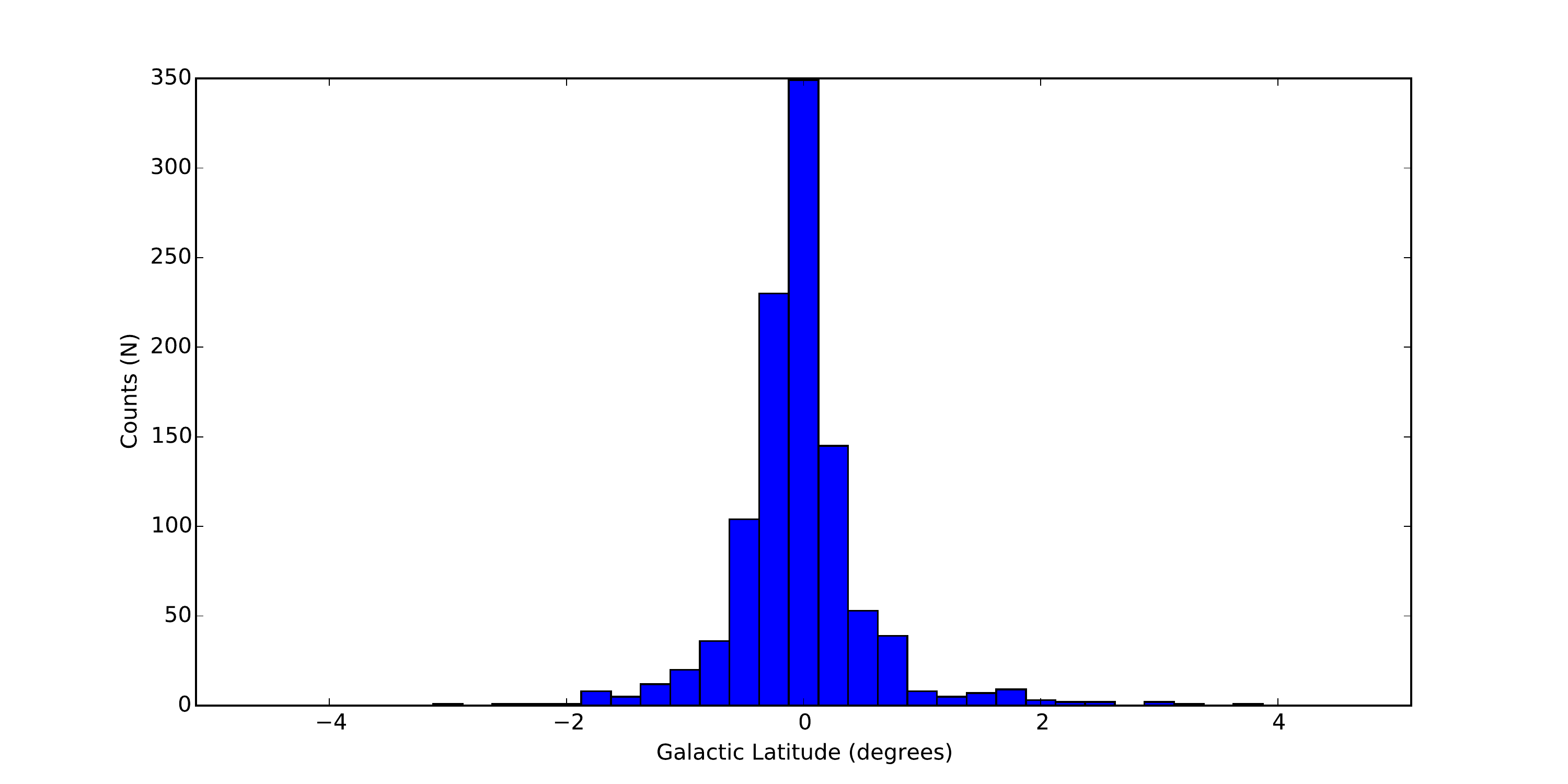}
\includegraphics[width=6cm]{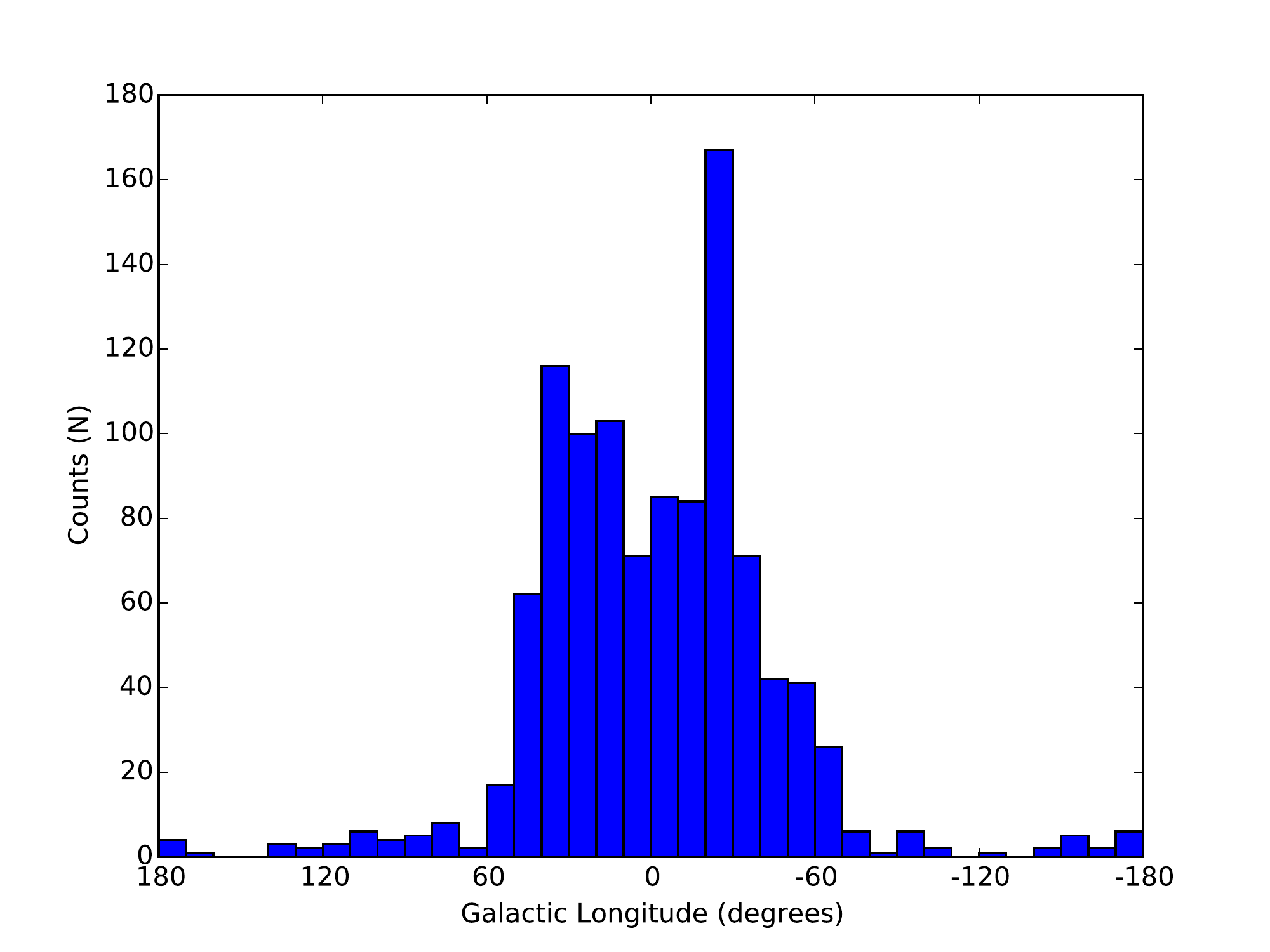}
\caption{Top: Distribution of 1085 6.7 GHz methanol maser sources. The red ``+" represents 32 newly detected sources in our survey. Middle: Distribution of the LSR velocities versus Galactic longitude. The blue dots represent the 224 detected sources in our survey and the red dots represent the 32 newly detected sources in our survey. Bottom: Source numbers versus galactic latitude, $|b| \le$ 5$^\circ$ (left panel), and galactic longitude (right panel).}
\end{figure*}
\clearpage

\newpage
\begin{figure*}
\centering
\includegraphics[width=15cm]{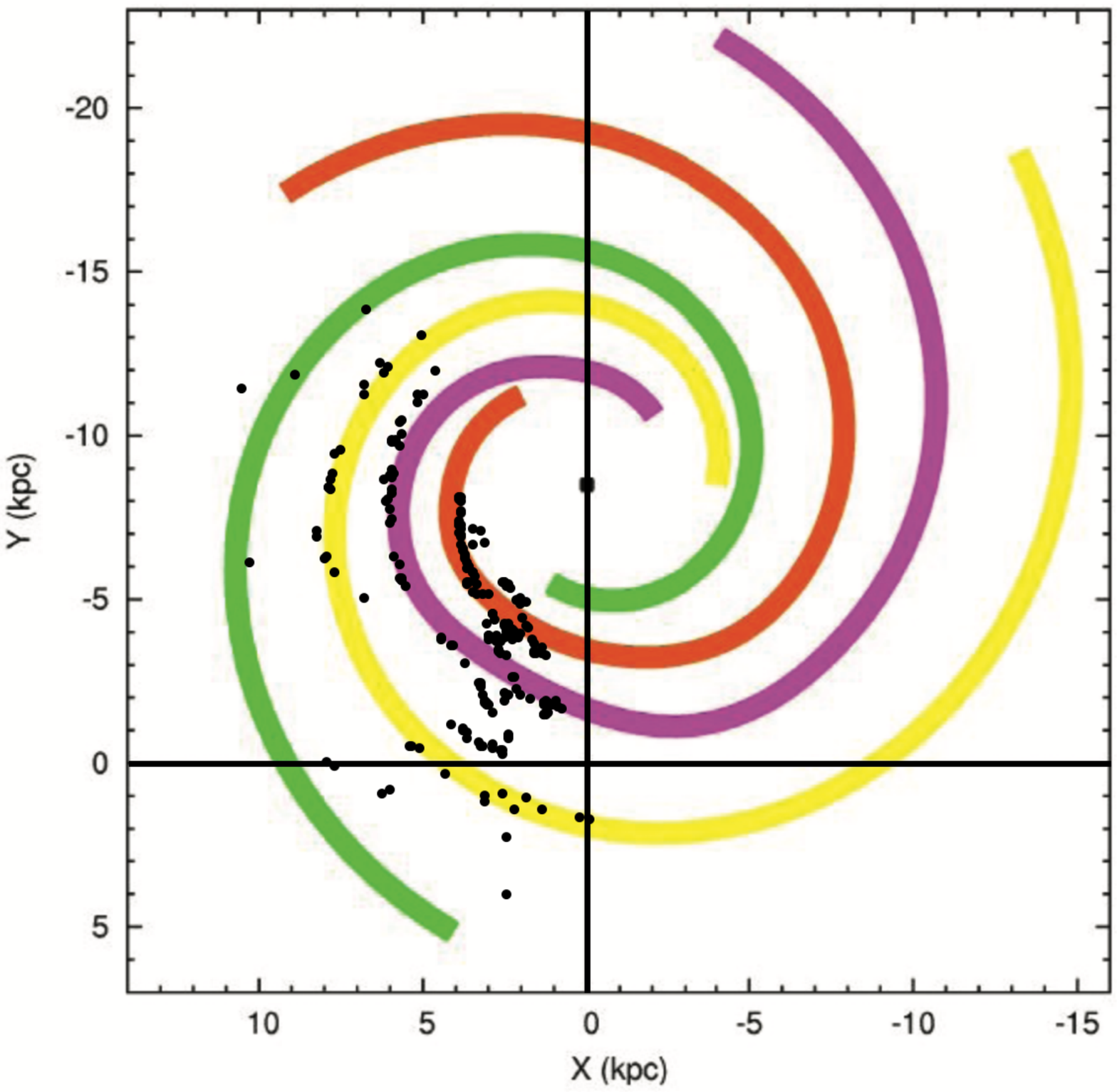}
\caption{The black dots represent the positions of the detected sources in our survey. The Sun locates at the origin point (0, 0). The spiral arms (yellow -- Perseus; purple -- Carina-Sagittarius; orange -- Crux-Scutum; green -- Norma) are illustrated same as \citet[]{Gre2010}. }
\end{figure*}
\clearpage


\begin{thebibliography}{99}

\bibitem[Araya et al.(2010)]{Ara2010} Araya, E.~D., Hofner, P., Goss, W.~M., et al.\ 2010, \apjl, 717, L133 

\bibitem[Bartkiewicz et al.(2016)]{Bar2016} Bartkiewicz, A., Szymczak, M., \& van Langevelde, H.~J.\ 2016, \aap, 587, A104

\bibitem[Batrla et al.(1987)]{Bat1987} Batrla, W., Matthews, H.~E., Menten, K.~M., \& Walmsley, C.~M.\ 1987, \nat, 326, 49

\bibitem[Breen et al.(2015)]{Bre2015} Breen, S.~L., Fuller, G.~A., Caswell, J.~L., et al.\ 2015, \mnras, 450, 4109

\bibitem[Caratti o Garatti et al.(2017)]{Car2017} Caratti o Garatti, A., Stecklum, B., Garcia Lopez, R., et al.\ 2017, Nature Physics, 13, 276 

\bibitem[Caswell(1996)]{Cas1996} Caswell, J.~L.\ 1996, \mnras, 283, 606

\bibitem[Caswell et al.(2010)]{Cas2010} Caswell, J.~L., Fuller, G.~A., Green, J.~A., et al.\ 2010, \mnras, 404, 1029

\bibitem[Caswell et al.(2011)]{Cas2011} Caswell, J.~L., Fuller, G.~A., Green, J.~A., et al.\ 2011, \mnras, 417, 1964

\bibitem[Caswell et al.(1995b)]{Cas1995b} Caswell, J.~L., Vaile, R.~A., \& Ellingsen, S.~P.\ 1995, \pasa, 12, 37

\bibitem[Caswell et al.(1995a)]{Cas1995} Caswell, J.~L., Vaile, R.~A., Ellingsen, S.~P., Whiteoak, J.~B., \& Norris, R.~P.\ 1995, \mnras, 272, 96

\bibitem[Chibueze et al.(2017)]{Chi2017} Chibueze, J.~O., Csengeri, T., Tatematsu, K., et al.\ 2017, \apj, 836, 59

\bibitem[Csengeri et al.(2017a)]{Cse2017a} Csengeri, T., Bontemps, S., Wyrowski, F., et al.\ 2017, \aap, 600, L10

\bibitem[Csengeri et al.(2017b)]{Cse2017b} Csengeri, T., Bontemps, S., Wyrowski, F., et al.\ 2017, \aap, 601, A60 

\bibitem[Csengeri et al.(2018)]{Cse2018} Csengeri, T., Bontemps, S., Wyrowski, F., et al.\ 2018, \aap, 617, A89 

\bibitem[Csengeri et al.(2014)]{Cse2014} Csengeri, T., Urquhart, J.~S., Schuller, F., et al.\ 2014, \aap, 565, A75 

\bibitem[Cyganowski et al.(2009)]{Cyg2009} Cyganowski, C.~J., Brogan, C.~L., Hunter, T.~R., \& Churchwell, E.\ 2009, \apj, 702, 1615 

\bibitem[Dame et al.(2001)]{Dame2001} Dame, T.~M., Hartmann, D., \& Thaddeus, P.\ 2001, \apj, 547, 792

\bibitem[Dong et al.(2018a)]{Don2018a} Dong, J., Fu, L., Liu, Q., \& Shen, Z. 2018, Experimental Astronomy, 45, 397

\bibitem[Dong et al.(2016)]{Don2016} Dong, J., Wu, Y.~J., Yuan, J., et al.\ 2016, Progress in Astronomy, 34, 212

\bibitem[Dong et al.(2018b)]{Don2018b} Dong, J., Zhong, W., Wang, J., Liu, Q., \& Shen, Z. 2018, IEEE Transactions on Antennas and Propagation, 66, 2044

\bibitem[Ellingsen(2006)]{Ell2006} Ellingsen, S.~P.\ 2006, \apj, 638, 241

\bibitem[Ellingsen(2007)]{Ell2007} Ellingsen, S.~P.\ 2007, \mnras, 377, 571

\bibitem[Ellingsen et al.(1996)]{Ell1996} Ellingsen, S.~P., von Bibra, M.~L., McCulloch, P.~M., et al.\ 1996, \mnras, 280, 378

\bibitem[Green et al.(2010)]{Gre2010} Green, J.~A., Caswell, J.~L., Fuller, G.~A., et al.\ 2010, \mnras, 409, 913

\bibitem[Green et al.(2012)]{Gre2012} Green, J.~A., Caswell, J.~L., Fuller, G.~A., et al.\ 2012, \mnras, 420, 3108

\bibitem[Hunter et al.(2017)]{Hun2017} Hunter, T.~R., Brogan, C.~L., MacLeod, G., et al.\ 2017, \apjl, 837, L29

\bibitem[Hunter et al.(2018)]{Hun2018} Hunter, T.~R., Brogan, C.~L., MacLeod, G.~C., et al.\ 2018, \apj, 854, 170

\bibitem[Inayoshi et al.(2013)]{Ina2013} Inayoshi, K., Sugiyama, K., Hosokawa, T., Motogi, K., \& Tanaka, K.~E.~I.\ 2013, \apjl, 769, L20

\bibitem[Law et al.(2008)]{Law2008} Law, C.~J., Yusef-Zadeh, F., \& Cotton, W.~D.\ 2008, \apjs, 177, 515

\bibitem[MacLeod et al.(1992)]{Mac1992} MacLeod, G.~C., Gaylard, M.~J., \& Nicolson, G.~D.\ 1992, \mnras, 254, 1P

\bibitem[Maswanganye et al.(2015)]{Mas2015} Maswanganye, J.~P., Gaylard, M.~J., Goedhart, S., Walt, D.~J.~v.~d., \& Booth, R.~S.\ 2015, \mnras, 446, 2730

\bibitem[Menten(1991)]{Men1991} Menten, K.~M.\ 1991, \apjl, 380, L75

\bibitem[Meyer et al.(2017)]{Mey2017} Meyer, D.~M.-A., Vorobyov, E.~I., Kuiper, R., \& Kley, W.\ 2017, \mnras, 464, L90

\bibitem[Minier et al.(2003)]{Min2003} Minier, V., Ellingsen, S.~P., Norris, R.~P., \& Booth, R.~S.\ 2003, \aap, 403, 1095

\bibitem[Olmi et al.(2014)]{Olm2014} Olmi, L., Araya, E.~D., Hofner, P., et al.\ 2014, \aap, 566, A18 

\bibitem[Pandian et al.(2007)]{Pan2007} Pandian, J.~D., Goldsmith, P.~F., \& Deshpande, A.~A.\ 2007, \apj, 656, 255 

\bibitem[Parfenov \& Sobolev(2014)]{Par2014} Parfenov, S.~Y., \& Sobolev, A.~M.\ 2014, \mnras, 444, 620

\bibitem[Pestalozzi et al.(2005)]{Pes2005} Pestalozzi, M.~R., Minier, V., \& Booth, R.~S.\ 2005, \aap, 432, 737

\bibitem[Reid et al.(2016)]{Reid2016} Reid, M.~J., Dame, T.~M., Menten, K.~M., \& Brunthaler, A.\ 2016, \apj, 823, 77

\bibitem[Reid et al.(2009)]{Reid2009} Reid, M.~J., Menten, K.~M., Zheng, X.~W., et al.\ 2009, \apj, 700, 137

\bibitem[Reid et al.(2014)]{Reid2014} Reid, M.~J., Menten, K.~M., Brunthaler, A., et al.\ 2014, \apj, 783, 130

\bibitem[Sanna et al.(2015)]{San2015} Sanna, A., Menten, K.~M., Carrasco-Gonz{\'a}lez, C., et al.\ 2015, \apjl, 804, L2

\bibitem[Schuller et al.(2009)]{Sch2009} Schuller, F., Menten, K.~M., Contreras, Y., et al.\ 2009, \aap, 504, 415 

\bibitem[Szymczak et al.(2018)]{Szy2018} Szymczak, M., Olech, M., Wolak, P., G{\'e}rard, E., \& Bartkiewicz, A.\ 2018, \aap, 617, A80

\bibitem[Szymczak et al.(2012)]{Szy2012} Szymczak, M., Wolak, P., Bartkiewicz, A., \& Borkowski, K.~M.\ 2012, Astronomische Nachrichten, 333, 634

\bibitem[Urquhart et al.(2009)]{Urq2009} Urquhart, J.~S., Hoare, M.~G., Purcell, C.~R., et al.\ 2009, \aap, 501, 539

\bibitem[Urquhart et al.(2014)]{Urq2014} Urquhart, J.~S., Moore, T.~J.~T., Csengeri, T., et al.\ 2014, \mnras, 443, 1555 

\bibitem[Urquhart et al.(2013)]{Urq2013} Urquhart, J.~S., Thompson, M.~A., Moore, T.~J.~T., et al.\ 2013, \mnras, 435, 400 

\bibitem[van der Walt et al.(2009)]{Wal2009} van der Walt, D.~J., Goedhart, S., \& Gaylard, M.~J.\ 2009, \mnras, 398, 961

\bibitem[van der Walt et al.(2016)]{Wal2016} van der Walt, D.~J., Maswanganye, J.~P., Etoka, S., Goedhart, S., \& van den Heever, S.~P.\ 2016, \aap, 588, A47

\bibitem[van der Walt et al.(1996)]{van1996} van der Walt, D.~J., Retief, S.~J.~P., Gaylard, M.~J., \& MacLeod, G.~C.\ 1996, \mnras, 282, 1085

\bibitem[Voronkov et al.(2014)]{Vor2014} Voronkov, M.~A., Caswell, J.~L., Ellingsen, S.~P., Green, J.~A., \& Breen, S.~L.\ 2014, \mnras, 439, 2584

\bibitem[Voronkov et al.(2010)]{Vor2010} Voronkov, M.~A., Caswell, J.~L., Ellingsen, S.~P., \& Sobolev, A.~M.\ 2010, \mnras, 405, 2471

\bibitem[Walsh et al.(1997)]{Wal1997} Walsh, A.~J., Hyland, A.~R., Robinson, G., \& Burton, M.~G.\ 1997, \mnras, 291, 261

\bibitem[Wright et al.(2010)]{Wright2010} Wright, E.~L., Eisenhardt, P.~R.~M., Mainzer, A.~K., et al.\ 2010, \aj, 140, 1868-1881

\bibitem[Xu et al.(2008)]{Xu2008} Xu, Y., Li, J.~J., Hachisuka, K., et al.\ 2008, \aap, 485, 729

\bibitem[Yang et al.(2017)]{Yang2017} Yang, K., Chen, X., Shen, Z.~Q., et al.\ 2017, \apj, 846, 160 (Paper \uppercase\expandafter{\romannumeral1})

\end{thebibliography}
\end{document}